\documentclass[fleqn,usenatbib]{mnras}

\usepackage{newtxtext,newtxmath}

\usepackage[T1]{fontenc}

\DeclareRobustCommand{\VAN}[3]{#2}
\let\VANthebibliography\thebibliography
\def\thebibliography{\DeclareRobustCommand{\VAN}[3]{##3}\VANthebibliography}

\usepackage{comment} 

\usepackage{graphicx}	
\usepackage{amsmath}	

\usepackage{svg}
\usepackage{ulem}
\usepackage{subcaption} 

\newcommand\lgal{\textsc{L-Galaxies}}
\newcommand\secref[1]{Sect.~\ref{#1}}
\newcommand\figref[1]{Fig.~\ref{#1}}
\newcommand\figsref[1]{Figs.~\ref{#1}}
\newcommand\equref[1]{Eq.~\eqref{#1}}

\newcommand\appref[1]{Appendix~\ref{#1}}
\newcommand\sfg{SFG}
\newcommand\qg{QG}
\newcommand\smf{SMF}
\defcitealias{Henriques2015}{H15}
\defcitealias{Henriques2020}{H20}
\defcitealias{Ayromlou2021}{A21}

\newcommand\hXV{\citetalias{Henriques2015}}
\newcommand\hXX{\citetalias{Henriques2020}}
\newcommand\aXXI{\citetalias{Ayromlou2021}}

\title[\textsc{L-Galaxies} and galaxy scaling relations]{Probing galaxy evolution from $z=0$ to $z\simeq10$ through galaxy scaling relations in three \textsc{L-Galaxies} flavours}
\author[Vani et al.]{%
Akash Vani$^{1,2}$ \thanks{E-mail: vani@mpa-garching.mpg.de},
Mohammadreza Ayromlou$^{3,4}$, Guinevere Kauffmann$^{1}$, and Volker Springel$^{1}$
\\%
$^{1}$Max Planck Institute for Astrophysics, Karl-Schwarzschild-Str. 1, 85741 Garching bei München, Germany\\%
$^{2}$Ludwig-Maximilians-Universität München, Geschwister-Scholl-Platz 1, 80539 München, Germany \\%
$^{3}$Universität Heidelberg, Zentrum für Astronomie, Institut für theoretische Astrophysik, Albert-Ueberle-Str. 2, 69120 Heidelberg, Germany\\
$^{4}$Argelander-Institut f\"ur Astronomie, Auf dem H\"ugel 71, D-53121 Bonn, Germany\\
}

\date{Accepted XXX. Received YYY; in original form ZZZ}

\pubyear{2024}

\begin{document}
\label{firstpage}
\pagerange{\pageref{firstpage}--\pageref{lastpage}}
\maketitle

\begin{abstract}
We present a comprehensive examination of the three latest versions of the \textsc{L-Galaxies} semi-analytic galaxy formation model, focusing on the evolution of galaxy properties across a broad stellar mass range ($10^7\:{\rm M}_{\odot}\lesssim{M_\star}\lesssim10^{12}\:{\rm M}_{\odot}$) from $z=0$ to $z\simeq10$. This study is the first to compare predictions of \lgal{} with high-redshift observations well outside the original calibration regime, utilising multiband data from surveys such as SDSS, CANDELS, COSMOS, HST, JWST, and ALMA. We assess the models' ability to reproduce various time-dependent galaxy scaling relations for star-forming and quenched galaxies. Key focus areas include global galaxy properties such as stellar mass functions, cosmic star formation rate density, and the evolution of the main sequence of star-forming galaxies. Additionally, we examine resolved morphological properties such as the galaxy mass-size relation, alongside core $(R<1\,{\rm{kpc}})$ and effective $(R<R_{\rm{e}})$ stellar-mass surface densities as a function of stellar mass. This analysis reveals that the \textsc{L-Galaxies} models are in qualitatively good agreement with observed global scaling relations up to $z\simeq10$. However, significant discrepancies exist at both low and high redshifts in accurately reproducing the number density, size, and surface density evolution of quenched galaxies. These issues are most pronounced for massive central galaxies, where the simulations underpredict the abundance of quenched systems at $z\geq1.5$, reaching a discrepancy of a factor of 60 by $z\approx3$, with sizes several times larger than observed. Therefore, we propose that the physical prescriptions governing galaxy quenching, such as AGN feedback and processes related to merging, require improvement to be more consistent with observational data.


\end{abstract}

\begin{keywords}
galaxies: formation –- galaxies: evolution -– galaxies: high-redshift -–
methods: analytical -– methods: numerical
\end{keywords}



\section{Introduction}
In recent years, significant advancements in telescope and detector technologies, coupled with improved computational capabilities, have led to significant progress in our understanding of galaxy formation and evolution.
Observations of galaxies in large surveys in the early 2000s allowed us to characterise global galaxy properties such as stellar mass 
\citep[e.g.,][]{Cole2001SMF, Kauffmann2003,Kauffmann2003_SFH_internalstruct, Bell2003_stellarmass, Drory2009, Marchesini2009, Peng2010, Whitaker2011_UVJ, Fontana2014} and
star formation rates \citep[e.g.,][]{Lilly1996_SFR, Madau1996_SFRH, Madau1998_SFRH, Kennicutt1998, BelldeJong2001, Brinchmann2004, Elbaz2007, Daddi2007, Santini2009, Evans2009_Spitzer, Speagle2014} for samples
of hundreds of thousands of galaxies at low redshifts and to study global galaxy scaling relations. More recent observations have allowed us not only to characterise the same global galaxy properties at high redshifts \citep[e.g.,][]{Muzzin2013, Ilbert2013_nuvrj,  Schreiber2015_SFMS, Stefanon2021_SMF, Santini2022, Weaver2023, Harikane2022_SFRH, Harikane2023_SFRH, Harikane2024_SFRH} but also resolved properties such as morphology and size, stellar surface mass density and star formation rate profiles as well as deviations from axisymmetry in these systems \citep[e.g.,][]{Wuyts2013, vanderWel2014b, SchreiberandWuyts2020_main, Tacconi2020GMF, Huertas-Company2024_Morphology, Conselice2024_morphology}. Recent studies have further delved into local Galactic properties, investigating different evolutionary processes such as stellar content, radial migration, star formation and chemical enrichment histories \citep[e.g.,][]{Snaith2015, Sysoliatina2021_MWmodel, Katz2023, GaiaDR32023, Golovin2023_CNS}, thereby providing a comprehensive view of galaxy evolution on all scales and epochs.

Galaxies have structure over a very wide range of physical scales, and the physical processes that determine these structures can be quite different on small and large scales. The field has seen the development of a variety of simulation techniques \citep[see][for a review]{Somerville2015_review} such as semi-analytical models \citep[e.g.,][]{Kauffmann1993, Somerville1999, DeLucia2007, Just_Jahreiss_2010, Guo2011, Croton2016_SAGE, Lagos2018_SHARK, Cora2018_SAGSAM, Henriques2020},  cosmological hydrodynamical models \citep[e.g.,][]{Vogelsberger2014, Vogelsberger2020_review,  Hopkins2014FIRE_HDsim, Hopkins2018, Dubois2014_horizonAGN_HDSim, Schaye2015, Schaye2023_FLAMINGO, McCarthy2017_BAHAMAS_HDsim, Pillepich2018, Villaescusa-Navarro2021_CAMELS, Bhagwat2024_SPICE} and zoom-in simulations of galaxies that aim
to resolve the detailed internal structure of galaxies
\citep[e.g.,][]{Naab2014, Wang_2015_NIHAO, Grand2017, Grand2024, Tremmel2019_ROMULUSC, Dubois2021_NEWHORIZON, Wetzel2023_FIRE2}.

The progress of these simulations has enabled the creation of highly accurate representations of the observable universe within the  $\Lambda{}\text{CDM}$ cosmological framework \citep{Planck_2018_2020}. Such recent successes are a testament to the realistic nature of these simulations as they now start to be able to reproduce not only global galaxy properties but also how the material is distributed within galaxies \citep[e.g.,][]{Pillepich2019IllTNG50, Henriques2020}.

To interpret new observations from the most modern galaxy surveys and telescopes, such as the Dark Energy Spectroscopic Instrument  \citep[DESI, ][]{DESI2016}, the Sloan Digital Sky Survey-V \citep[SDSS-V, ][]{Kollmeier2017_SDSS_V}, the 4-metre Multi-Object Spectroscopic Telescope \cite[4MOST, ][]{deJong2019_4MOST}, the JWST Advanced Deep Extragalactic Survey \citep[JADES, ][]{Eisenstein2023_JADES} with the James Webb Space Telescope (JWST), and the Cosmic Evolution Early Release Science Survey \citep[CEERS, ][]{Finkelstein2023_CEERS}, theoretical models based either on hydrodynamical or semi-analytical models must refine their capabilities. This is crucial for maintaining consistency with the local and global galaxy properties of observations throughout the evolution of time.

While offering high accuracy and realism, cosmological hydrodynamical simulations demand substantial computational resources, making it difficult to explore large volumes at high resolution or navigate effectively the multidimensional parameter space of uncertainties in the physical treatment of processes such as star formation, supernova (SN) and active galactic nuclei (AGN) feedback \citep{Pillepich2019IllTNG50, Nelson2019_IllustrisTNG50, Bhagwat2024_SPICE}. These challenges are addressed by semi-analytical models, which significantly accelerate the model's execution time by effectively modelling key galaxy evolution physical processes \citep{Kauffmann1993, Somerville1999} in parameterised form instead of solving the equations of hydrodynamics for very large systems of resolution elements.

A semi-analytical approach to galaxy evolution incorporates fundamental baryonic processes that govern galaxy formation and evolution such as the infall of the gas into the dark matter haloes, gas cooling, star formation and stellar feedback, chemical evolution, supermassive black hole (SMBH) formation and growth,  AGN feedback, and so on. This method builds upon the foundation of dark matter halo merger trees, which are either constructed through Monte Carlo realisations or derived from particle data of dark matter-only simulations \citep[e.g.,][]{Kauffmann1993, Kauffmann1999, Somerville1999, Springel2001Subfind, DeLucia2006}. Lacking a direct implementation of hydrodynamical processes, semi-analytical models realise a parameterised modelling of gas-physical processes on top of dark matter halo merger trees, necessitating analytical approximations. Consequently, the development of such models relies on various simplifying assumptions that can introduce modelling uncertainties in the outcomes. For example, earlier semi-analytical models have assumed that the entire hot gas reservoir of a galaxy is stripped upon crossing the virial radius of the halo \citep[e.g.,][]{Kauffmann1999, Croton2006}. In contrast, modern semi-analytical models, like the \lgal{} model, also known as the Munich model \citep{Ayromlou2021}, propose a gradual and time-dependent gas stripping mechanism, enabling satellite galaxies to retain a portion of their gas for extended periods.

Galaxies are generally classified into two groups: star-forming galaxies (\sfg{}s) and passive or quiescent galaxies (\qg{}s), based on their star formation activity. Various quenching mechanisms that can turn a galaxy into a passive system have been identified, including intrinsic quenching due to stellar and AGN feedback \citep[][]{Kauffmann2003, DiMatteo2005, Elbaz2007, Beckmann2017}, and environmental quenching \citep[][]{Gunnand_gott1972, Springel2005_mergers, Peng2010, Peng2015}. The diverse array of galaxies, varying in size, morphology, and other characteristics, plays an essential role in our empirical understanding of galaxy evolution \citep[e.g.,][]{Conselice2014_review, Naab2017_newview}. Notably, the study of galaxy scaling relations, reflecting the complex interplay between attributes like mass, size, luminosity, colour, star formation rate, and several other galaxy properties, has long been a fundamental cornerstone in comprehending the complex interplay between various galaxy attributes as a function of cosmic time \citep[see ][for a review]{Stark2016, SchreiberandWuyts2020_main, DOnofrio2021}. Such correlations along with their slope, zero point and scatter \citep[e.g.,][]{Stone2021, Popesso2023} can help us uncover the complex physics behind the formation and evolution of \sfg{}s and \qg{}s. Moreover, galaxy scaling relations can be used to inform and enrich new generations of galaxy formation models. 

This work provides a comprehensive and contemporary examination of the three recent versions or flavours of the \lgal{} semi-analytic model. These correspond to the most recent and independently calibrated versions of the \lgal{} framework, namely: \citet{Henriques2015}, \citet{Henriques2020}, and \citet{Ayromlou2021}, all of which were designed as enhancements over their corresponding predecessors. 

Our objective is to conduct an extensive comparison of the results generated by these models across a wide range of masses and redshifts, to see how well
they agree with recent observational scaling relations, particularly at high redshift ($z\gtrsim2$). 
Our focus includes exploring the stellar mass functions, the cosmic star formation density, the main sequence of \sfg{}s (i.e., the ${\rm SFR}-M_{\star}$) relation, the galaxy mass-size relation and the evolution of the core ($R<1\:{\rm kpc}$) and effective ($R<R_{\rm e}$) stellar mass surface density. These relations are analysed for both \sfg{}s and \qg{}s independently. While some of these relations have been thoroughly studied at low redshift \citep{Croton2016_SAGE, Lagos2018_SHARK, Ayromlou2021}, a comprehensive exploration across a wide mass and redshift range is yet to be realised for a semi-analytical model implementation. Furthermore, we supplement our analysis with an examination of baryonic processes that influence galaxy quenching, using the latest version of the \lgal{} model \citep[][]{Ayromlou2021}. This includes an analysis of the distribution of baryons and gas content within haloes, as well as the relative distribution of black holes, hot gas, and cold gas in quenched and star-forming galaxies. This comprehensive comparative analysis serves a dual purpose: not only does it emphasise important benchmarks of our current understanding of the \lgal{} model, but it also pinpoints the specific areas where model enhancements are needed. By extending our investigation to encompass a broader range of mass and redshift, this work identifies key areas for improvement. 
The refinement and further development of the \lgal{} semi-analytical model are planned for future projects, which will be informed by the findings of this study. These efforts will focus on addressing the discrepancies identified here and improving the implementation of physical processes in the model to achieve better consistency with observational data.

In this paper, we assume a $\Lambda{}\text{CDM}$ cosmology with $\Omega_{\Lambda}=0.69$, $\Omega_{\rm b}=0.049$, $\Omega_{\rm m}=0.315$ and $\mathrm{H_{0}=67.3\;km\;s^{-1}\;Mpc^{-1}}$ \citep{Planck2016}. All magnitudes are in the AB photometric system \citep{OkeandGunn1983_ABphoto}, and a \citet{Chabrier2003_IMF} initial mass function (IMF) is used throughout this analysis unless stated otherwise.

This paper is organised as follows. Section~\ref{Sec:Observation_summary} summarises the different observational low and high redshift galaxy scaling relations used in this work. In \secref{Sec:Model}, we outline the framework of the \lgal{} semi-analytical and briefly describe the key baryonic processes relevant to this study. Section~\ref{Sec:Results} presents the comparison of the model results with the observational galaxy scaling relations for star-forming and quenched systems. Section~\ref{Sec:Baryonic_features} focuses on baryonic physics which controls galaxy quenching and \secref{sec:discussion} offers a discussion of these findings and similar works. Lastly, \secref{sec:summary_conclusion} summarises our study's results and presents our conclusions.

\section{Summary of Observational Data Compilations}
\label{Sec:Observation_summary}

In this section, we provide a concise overview of the low to high-redshift observational metrics and galaxy scaling relations explored in this paper. These correlations serve as key benchmarks, offering insights into the physical processes within galaxies and guiding the refinement of galaxy evolution models.

\subsection{Stellar mass functions}

The galaxy stellar mass function (\smf{}) is defined as the number of galaxies per unit volume in stellar mass bins at a given redshift.
Understanding its evolution helps us understand various effects of physical processes that govern galaxy evolution \citep[e.g.,][]{Li2009, Peng2010}. Much progress has been made recently in our understanding of this relation thanks to deep surveys probing the stellar mass function down to lower masses and higher redshifts  \citep[e.g.,][]{Ilbert2013_nuvrj, Muzzin2013, Tomczak2014, McLeod2021_SMF, Santini2022, Weaver2023}.

Galaxies can generally be classified as star-forming galaxies (\sfg{}s) which boast active star formation, or quiescent galaxies  (\qg{}s) which have little to no star formation activity. The quenching mechanism of a galaxy can be due to the heating or removal of gas by an AGN or SN feedback. This is sometimes called ``mass quenching'' because of the strong mass-dependent effects of these two processes. Removal of gas by SN-driven winds is more efficient in low-mass galaxies, whereas the most massive black holes (BH) that give rise to the most powerful AGN are located in the most massive galaxies.  Alternatively, quenching can be due to environmental effects such as mergers, tidal interactions and ram-pressure stripping \citep[see ][]{Peng2010, Wetzel2013, ManandBelli2018_sf_quenching}.

To gain deeper insights into how diverse processes impact the structure and evolution of the \smf{}, we analyse the star-forming galaxy stellar mass function (\sfg{} \smf{}), the quiescent galaxy stellar mass function (\qg{} \smf{}) and the combined representation of the two, the total \smf{}. 

Studies have found that the galaxy stellar mass function is well described by the empirical Schechter function \citep{Schechter1976} with an exponential downturn at a characteristic mass and a power law at low masses with a slope that has little redshift evolution \citep{Ilbert2013_nuvrj, Tomczak2014, Weaver2023}. However, the influence of physical processes on the shape of the \smf{} is still unclear. \citet{Peng2010} suggest that a single Schechter shape is due to environmental effects on galaxies in the local universe, while a double Schechter curve can arise due to a combination of both, environmental and mass quenching processes. At $z\gtrsim7$, deviations from the Schechter function have been observed, with the \smf{} being better described by a power law \citep[e.g., ][]{Stefanon2021}. Understanding the evolution of the observed \smf{} is critical, as it represents one of the most reliable observables used in galaxy formation simulations. The \lgal{} models, for example, employ the \smf{} at $z=0$, $1$, and $2$ for calibration purposes.

In this work, we utilise the observed \smf{} from \citet{Weaver2023}, where the authors employ the COSMOS2020 photo-$z$ catalogue \citep{Weaver2022COSMOS2020} to present an extensive study on the assembly and quenching of star formation in galaxies. This work provides parametrisations of the shape and evolution of the total \smf{} up to $z\sim7.5$, and \sfg{} and \qg{} \smf{} up to $z\sim5.5$. Galaxies are divided into these two classes using the NUVrJ colour-colour selection criterion \citep{Ilbert2013_nuvrj}. The study also highlights challenges in measuring rest frame J-band fluxes beyond $z\sim3$ in Spitzer/IRAC bands, making these measurements reliant on the modelling. By $z\sim5$, the rest-frame r-band requires extrapolation, making it statistically difficult to separate \qg{}s and dusty \sfg{}s. Therefore, it is crucial to acknowledge that the selection of \qg{}s between $3 < z \lesssim 5.5$ is fraught with a higher degree of uncertainty.

The \citet{Weaver2023} study reports a smooth, monotonic evolution in the galaxy \smf{} at all masses since $z \sim7.5$ with consistent growth in the number of star-forming systems (since $z\sim5.5$) and a shift towards a larger number of \qg{} systems at lower redshifts. The \smf{} of the quenched systems also illustrates a buildup of low mass galaxies up to $z\sim1.5$, along with a higher abundance of massive quenched systems at lower redshifts compared to previous reports \citep{Ilbert2013_nuvrj, Davidzon2018}. While the \sfg{}s display a double Schechter form at lower redshifts, flattening into a power-law-like form at higher redshifts, the \qg{}s show a more rapid decline in their normalisation and a steepening low-mass end slope over time.

It is important to note that the derived value of the low mass end of the \smf{}s, especially for the quenched systems, can be in error because of mass incompleteness. Additionally, the \citet{Weaver2023} study also addresses the presence and evolution of low-mass quiescent systems at redshifts $z > 1.5$. Although COSMOS2020 data suggests that the low mass \qg{}s appear to effectively vanish at $z > 1.5$, evidence from \citet{Santini2022} points to the existence of $M_\star < 10^9 \, {\rm M}_{\odot}$ quiescent systems beyond this redshift. Their study is from a significantly smaller effective area compared to the COSMOS2020 study. 
We note that the total \smf{} by \citet{Weaver2023} is consistent with previous work done by \citet{Ilbert2013_nuvrj, Tomczak2014, Grazian2015} and \citet{Stefanon2021_SMF}. It has also been compared with predictions from a number of cosmological galaxy formation simulations, for example, EAGLE \citep{Furlong2015_EAGLE}, SHARK \citep[][]{Lagos2018_SHARK} and Illustris TNG \citep{Pillepich2018}.

\subsection{Cosmic star formation rate density}
The cosmic SFR density (CSFRD) quantifies the mass of stars formed per unit volume and unit time at a given redshift. It serves as a fundamental metric for understanding the underlying processes regulating star formation and galactic evolution. This metric's significance lies in its direct linkage to the stellar mass and the gas reservoirs within galaxies. In this work, we compare our findings to those from \citet{Madau2014}, which used data from various publications spanning 2006-2013, derived from different instruments from $z\sim 0$ up to ${z\sim 8}$. Additionally, we incorporate insights from \citet{Harikane2022_SFRH}, who employ data from the Subaru/Hyper Suprime-Cam survey \citep{Aihara2018_HSC} and the CFHT Large Area U-band Survey \citep{Sawicki2019_CFHT}, focusing on $z=2-7$.

From both studies, a consistent picture emerges where the SFR density peaks approximately at redshift $z\sim 1.9$ before entering an exponential decline. It should be noted that the exact slope of the high redshift decline is still debated \citep[see][]{Bouwens2012_SFH, Bouwens2020_CSFRD, Rowan-Robinson2016_SFH, Driver2018_SFH, Donnan2023_UV_JWST, Harikane2023_SFRH, Harikane2024_SFRH, Kim2024_SFH}.
\citet{Harikane2022_SFRH} highlight that the star formation efficiency, i.e.~the rate at which galaxies transform gas into stars, exhibits minimal evolution across a broad redshift range. This finding suggests that the evolution of the CSFRD is predominantly influenced by the increase in halo number density due to structure formation and the diminishing accretion rate attributed to cosmic expansion. Furthermore, considering the elevated number density of galaxies and the high SFR density for $z >10$, \citet{Harikane2024_SFRH} suggests that this might be indicative of high-efficiency star formation regimes, AGN activity, and a top-heavy IMF possibly from Population-III like stars.

\subsection{Star-forming main sequence galaxies}
The main sequence of star-forming galaxies refers to the tight relation between the star formation rate (SFR) and the stellar mass (${M_\star}$) of galaxies. This relationship is one of the important galaxy scaling relations, and the evolution of its slope and scatter has been extensively studied over the last decade \citep[see ][]{Speagle2014, Popesso2023}.
A consensus among various studies suggests a power law relationship, ${\rm SFR}\propto M_{\star}^{\alpha}$ consistent across both low and high redshifts \citep{Peng2010, Speagle2014}. However, other recent works indicate a deviation from this trend, specifically a bending at the high mass end across different redshifts \citep{Lee2015, Leja2022SFRM, Tomczak2016, Popesso2023, Merida2023_MSSFR}.
This relation can potentially help us understand the physics governing star formation and its cessation, shedding light on processes such as cold gas accretion, feedback from AGN, and the influence of galaxy interactions and mergers on SFR \citep{Pontzen2017_AGN_Merger_SF}.

In this work, we draw primarily on the findings from \citet{Popesso2023}, who present a comprehensive analysis of the evolution of the main sequence over a wide range of redshifts (${0\lesssim z\lesssim6}$) and stellar masses ${10^{8.5}\:\lesssim {\rm M}_\star/M_{\odot}\:\lesssim10^{11.5}}$. The study compiles data from numerous papers between 2014-2022, standardising the calibrations between observables, such as the stellar mass and SFR, to ensure consistency across different datasets. This standardisation allows for the quantification of variations in the MS shape and normalisation over a broad interval of cosmic time. The results reveal similar curvature towards massive galaxies at all redshifts, with the MS shape primarily influenced by the turnover mass, which exhibits marginal temporal evolution, leading to a slightly steeper MS towards ${z\sim4-6}$. Below the turnover mass, the specific star formation rate (sSFR) is nearly constant for all galaxy masses, whereas above this mass the sSFR is suppressed. This study further argues that the MS bending over with time is caused by the reduced availability of cold gas in haloes entering the hot accretion phase where BHs can operate more efficiently.

Additionally, we reference the work of \citet{Speagle2014} to provide a broader perspective. This earlier study compiles various literature sources between 2007-2014 and finds a log-linear relationship between the SFR and galaxy stellar mass. According to \citet{Popesso2023}, the linear trend identified in \citet{Speagle2014} could be attributed to the limited mass range used in their analysis. Moreover, measuring the SFR in high-mass galaxies also presents several challenges, such as dust attenuation \citep{Boselli2009_HMGalaxies}, variations in star formation efficiency, and influences from morphology and environmental factors \citep{Elbaz2007}.

\subsection{Galaxy mass-size relations}
Galaxy size, commonly represented by the half-light radius or effective radius (${R_{\rm e}}$), which encompasses half of the galaxy's total emitted flux, is another key property reflecting a galaxy's evolutionary history and its connection to its dark matter halo \citep{Kormendy1977Kormendy, Mo1998, Kravtsov2013}. The degree to which angular momentum is conserved when the gas cools and condenses within dark matter haloes and the efficiency with which stars form within a galaxy significantly influence its size evolution. It was initially thought that passive galaxies exhibit minimal growth in both size and mass, primarily evolving through the ageing of their stellar populations, but modern theories predict more substantial size growth due to so-called dry (gas-poor) major and minor mergers \citep[e.g., ][]{Naab2009}. Conversely, gas-rich ``wet'' mergers can trigger starburst events resulting in dense central cores of stars in the remnant galaxies  \citep{Hernquist1989, Robertson2006}.

The relationship between size and mass, often expressed as ${R_{\rm e}\propto M^{\alpha}}$, has been subject to extensive study across various epochs for both \sfg{}s and \qg{}s. For example, \citet{Shen2003} used Sloan Digital Sky Survey \citep[SDSS,][]{York2000_SDSS} data for galaxies in the local universe to derive a size-mass relations for late-type and early-type galaxies. Extending the study using galaxies with $M_\star>10^9 \, {\rm M}_{\odot}$ up to redshifts of $z=3$ with data from CANDELS \citep{Grogin2011_CANDELS, VanDerWel2012} and 3D-HST data \citep{Brammer2012_3dHST}, \citet{VanDerWel2014_MSR} found consistency in the slopes of the size-mass relations with those reported for local early and late-type galaxies. However, the galaxy sizes at higher redshifts ($z\sim3$) were around 2 and 4 times smaller for late and early types, respectively. This suggests that
the sizes of galaxies evolve with time. These findings have been corroborated and further expanded upon by studies such as \citet{Mosleh2012, Morishita2014, Allen2017, Mowla2019, Nedkova2021, Ormerod2024_masssize_JWST, Ward2024, vanderWel2024_JWST_MSR}. Furthermore, it is important to note that galaxy size measurements can be significantly biased by the wavelength of observation \citep[e.g., ][]{Vulcani2014, Suess2019, Martorano2023_MSR, Nedkova2024arX_UVsizes, Nedkova2024arXiv_bulge_disc}.

In this work, we primarily utilise the galaxy mass-size relation provided by \citet{VanDerWel2014_MSR}, which employs the GALFIT package \citep{Peng2010_GALFIT} for fitting single-component S\'ersic profiles to two-dimensional light distributions, alongside using custom PSF models \citep{VanDerWel2012}. The authors provide rest-frame $5000\,\text{\AA}$ sizes by applying corrections for the wavelength dependency of galaxy sizes. The work by \citet{VanDerWel2014_MSR} spans a redshift range from $ 0< z<3$. The study underscores that early-type or quenched galaxies consistently exhibit smaller sizes compared to their late-type or star-forming counterparts across all redshifts. Additionally, the rate at which galaxy sizes evolve is significantly different between these two classes. Quenched systems exhibit a rapid growth proportional to $(1+z)^{-1.48}$, indicative of substantial structural transformations during their evolution. Conversely, the \sfg{}s display a more modest size evolution, proportional to $(1+z)^{-0.75}$. Moreover, \qg{}s are characterised by a steep size-mass relationship, where ${R_{\rm e} \propto M_\star^{0.75}}$, in contrast to the shallower slope of ${R_{\rm e} \propto M_\star^{0.22}}$ observed in \sfg{}s.  The study also notes an increase in the number density of compact, massive early-type galaxies from $z=3$ to $z=1.5-2$, followed by a subsequent decrease, suggesting early formation and rapid evolutionary changes. On the other hand, the size distribution of late-type galaxies is less skewed, with more gradual changes likely driven by steady gas accretion and ongoing star formation \citep{Kauffmann2006, Delgado2015}. To supplement \citet{VanDerWel2014_MSR}'s findings, we also incorporate the results from \citet{Mowla2019}, which apply a similar methodology to the COSMOS-DASH dataset.

\subsection{Central stellar mass surface density}
\label{sec:SMSD}

Stellar mass surface density ($\Sigma$) serves as a critical spatially resolved parameter, quantifying the amount of stellar mass per unit area within galaxies and offering insights into their formation histories and structural dynamics \citep[e.g., ][]{Cheung2012_SMSD, Barro2017, Whitaker2017_MSR}. 
This highlights two formation scenarios; one suggests compaction \citep[e.g., ][]{Naab2009} which in turn feeds the AGN and the other suggests an inside-out growth mechanism \citep[e.g., ][]{VanDerWel2012, Patel2013, Woo2019}.
Thus, this measure is important for understanding the dynamics of star formation and the relative impact of in situ versus ex-situ mass assembly, and for distinguishing between different quenching mechanisms within galaxies \citep{Conselice2014_review, Conselice2024_morphology}.
High $\Sigma$ values correlate with the development of a galaxy's core or bulge, which in turn influences the activity of SMBHs that regulate star formation through AGN feedback mechanisms \citep{Fang2013, Parsotan2021}. Several studies indicate that \qg{}s typically exhibit higher $\Sigma$, are more compact and have a greater prevalence of AGN compared to their star-forming counterparts \citep{vanDokkum2008, Cheung2012_SMSD, Tacchella2016, Barro2017}. Additionally, these galaxies with higher $\Sigma$ tend to remain quiescent across all redshifts \citep[e.g., ][]{Williams2010}, reinforcing the significance of central density measurements in tracing evolutionary pathways.

The analysis of $\Sigma$ often focuses on either the core 1~kpc region or the effective radius, termed as core 1 kpc stellar densities ($\Sigma_{\rm 1\: kpc}$) or effective densities ($\Sigma_{R_{\rm e}}$), respectively, where both form a log-linear relationship with the galaxy's stellar mass. Some studies such as \citet{Fang2013, Tacchella2016, Barro2017, Woo2019, Suess2021, Abdurrouf2023} and \citet{Matharu2024} advocate the use of $\Sigma_{\rm 1\: kpc}$ due to its lower susceptibility to observational biases and its strong correlation with SMBH properties and core concentration \citep{Bezanson2009, Szomoru2012, McDermid2015, Graham2013, vanderWel2014b}, while other works focus on the effective densities \citep[e.g., ][]{Kim2018, Clausen2024, Ji2024}, which is representative of the central bulge region.

Here, we use the results from \citet{Barro2017}, who established the scaling relations for the core $1\,{\rm kpc}$ and $R_{\rm e}$ stellar mass surface density of galaxies, defined as $\Sigma_{\rm 1\: kpc} = M_\star(<1\,{\rm kpc})/(\pi(1\,{\rm kpc})^2)$ and $\Sigma_{R_{\rm e}}=0.5\,M_\star(<R_{\rm e})/(\pi R^2_{\rm e})$, respectively. They analysed data from the HST and Spitzer/IRAC selected catalogue for the CANDELS GOODS-S field \citep{Guo2013_CANDLES} across a broad mass range and redshifts ($0.5\lesssim z\lesssim3$). Their analysis indicates that while the best-fit slopes of the $\Sigma \propto M_\star$ relation remain constant over time, the zero points show a temporal evolution.
For \sfg{}s, the zero point of both $\Sigma_{\rm 1\: kpc}$ and $ \Sigma_{R_{\rm e}}$ decreases by approximately 0.3~dex from $z=3$ to $z=0.5$. In contrast, \qg{}s exhibit a more pronounced reduction in both $\Sigma_{\rm 1\: kpc}$ and $\Sigma_{R_{\rm e}}$, decreasing by about 0.3~dex and nearly 1~dex, respectively. \citet{Barro2017} highlights an empirical relation between $\Sigma_{\rm 1\: kpc}$ and $\Sigma_{R_{\rm e}}$ relations modelled as, $\log \Sigma_{\rm 1\:kpc} \propto \log \gamma(R_{\rm e}^{1/n})$, where $\gamma$ is the incomplete gamma function which is dependent on the effective radius and S\'ersic index $n$. Due to this, for quenched systems, a steep slope in the $R_{\rm e}-M_\star$ plane results in a negative slope in the $\Sigma_{R_{\rm e}}-M_\star$ plane.
Furthermore, it is important to note that their quenched sample contained limited \qg{}s in the $2.2\lesssim z \lesssim3$ range, necessitating a fixed slope for their analysis. The reported increased surface densities align with the process of bulge formation and the growth of central SMBHs, which are critical in the feedback processes that quench star formation \citep[e.g., ][]{Kauffmann2012}.

\section{Simulations and galaxy formation model}
In this section, we provide an overview of the \lgal{} model and outline the different model flavours used in this study.

\label{Sec:Model}
\subsection{Simulations and subhalo identification}
\label{sec:sim&subhalo}
 In this work, we use the dark matter halo merger trees from the Millennium-I (MRI) and Millennium-II (MRII) simulations \citep{Springel2005MilI, Boylan-Kolchin2009MilII} which use a box size of $480.3\,h^{-1}{\rm Mpc}$ and $96.1\,h^{-1}{\rm Mpc}$, respectively, and provide a dynamic range of five orders of magnitude in stellar mass (${10^{7}\:{\rm M}_{\odot}\lesssim M_{\star} \lesssim 10^{12}\: {\rm M}_{\odot}}$) for the semi-analytic galaxies. These simulations trace $2160^3$ particles from $z=127$ to $z=0$ with 64 snapshots in MRI and 68 snapshots in MRII. Furthermore, both of these simulations assume a $\Lambda{}\text{CDM}$ cosmology with the parameters scaled to the Planck Cosmology \citep{Planck2016} using the conversion from \citet{Angulo2010} and \citet{Angulo2015}. Note that this scaling leads to effective box sizes (as quoted above) that are slightly different from their original values. The dark matter haloes are identified using the Friends Of Friends (FOF) algorithm \citep{Davis1985}, while the sub-haloes (self-bound substructures in each FOF group) are detected using the \textsc{Subfind} algorithm \citep{Springel2001Subfind}. Further details can be found in the recent update of the model in \citet{Ayromlou2021} and the references therein.

\subsection{\lgal{} model}
\label{Sec:Lgal_model}
\begin{figure}
    \centering
    \includegraphics[width=1\linewidth]{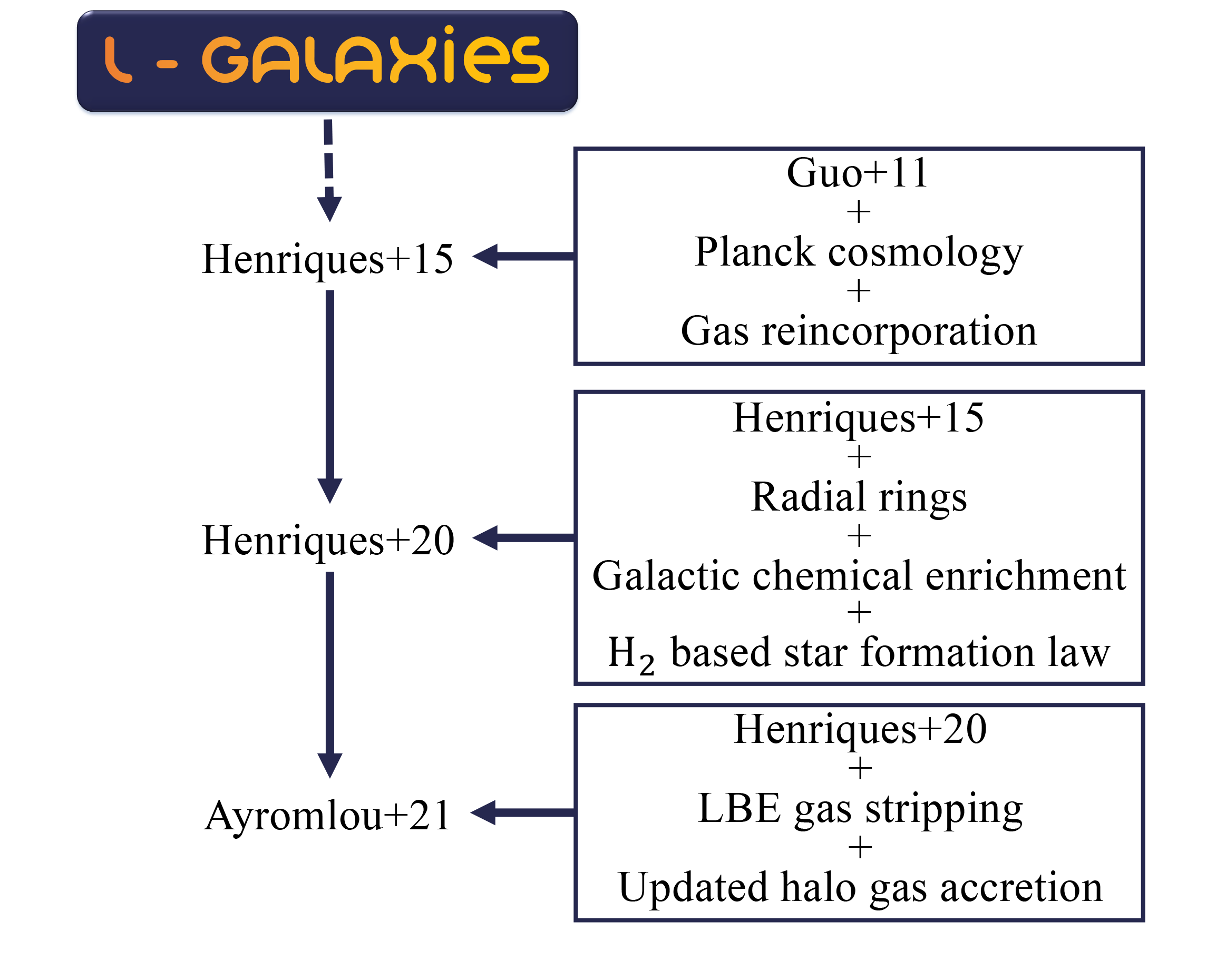}
    
    \caption{Overview of the three calibrated \lgal{} flavours: \hXV{}, \hXX{}, and \aXXI{}, illustrating key updates and new prescriptions implemented in each of them.
    }
    \label{fig:LgalFlowChart}
\end{figure}

\lgal{}\footnote{The models can be found at \url{https://lgalaxiespublicrelease.github.io/}, which also includes complete documentation for all three model flavours.} is a semi-analytical galaxy formation model that uses a set of equations to model baryonic physics on top of dark matter halo merger trees. It has evolved and expanded its capabilities over the course of three decades \citep{White1978, White1991, Kauffmann1993, Kauffmann1999, Springel2001Subfind, Springel2005MilI, Guo2011, Guo2013, Henriques2015, Henriques2020, Ayromlou2021}, with additional versions incorporating modifications such as BH physics, galactic chemical evolution, and binary stellar evolution \citep{Yates2021_LGAL,Yates2023_LGal, IzquierdoVillalba2021_LGAL_BH, Spinoso2022_LGAL_BH, Barrera2023_LGAL_MTNG}, which are based on \citet{Henriques2015, Henriques2020} and \citet{Ayromlou2021}.
In this work, we focus on the three most recent, fully calibrated, and publicly released flavours of \lgal{}, namely:  \citet[][]{Henriques2015} (hereafter \citetalias{Henriques2015}),  \citet[][]{Henriques2020} (hereafter \citetalias{Henriques2020}), and \citet[][]{Ayromlou2021} (hereafter \citetalias{Ayromlou2021}). A summary of the new implementations is depicted in \figref{fig:LgalFlowChart}.

\citetalias{Henriques2015} builds upon the foundation laid by \citet{Guo2011} by introducing substantial enhancements to it. These enhancements include an improved representation of the build-up of the galaxy population accomplished by updating the underlying cosmology to that of the Planck-1 cosmology \citep{Planck2016}, as well as refinements in the handling of gas reincorporation timescales and the virial mass threshold that governs ram pressure stripping effects on satellite galaxies. The model's underlying parameters were fine-tuned by using a robust Markov Chain Monte Carlo (MCMC) method, in order to match a comprehensive set of observational data across redshifts $z=0$, $1$, $2$, and $3$. This calibration was based on the stellar mass function and the fraction of \qg{}s at these epochs. These updates also effectively addressed issues related to the star formation threshold and the efficiency of the radio mode feedback mechanism from AGN. 

\citetalias{Henriques2020} represents a significant advancement over former models, incorporating a more detailed treatment of galactic discs with the ability to resolve radial properties of galaxies. This allows direct comparison with observation at kpc-scales of different galaxy properties. This was achieved by dividing the stellar and gaseous disc into a series of concentric `rings' following the methodology of \citet{Fu2013_rings}. Due to this upgrade, the gas cooling and gas inflow laws had to be updated as well. Furthermore, this model also adds a $\text{H}_2$-based star formation law \citep[see ][]{Fu2013_rings, Krumholz2009HI_H2} and a comprehensive chemical enrichment model  \citep{Yates2013_chemicalenrichment} that explicitly accounts for mass-dependent delay times associated with SN type-II, SN type-Ia, and AGB stars. Like its predecessor, \citetalias{Henriques2020} underwent calibration using the same MCMC-based method at $z=0$ and $z=2$. This calibration is based on the stellar mass function and the fraction of \qg{}s, as well as the HI mass function at $z=0$, ensuring its compatibility with similar observational constraints.

Finally, the latest version of \lgal{}, \citetalias{Ayromlou2021}, built on top of \citetalias{Henriques2020} introduced a sophisticated methodology to represent gas-stripping processes within and beyond the halo boundary. This update also includes a realistic ram-pressure stripping mechanism for all galaxies from \citet{Ayromlou2019_rampressure}. This was achieved using the local background environment (LBE) of galaxies. This model was also calibrated using an MCMC approach with recent data on the stellar mass function and the fraction of \qg{}s at $z=0$, $1$, and $2$.

\subsection{Galaxy properties in the model}

In this subsection, we briefly describe the implementation of physical processes most relevant to our study. A comprehensive model description, including the detailed input physics, can be found in the supplementary material of \citetalias{Ayromlou2021}.

\subsubsection{Cooling modes}
Infalling gas shock heats upon entering a halo, with the shock location varying based on halo mass and time. Low-mass haloes at early times have shocks close to the central object, allowing rapid settling onto the cold gas disc. Higher mass haloes form a quasi-static hot atmosphere farther from the centre, with the transition mass at around $M_{\star}\sim 10^{12}\: \text{M}_{\odot}$ \citep{White1978, White1991}. The gas cooling mechanism has been unchanged for the three models. Following the frameworks established by \citet{White1991} and \citet{Springel2001Subfind}, the model presumes that within this quasi-static regime, the gas cools from an isothermally distributed hot atmosphere. The cooling time is defined as the ratio of the gas's thermal energy to its cooling rate per unit volume. This leads to cold gas accretion onto the halo, with a portion subsequently incorporated into the central galaxy. 

\subsubsection{Star formation}
Stars are assumed to be formed from the accreted cold gas within the disc of each galaxy. \citetalias{Henriques2015} adopts a fixed ratio of atomic and molecular hydrogen, with the star formation prescription based on the total interstellar medium (ISM) gas content. In the later updates of the model, \citetalias{Henriques2020} and \citetalias{Ayromlou2021} introduced a $\text{H}_2$-based star formation law, which assumes that the star formation surface density is proportional to the $\text{H}_2$ surface density \citep[see ][]{Fu2013_rings, Krumholz2009HI_H2}. These model flavours also include an inverse dependence on dynamical time, making star formation efficient at early times when the dynamical time scales are shorter. Furthermore, star formation is also triggered during galaxy mergers, following the formulation of \citet{Somerville2001}.

\subsubsection{Supernova feedback}
As massive, short-lived stars approach their death, they release a large amount of mass and energy in the form of supernova (SN) explosions and stellar winds. \citetalias{Henriques2015} assumes an instantaneous recycling approximation, in which the entirety of metals and energy generated during an episode of star formation is released instantaneously. The injected energy impacts cold and hot interstellar gas, reheating and transferring it into the atmosphere or out of the galaxy as winds. \citetalias{Henriques2015} introduced two efficiency factors, one governs how much SN energy contributes to long-term changes in the galaxy, while the other determines how much of this energy reheats cold gas and injects it into the hot atmosphere. In this model, the energy available from SN and winds is linked to the mass of stars formed. 

On the other hand, \citetalias{Henriques2020} and \citetalias{Ayromlou2021} modified this mechanism and propose that the energy from SN and winds is returned to the ISM, enabling a time-dependent return of mass into the system.  In all three models, any excess SN energy not utilised is used to expel the gas into an ejecta reservoir, where it is not available for cooling.
A fraction of this ejected gas is reincorporated by adopting the implementation outlined by \citet{Henriques2013winds}, where the timescale of reincorporation inversely correlates with the mass of the host halo. 

\subsubsection{Black hole related processes}
\label{sec:BHprocesses}

In all three model flavours, supernovae and stellar winds impact low-mass galaxies significantly, but they are inefficient in curtailing cooling in more massive systems. To address this, all models follow \citet{Croton2006} along with the methodology stated in \citet{Henriques2013winds} in suggesting that the central SMBHs are primarily responsible for halting galaxy growth in massive haloes through two main mechanisms: ``quasar mode'' and ``radio mode''.

In the ``quasar mode'', these BHs form and grow by accreting cold gas funnelled towards the halo centre during galaxy mergers. This mode of growth, while significant for mass accumulation, does not contribute further feedback beyond the starbursts triggered during such gas-rich mergers. The amount of gas accreted ($\Delta M_{\text{BH, Quasar}}$) in this mode is dependent on the properties of the merging galaxies (i.e. the total baryon mass, $M_{\rm sat}$ and $M_{\rm cen}$), specifically their total cold gas content ($M_{\rm cold}$) and the virial velocity of the central halo ($V_{200c}$), calculated as:
\begin{equation}
    \Delta M_{\text{BH,Quasar}} = \frac{f_{\text{BH}} \left({M_{\text{sat}}}/{M_{\text{cen}}}\right) M_{\text{cold}}}{1 + \left({V_{\text{BH}}}/{V_{200c}}\right)^2},
\end{equation}
where $f_{\text{BH}}$ and $V_{\text{BH}}$ represent the efficiency of BH growth and the velocity scale for quasar growth, respectively, both of which are free parameters.

In contrast, the ``radio mode'' involves, continuous low-level accretion of gas from the hot gas atmosphere surrounding the galaxy into the central SMBH. 
The implementation is designed to capture the effects of jets and bubbles injecting energy into the surrounding hot gas halo and thereby suppressing further cooling to the cold disc, leading to a decrease in the effective cooling rate. The BH accretion rate is governed by the equation:
\begin{equation}
\label{eq:RadioMode}
    \dot{M}_{\text{BH}} = \kappa_{\text{AGN}} \left(  \frac{M_{\text{hot}}}{10^{11} \, {\rm M}_\odot} \right) \left( \frac{M_{\text{BH}}}{10^8 \, {\rm M}_\odot} \right),
\end{equation}
where $\kappa_{\text{AGN}}$ is the efficiency of feedback in the radio mode, which is a free parameter.  This formalism, similar to that in \citet{Croton2006}, has been modified in all three of the models by introducing a new normalisation factor that enhances the efficiency of accretion at lower redshifts.

The energy released during the ``radio mode'' depends on the accretion rate onto the BH as defined by:
\begin{equation}
    \dot{E}_{\text{radio}} = {{\eta} \:\dot{M}_{\text{BH}} \:{ c^2}},
\end{equation}
where $\eta=0.1$ represents the efficiency of converting accreted mass into energy, and $c$ is the speed of light. This energy effectively suppresses the cooling of the surrounding hot gas, moderating the inflow of material onto the galactic disc and regulating star formation processes.

\subsubsection{Environmental processes}
\label{sec:env_process}
Galaxies are influenced by tidal forces, hot gas interactions, and encounters with other galaxies, altering their structure and evolution, sometimes leading to complete disruption. The modern \lgal{} versions allow for gradual hot gas stripping, thereby allowing satellite galaxies to retain some hot gas. \citetalias{Henriques2015} and \citetalias{Henriques2020} implement tidal stripping for satellite galaxies within the virial halo boundary. Moreover, ram-pressure stripping is limited to satellites within the virial radius of massive galaxy clusters above a certain threshold mass in both  \citetalias{Henriques2015} and \citetalias{Henriques2020}. This artificial threshold was introduced to prevent an excessive number of quenched low-mass galaxies. However, in \citetalias{Ayromlou2021}, this limitation was removed. In this model flavour, ram-pressure stripping has been extended to include all satellite and central galaxies within the simulation, employing the local background environment (LBE) measurements by \citet{Ayromlou2019_rampressure}. This update positions the new version of \lgal{} as the only semi-analytical model that captures environmental effects across the entire simulation. Furthermore, \citetalias{Ayromlou2021} has extended tidal stripping to encompass all satellite galaxies, both within and beyond the virial radius. Furthermore, in all three flavours, the tidal disruption of stellar and cold gas remains unchanged from \citet{Guo2011}. In this framework, the model considers disruption only for orphan galaxies, which have already lost their dark matter subhalo and, consequently, their hot gas.

\subsubsection {Post-processing of colours}
First implemented in \citetalias{Henriques2015}, all three models store star formation and metal enrichment histories using the approach described in  \citet{Shamshiri2015color}. This enables the models to compute luminosities, colours and spectral properties of galaxies in post-processing using any stellar population synthesis model. In this case, the \lgal{} models use \citet{Maraston2005_stellar_pop_syn} with a \citet{Chabrier2003_IMF} IMF as our default stellar population model. This approach allows the model snapshots to contain emissions in 40 bands, all calculated in post-processing. Effects of dust extinction from the diffuse ISM and molecular clouds are added on top of these dust-free colours based on \citet{Devriendt1999_extinction} and \citet{Charlot2000extinction}, respectively.

\section{Results}
\label{Sec:Results}
Throughout this analysis, we switch between the results of the MRI and MRII simulations at $M_\star = 10^{9.5}\:{\rm M}_{\odot}$, the stellar mass at which galaxy stellar mass functions and other galaxy properties converge the best between the two runs. Thus, the results of MRI cover the mass range  $M_\star \gtrsim \:10^{9.5} {\rm M}_{\odot}$, while the higher resolution MRII is used for $M_\star < 10^{9.5} \: {\rm M}_{\odot}$, together covering a total stellar mass range of $10^{7}\: {\rm M}_{\odot} \lesssim M_\star \lesssim 10^{12} \: {\rm M}_{\odot}$.

For simplicity of presentation, and because most results do not differ substantially between the different model flavours, we adopt the most complete model, \citetalias{Ayromlou2021} as the default model, unless stated otherwise. For reference, specific results from the older model variant \hXV{} are detailed in \appref{app:H20_H15}. The results from \hXX{} are not shown, as they closely resemble those of \aXXI{}. This similarity arises because the primary difference between \aXXI{} and \hXX{} lies in the environmental gas stripping prescription which affects low-mass galaxies (as described in \secref{sec:env_process} and \figref{fig:LgalFlowChart}).

\subsection{Methods for selecting galaxies from simulations}
\label{sec:nuvrj}
The distinction between \sfg{}s and their quiescent counterparts can be made using various methods. One common approach is applying a specific SFR (sSFR) cut, identified from fitting broad band photometry to a spectral energy distribution (SED) template \citep[e.g., ][]{Franx2008_sSFRcut, Salim2018_ssfr, Davidzon2018}. Alternatively, rest-frame colour-colour criteria offer a different means of differentiation, using the idea that \sfg{}s are typically bluer due to ongoing star formation, while \qg{}s are redder due to a lack of young stars. In such a case one can use the $\rm U-V$ and $\rm V-J$ (also called $\rm UVJ$) colours \citep[e.g., ][]{Whitaker2011_UVJ}, the $\rm NUV-r$ and $\rm r-J$ ($\rm NUVrJ$) colours \citep[e.g., ][]{Ilbert2013_nuvrj}, or the $\rm NUV-U $ and $\rm V-J$  colours ($\rm NUVUVJ$) \citep[e.g.,][]{Gould2023_NUVUVJ}  as a separating criterion. 

In this work, we adopt the rest frame $\rm NUV-r$ and $\rm r-J$ colour-colour selection criterion (hereafter referred to as $\rm  NUVrJ$) given by \citet{Ilbert2013_nuvrj}:
\begin{equation}
\label{eq:NUVrJ}%
\rm (NUV-r)>3(r-J)+1 \;\text{and}\; (NUV-r)>3.1.
\end{equation} 
This criterion, which employs rest frame near-ultraviolet ($\rm NUV$) and optical ($\rm r$ and $\rm J$) colours, is particularly sensitive to UV emission primarily originating from young, hot stars in \sfg{}s. This selection method is especially adept at isolating dusty \sfg{}s from the \qg{}s, with the slant line of the criterion designed to be perpendicular to increasing sSFR and parallel to the axis of increasing dust attenuation. 
The $\rm NUVrJ$ criterion approximates an ${\rm sSFR} \lesssim 10^{-11}\:[{\rm year}^{-1}]$ at $z=0$ \citep{Davidzon2018}, and by extending into the UV regime, it offers enhanced sensitivity and a strong correlation with sSFR compared to the traditional $\rm UVJ$ criterion \citep{Leja2019_UVJ_NUVRJ}.
We note that the choice of selection criteria can influence the characteristics of the galaxy scaling relations \citep[e.g., ][]{Henriques2017_quenching_colorcuts, Leja2019_UVJ_NUVRJ, Popesso2023, Pearson2023, Gould2023_NUVUVJ}. 

\begin{figure*}
    \centering
    \includegraphics[width=1\linewidth]{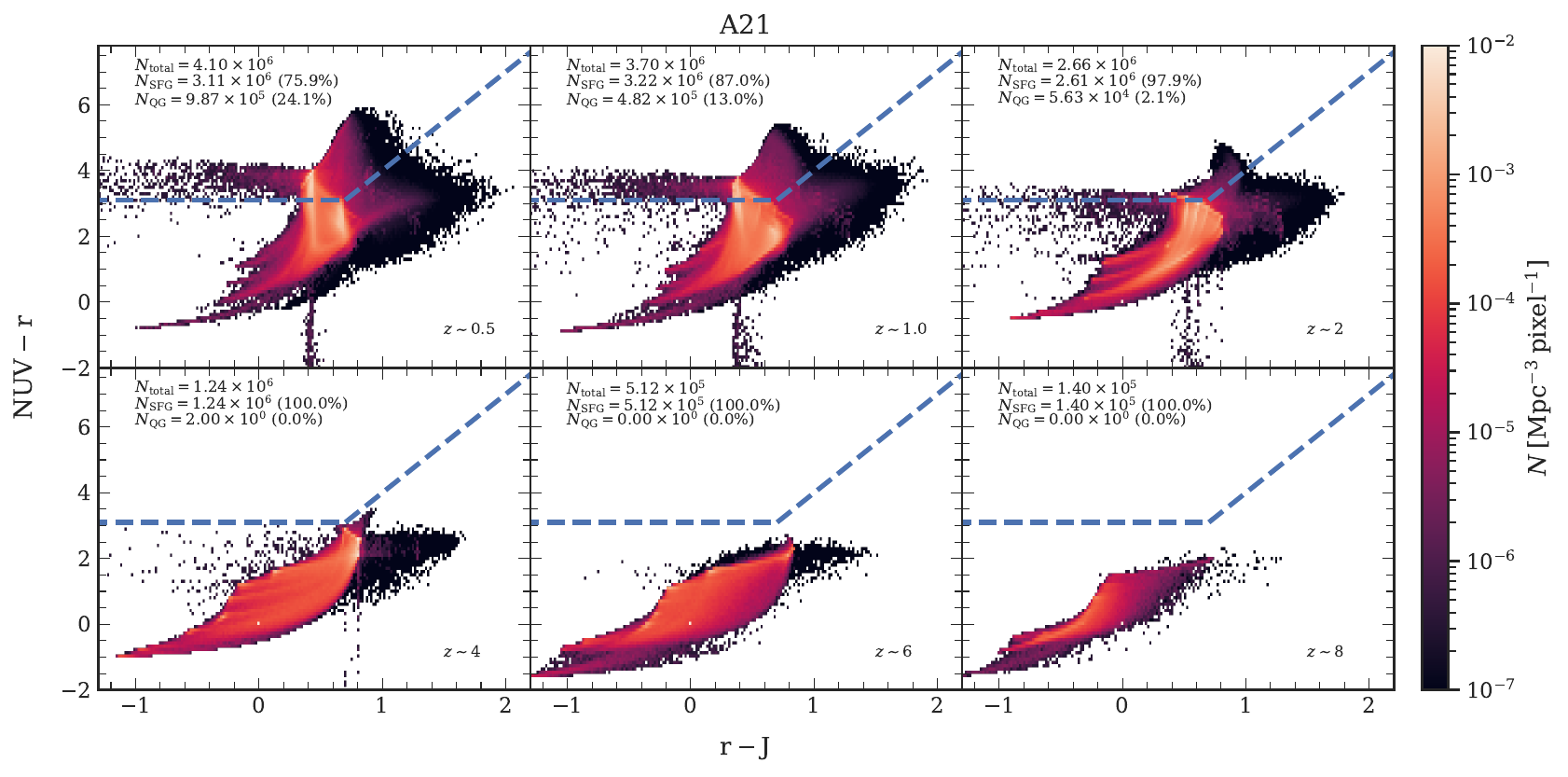}
    \caption{
    $\rm NUVrJ$ colour-colour plot: The plot, combined for MRI and MRII, illustrates the $\rm NUVrJ$ colour-colour selection criteria used to classify galaxies as either star-forming or quiescent across various redshifts. The classification is based on rest-frame $\rm NUV-r$ and $\rm r-J$ colours, following the methodology outlined by \citet{Ilbert2013_nuvrj} (blue dashed line). Galaxies above this relation are classified as quenched, while those below are considered star-forming. The number statistics for each redshift bin are also displayed in each subplot. Here, ${N_{\text{QG}}}$ represents the number of quiescent galaxies, and ${N_{\text{SFG}}}$ represents the number of star-forming galaxies in each subplot separated as per the $\rm NUVrJ$ criterion.  Furthermore, ${N_{\text{total}}}$ is the total number of galaxies at the represented redshift. The fraction (in \%) of quenched and star-forming systems per total number of galaxies at each redshift is shown in the brackets.  The plot shows the results from \citetalias{Ayromlou2021}, and the results from \hXV{} are given in \figref{fig:NUVRJ_H20_H15}. }
    \label{fig:NUVRJ}
\end{figure*}

To visually represent our selection, \figref{fig:NUVRJ} illustrates the segregation of \sfg{}s and \qg{}s in the \aXXI{} model at six different redshifts (see \figref{fig:NUVRJ_H20_H15} for \hXV{}). The colour bar represents the number of galaxies per cubic Mpc per pixel, with a pixel size of $0.02 \: [{\rm mag}] \times 0.1 \: [{\rm mag}]$. The galaxies in the upper left quadrant are classified as \qg{}s. The plot also displays the total number of galaxies in the simulation at each redshift, followed by the number of \sfg{}s and \qg{}s selected as per \equref{eq:NUVrJ}, with the brackets indicating the fraction (in \%) of galaxies in relation to the total number of galaxies. For higher redshifts ( $z\gtrsim 2-3$), all three model flavours are devoid of \qg{}s (including misclassified dusty \sfg{}s). This deficit is discussed in \secref{sec:QG_smf} and \ref{Sec:Baryonic_features}.

\begin{figure}
    \centering
    \includegraphics[width=1\linewidth]{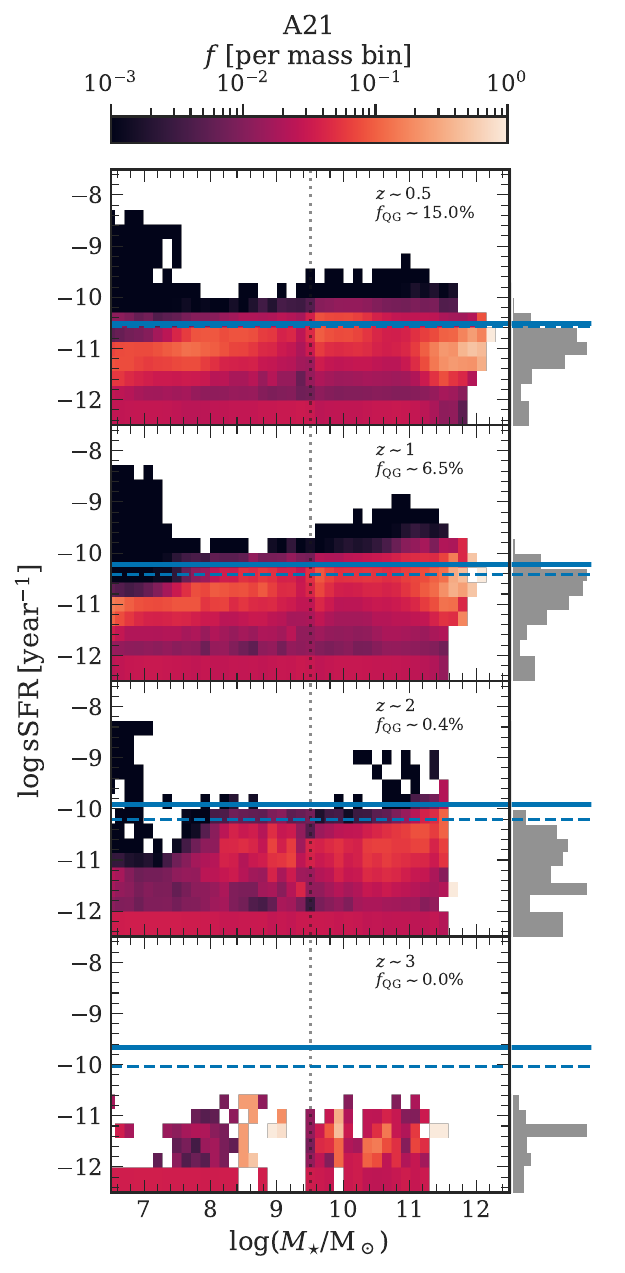}
    \caption{${\rm sSFR}-M_\star$ plot for \qg{}s: The plot illustrates the occupation of the ${\rm sSFR}-M_\star$ plane by the $\rm NUVrJ$ selected \qg{}s across different redshifts produced by \citetalias{Ayromlou2021}.
    Two lines delineate the threshold sSFR values used to distinguish \qg{}s from star-forming ones: the dashed line represents the sSFR cut-off according to \citet{Franx2008_sSFRcut}, while the solid line reflects the threshold from \citet{Henriques2020}. The 2D histogram is normalised by the sum of each column, and the colour bar represents the fraction of galaxies per mass bin, with a bin width of 0.14~dex.
    In each subplot, $f_{\text{QG}}$ represents the fraction (in \%) of quiescent galaxies, for the whole mass range, above the \citet{Henriques2020} threshold. These are `falsely' identified quenched systems by \equref{eq:NUVrJ} having a high sSFR. The vertical grey dashed line separates the two boxes of the simulation at ${M_\star = 10^{9.5}\:{\rm M}_{\odot}}$ and cells with ${\rm sSFR} \lesssim 10^{-12}$ are binned as a single column. The grey histogram on the right axis shows the probability density distribution of the sSFR.
    Results from \hXV{} are shown in \figref{fig:sSFR_H20_H15}.
    } 
    \label{fig:sSFR}
\end{figure}

Additionally, \figref{fig:sSFR} serves as a consistency check, displaying the sSFRs for the $\rm NUVrJ$ selected \qg{} population in \aXXI{} (see \figref{fig:sSFR_H20_H15} for \hXV{}). The 2D histogram is normalised by the sum of each column, so the colour coding represents the fraction of galaxies in each mass bin of size 0.14~dex with a pixel size of $0.14 \:[\log({\rm M}_{\odot})] \times 0.29 \:[\log({\rm year}^{-1})]$. The histogram on the right axis shows the probability distribution function (PDF) of the sSFR at each redshift. Cells with ${\rm sSFR} \lesssim 10^{-12}$ are binned as a single column,
because it is not normally possible to detect residual star formation in galaxies below this level.

This figure also displays two time-dependent sSFR cuts, one from \citet{Franx2008_sSFRcut} (dashed line), which selects \qg{}s having a ${\rm sSFR}\lesssim0.3/t_{{\rm H}(z=0)}$\footnote{$t_{{\rm H}(z=0)}$ is the Hubble time at $z=0$.} and another from \citet{Henriques2020} (solid line), which selects \qg{}s as 1~dex below of the main sequence given by ${\rm sSFR}=2\times(1+z)^2/t_{{\rm H}(z=0)}$. 
Additionally, ${f}_{\text{QG}}$ represents the fraction of galaxies above the \citet{Henriques2020} threshold. These galaxies have a higher sSFR and thus are effectively star-forming systems which are `misclassified' as quenched by the $\rm NUVrJ$ criterion.

Although \aXXI{} exhibits a small fraction of these misclassified systems, their relatively low numbers do not compromise the unified methodology applied across the models and observations, ensuring the robustness of our results.
Most of these galaxies above the threshold are found to be massive satellite galaxies. In contrast, the centrals are below the threshold. This fraction of galaxies above the threshold, ${f}_{\text{QG}}$, also decreases sharply for $z\gtrsim1$. The PDF further illustrates that the bulk of the distribution remains well below these thresholds. Parallel results from \hXV{} display similar trends, as shown in \figsref{fig:NUVRJ_H20_H15} and \ref{fig:sSFR_H20_H15}.

\subsection{Evolution of global galaxy properties to very high redshift}

\begin{figure}
    \centering
    \includegraphics[width=1\linewidth]{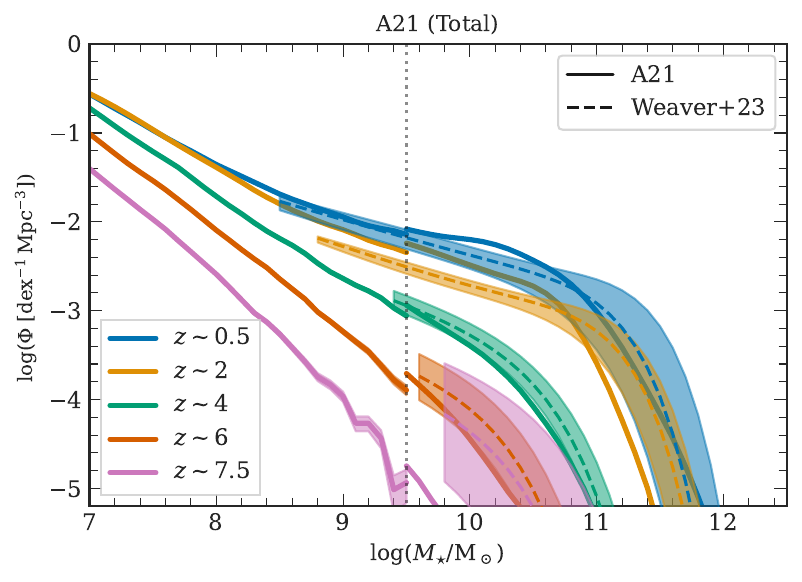}
    \includegraphics[width=1\linewidth]{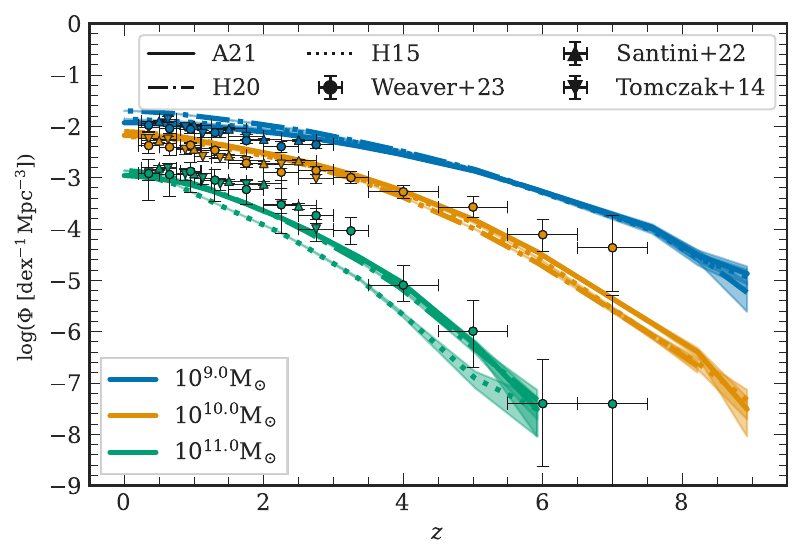}
    
    \caption{Total stellar mass function (\textit{top}) and total galaxy number density as a function of redshift  (\textit{bottom}): 
    In the top panel, the solid line shows the evolution of the stellar mass functions across 5 redshifts for the total (\sfg{} and \qg{}) galaxy sample from \aXXI{}. The dashed line represents the observational findings from \citet{Weaver2023}, with the accompanying shaded area indicating their errors. The total \smf{} from \hXV{} are shown in \figref{fig:SMFs_H20_H15}.
    The bottom plot showcases the evolution of the number density of galaxies in 3 mass bins of 0.2~dex width as a function of redshift for the total galaxy sample.  Each line style illustrates the results from distinct \lgal{} flavours: the solid line corresponds to \aXXI{}, the dot-dashed to \hXX{}, and the dotted to \hXV{}. The coloured circular points mark the observational results from \citet{Weaver2023}, \citet{Tomczak2014} and \citet{Santini2022}, with the corresponding error bars indicated for each mass bin.
    In both plots, Poisson errors are represented by the similarly coloured shaded region. The data spans the mass range covered by the MRI and MRII simulations, subject to a transition region at $M_\star = 10^{9.5}\:{\rm M}_{\odot}$.
    }
    \label{fig:SMF_tot_SMF_z} 
\end{figure}

Assessing the agreement of global galaxy properties between simulations and observations is a crucial first step for validating and refining cosmological models. In this section, we evaluate the global consistency of the three \lgal{} flavours against observations.

\subsubsection{The stellar mass function}
\label{sec:TotalSMF}
We calculate the \smf{} for the combined total sample of galaxies (\sfg{}s and \qg{}s) across five redshifts. To compute the \smf{}, at a given redshift, the number of galaxies within each 0.2~dex mass bin is divided by the volume of the corresponding simulation (MRI or MRII, detailed in \secref{sec:sim&subhalo}) and is normalised by the bin width. The \smf{} is derived using a sliding bin approach with step size of 0.1~dex. Furthermore, we impose a minimum threshold of 20 galaxies per mass bin. If this threshold is not met, the bin size is increased by 0.1~dex iteratively until it is satisfied. Our primary point of comparison is with the findings of \citet{Weaver2023} (dashed lines), as presented in \figref{fig:SMF_tot_SMF_z} (\textit{top}) for \aXXI{} (see \figref{fig:SMFs_H20_H15} for \hXV{}). We follow the mass completeness limits outlined by \citet{Weaver2023}, incorporating their error estimates represented by shaded areas. The \smf{}s of the simulated galaxies are presented with solid lines, each accompanied by their respective one-sigma Poisson noise.

We note that the parameters of all the models were tuned to fit the stellar mass function at $z=0$ as well as its evolution out to $z=2$. Model comparisons in the low redshift ranges do thus not represent a strong test of the input physics, so we will concentrate on higher redshift ranges in this paper. Comparisons for lower redshifts have been previously addressed in the respective publications detailing each model.

In~\figref{fig:SMF_tot_SMF_z} (\textit{top}), the \aXXI{} flavour exhibits qualitative agreement in shape and evolutionary trends of the \smf{} across lower redshifts $0\lesssim z \lesssim2$, albeit with small noticeable discrepancies. Similar to the findings in \citet{Weaver2023}, we note that the \smf{} decreases monotonically with mass at $ z\sim 0.5 - 2$. It tends to flatten out at lower redshifts before sharply falling at $M_\star \sim 10^{11}\: {\rm M}_{\odot}$.  The simulation suggests that the slope of the low-mass end steepens into a power-law shape, whereas the high-mass end after the `knee' falls off steeply, consistent with the finding of \citet{Weaver2023}.

At high  redshifts ($z\gtrsim2$), a marked deficiency in high-mass galaxies ($M_\star \gtrsim 10^{9.5}\:{\rm M}_{\odot}$) is notable in all three flavours. The discrepancy between simulations and observations is most apparent at $z\sim2$ in all three flavours by displaying an excess of $M_\star \lesssim 10^{10.8}\:{\rm M}_{\odot}$ galaxies and an underestimation of the \smf{} for more massive galaxies compared to observations. For even larger redshifts ($z \gtrsim 5$), the observations suggest a higher number density, but the growing uncertainty in the observational data complicates a clean separation of systematic errors and physical effects affecting the \smf{}'s shape.

In \figref{fig:SMF_tot_SMF_z} (\textit{bottom}), we illustrate the number density of galaxies per mass bin as a function of redshift categorised into three mass bins each spanning 0.2~dex. The line styles --- solid, dashed and dot-dashed --- represent the different \lgal{} flavours, while the different markers indicate the observational data from \citet{Tomczak2014, Santini2022} and \citet{Weaver2023}, along with their associated errors. These data points are derived from the corresponding Schechter function fits. Observational mass completeness limits are also depicted in this plot. 
The conclusions drawn from \figref{fig:SMF_tot_SMF_z} (\textit{bottom}) indicate subtle variations across the different \lgal{} flavours, with the \hXX{} showing a slight excess of $M_\star \sim 10^9 \: {\rm M}_{\odot}$  galaxies, a trend not observed in other flavours. Additionally, in all three models, there is a mild excess of $M_\star \sim 10^{10} \: {\rm M}_{\odot}$ galaxies at cosmic noon ($z\sim 2$) and a slight deficiency of $M_\star \sim 10^{11} \: {\rm M}_{\odot}$ galaxies at the same epoch, with \hXV{} underestimating the number density in this heaviest mass bin.  The models also show different growth rates of the most massive and least massive mass bins, consistent with the data. 
Overall, the \aXXI{} flavour offers improved performance across the three mass bins and redshifts, motivating the choice of this model as the fiducial model in this paper.

The evolution of the \smf{} indicates that the early universe predominantly contained small galaxies and that high-mass galaxies have not yet assembled.  We observe a substantial mass buildup of approximately $10^4\; [{\rm dex}^{-1}\:{\rm Mpc}^{-1}]$ between $2 \lesssim z \lesssim6$ for $M_\star \sim 10^{11} \:{\rm M}_{\odot}$ galaxies, and about $10^5 \;[{\rm dex}^{-1}\:{\rm Mpc}^{-1}]$ between $ 2 \lesssim z \lesssim 8$ for $M_\star \sim 10^{10} \:{\rm M}_{\odot}$ galaxies. Conversely, the number density appears to plateau at low-$z$ for all three mass bins, suggesting a slow down in star formation. The turnover point in the \smf{} at around the Milky Way mass ($ M_\star \sim 10^{10.8}\:{\rm M}_{\odot}$), above which the number density of galaxies sharply declines, provides critical insights into the efficiency of galaxy mergers and the feedback mechanisms from SN and AGN \citep[e.g.,][]{Graham2023_AGN_SN}.

\subsubsection{Evolution of the cosmic star formation rate density}
\label{subsec:CSFRd}
\begin{figure}
    \centering
    \includegraphics[width=1\linewidth]{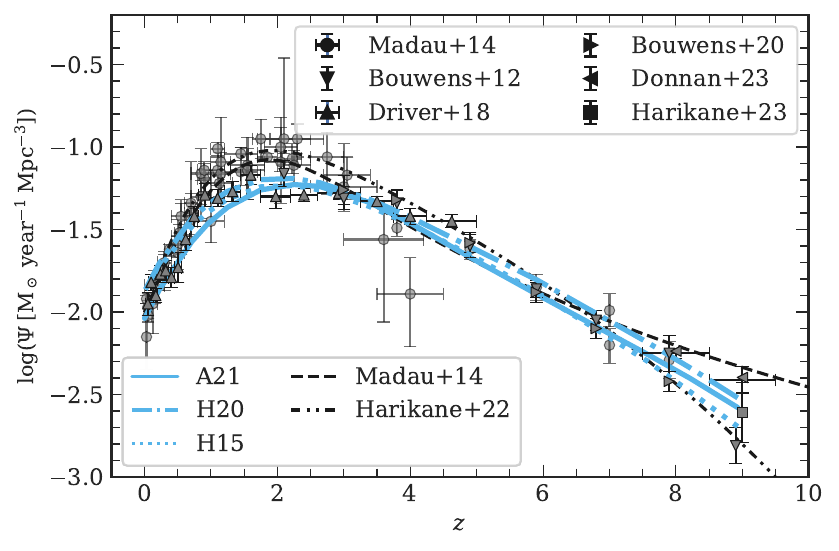}
    \caption{Cosmic SFR density as a function of redshift: The results from the different \lgal{} flavours are shown in different line styles in cyan. The observational data is obtained from \citet{Madau2014, Bouwens2012_SFH, Bouwens2020_CSFRD, Driver2018_SFH, Donnan2023_UV_JWST} and \citet{Harikane2023_SFRH}. The black dashed line and the dot-dot-dashed line are from \citet{Madau2014} and \citet{Harikane2022_SFRH}, respectively, representing the observational fits.}
    \label{fig:CSFR}
\end{figure}
In~\figref{fig:CSFR}, we compare the cosmic SFR density (CSFRD) in the three \lgal{} flavours with observational data from \citet{Madau2014, Bouwens2012_SFH, Bouwens2020_CSFRD, Driver2018_SFH, Donnan2023_UV_JWST} and \citet{Harikane2023_SFRH}, accompanied by the fits provided by \citet{Madau2014} and \citet{Harikane2022_SFRH}. The SFRs of all galaxies 
are summed up and normalised by the volume of the corresponding box (detailed in \secref{sec:sim&subhalo}). We convert all results from the \citet{Salpeter1955_IMF} IMF to the \citet{Chabrier2003_IMF} IMF using the conversion $M_{\star,{\rm Chabrier}}=0.64\times\;M_{\star,{\rm Salpeter}}$.

While all three \lgal{} flavours exhibit a good qualitative agreement with the data between $z\lesssim 1$ and $2.5\lesssim z \lesssim 10$, it is noteworthy that they tend to underestimate the CSFRD by approximately 0.15~dex, particularly at cosmic noon ($1.5 \lesssim z \lesssim2$) when compared with the fit from \citet{Harikane2022_SFRH}. This underestimation coincides with significant discrepancies in the stellar mass functions (see \figref{fig:SMF_tot_SMF_z}), suggesting a potential decrease in star formation efficiency possibly due to inhibited gas cooling or insufficient gas density to sustain star formation.
At the high-$z$ end, \lgal{} agrees well with the results of \citet{Harikane2022_SFRH} while having a steeper slope compared to \citet{Madau2014}, suggesting rapid galaxy mass buildup with a reasonably constant star formation efficiency \citep{Harikane2022_SFRH, Harikane2023_SFRH}. Among the model flavours, \hXV{} and \hXX{} exhibit a slightly better agreement with the observed data, whereas \aXXI{} shows a suppression in the CSFRD at lower redshift ($z\lesssim 2$).

\subsection{Evolution of the star-forming galaxy population}
\label{sec:SFG_SMF}
\begin{figure}
    \centering
    \includegraphics[width=1\linewidth]{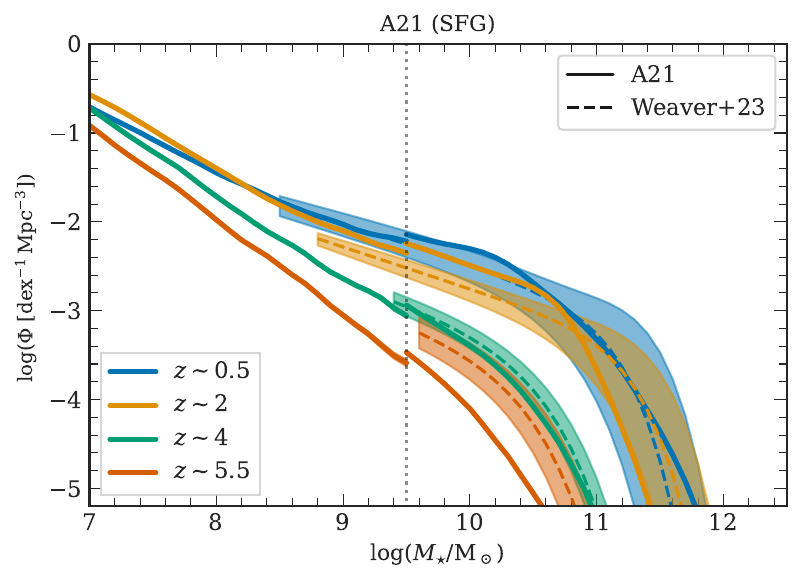}
    \includegraphics[width=1\linewidth]{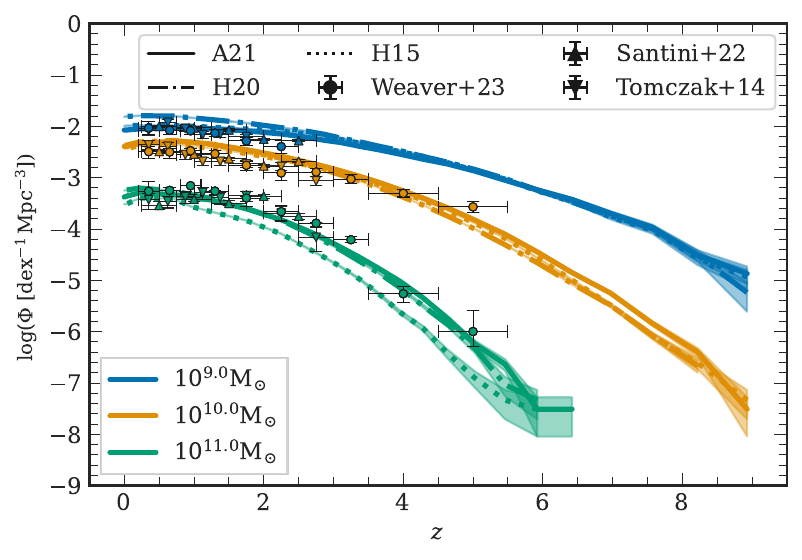}
    \caption{Stellar mass function of \sfg{}s (\textit{top}) and \sfg{} number density as a function of redshift (\textit{bottom}): Similar to \figref{fig:SMF_tot_SMF_z}, the top panel depicts the \smf{}s for the sample of \sfg{}s at four redshifts for the \aXXI{} flavour. The \smf{} from the \hXV{} flavour is shown in \figref{fig:SMFs_H20_H15}. The bottom panel shows the galaxy number density per mass bin to redshift relation for the sample of \sfg{}s in all three flavours along with the observed data. }
    \label{fig:SMF_SFG_SMFz_SFG}
\end{figure}

Similarly to the previous section, we calculate the \smf{} for the \sfg{} sample separated using the $\rm NUVrJ$ criterion (Eq. \ref{eq:NUVrJ}) and show the results in \figref{fig:SMF_SFG_SMFz_SFG} (\textit{top}). The \smf{} is plotted using a bin width of 0.2~dex, with a sliding step size of 0.1~dex and a minimum requirement of 20 galaxies per mass bin. We compare the simulation results with the observational findings of \citet{Weaver2023} including their one-sigma errors represented as the shaded regions. 

The comparison between observations and simulations of the \sfg{} \smf{} reveals similar discrepancies as those identified in the total \smf{} (see \figref{fig:SMF_tot_SMF_z}, \textit{top} panel). At $z\sim 2$ all three flavours display a lower turnover mass in the \sfg{} \smf{} --  $ 10^{10.8}\;{\rm M}_{\odot}$ compared to the observed turnover at about  $ 10^{11.1}\;{\rm M}_{\odot}$. This trend persists at other low redshifts. At larger redshifts ($z\sim5.5$), all model flavours underestimate the abundance of star-forming systems. Furthermore, \hXV{} exhibits a deficiency of \sfg{}s with $M_\star \gtrsim 10^{10.6} \:{\rm M}_{\odot}$ around $z\sim4$, a limitation not observed in \aXXI{} or \hXX{}.

Overall, the \smf{}s of \sfg{}s present varying shapes across different epochs, reflecting a double Schechter-like profile between  \mbox{$0\lesssim z \lesssim1.5-2$} with a `knee' (characteristic mass) between \mbox{$M_{\star}\sim10^{10.5}-10^{11.3} \: {\rm M}_{\odot}$}, which transitions to a power-law-like feature with a steep slope by $z\sim 3$.

In~\figref{fig:SMF_SFG_SMFz_SFG} (\textit{bottom}), we show the redshift evolution of the number density per mass bin of \sfg{}s compared to observational data from \citet{Tomczak2014, Santini2022} and \citet{Weaver2023}. The simulations slightly overpredict the number density of low mass $M_{\star}\sim10^{9} \:{\rm M}_{\odot}$ systems at $ z\sim2$ by about 1.5 times, and underpredict the massive $M_{\star}\sim 10^{11} \:{\rm M}_{\odot}$ systems at $z\gtrsim3$ by a factor of 2. For $M_{\star}\sim 10^{10} \:{\rm M}_{\odot}$ systems, an excess of simulated galaxies is seen at cosmic noon ($1.5 \lesssim z \lesssim 2$). This could be an indication of incompleteness or biases in the observations suggesting a non-physical origin \citep{Ilbert2013_nuvrj, Weaver2023}.

Among the model flavours, \aXXI{} agrees best with the observational data, whereas \hXX{} tends to overestimate the numbers in the low mass bins and \hXV{} underestimates the high mass bins by about 0.2~dex. This plot further illustrates the rapid mass assembly of massive galaxies through mergers, while lower-mass galaxies exhibit a slower, more gradual increase in number density across cosmic epochs. We note a moderate growth of \sfg{}s of about 1.5 times since $z\sim 2$, while a growth by about a factor 15 is observed for $M_{\star}\sim 10^{10}\: {\rm M}_{\odot}$ galaxies from $ 2\lesssim z \lesssim4$. The most massive galaxies grow by a factor of 40 over the same redshift range. The differences between the simulations and the observations underscore potential issues in the galaxy quenching mechanisms. Our results also demonstrate that the discrepancies with the data seen in the \smf{} arise from the star-forming population, which dominates the total population by number across the entire redshift range (see~\figref{fig:NUVRJ}).

\subsection{Evolution of the quiescent galaxy population}
\label{sec:QG_smf}

\begin{figure}
    \centering
    \includegraphics[width=1\linewidth]{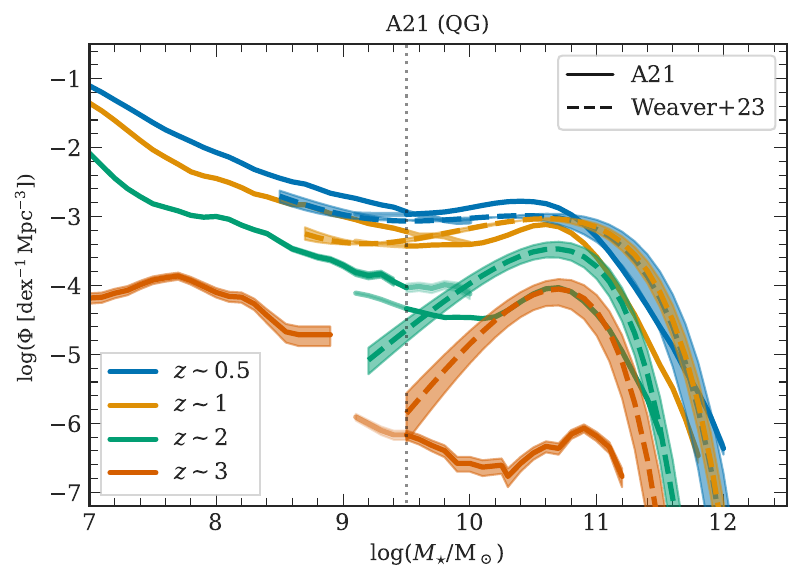}
    \caption{Stellar mass function of \qg{}s: 
    This figure displays the \qg{} \smf{} at four different redshifts within the \aXXI{} flavour alongside observational fits from \citet{Weaver2023}. Simulation results from MRI for $M_\star\lesssim10^{9.5}\:{\rm M}_{\odot}$ and MRII for $M_\star\gtrsim10^{9.5}\:{\rm M}_{\odot}$ are presented using fainter solid lines to indicate the transition region between the two runs (refer to \secref{sec:sim&subhalo}).
    The results from the other \lgal{} flavours are shown in \figref{fig:SMFs_H20_H15} and \ref{fig:SMF_QG_H20}.  }
    \label{fig:SMF_QG}
\end{figure}
\begin{figure*}
    \centering
    \includegraphics[width=0.45\linewidth]{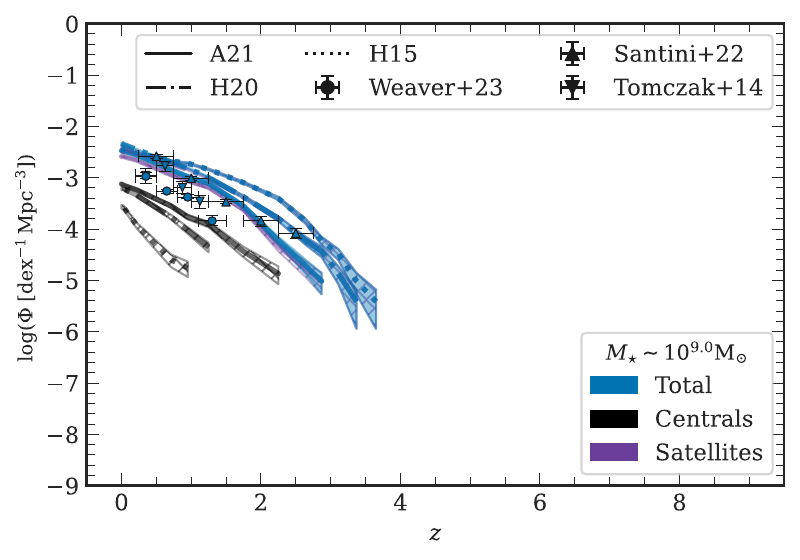}
    \includegraphics[width=0.45\linewidth]{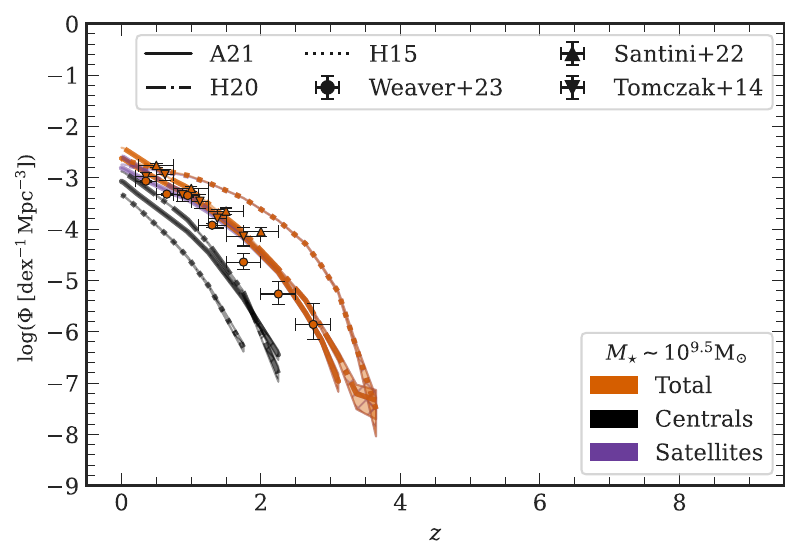}
    \includegraphics[width=0.45\linewidth]{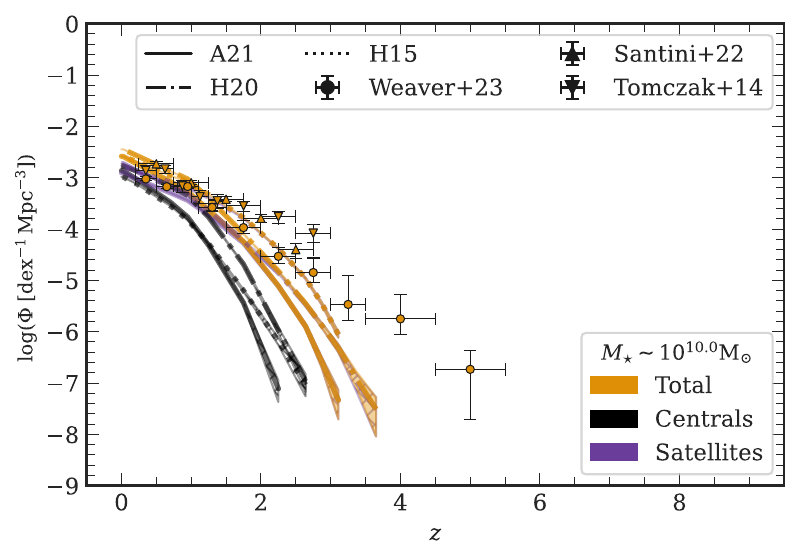}
    \includegraphics[width=0.45\linewidth]{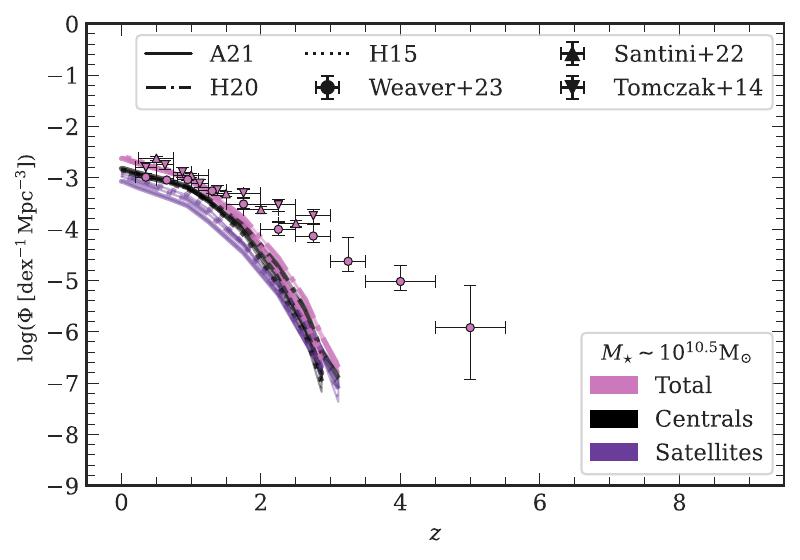}
    \includegraphics[width=0.45\linewidth]{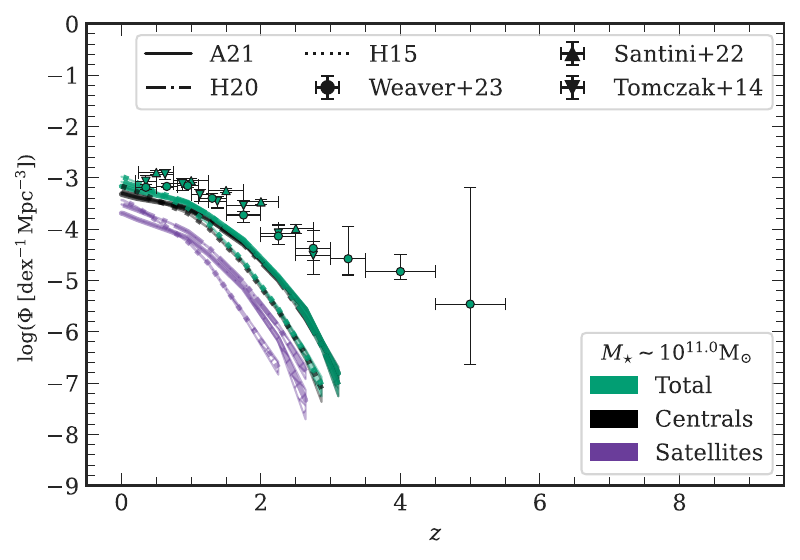}
    \includegraphics[width=0.45\linewidth]{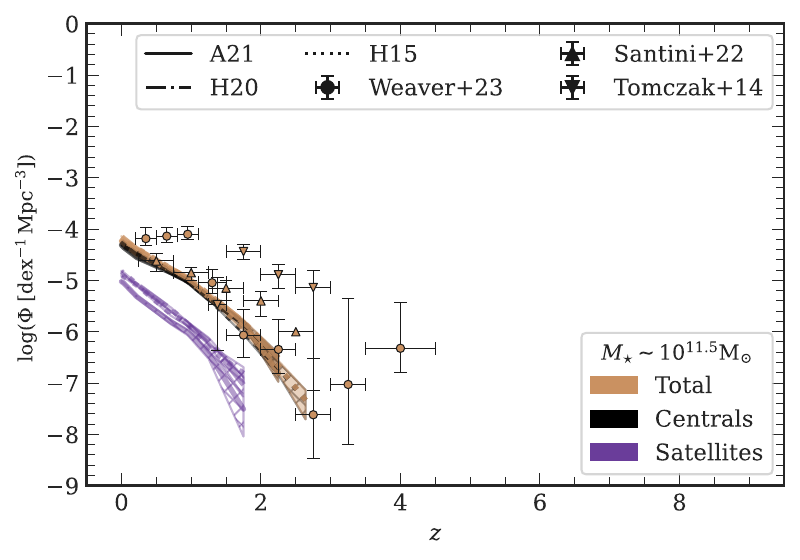}
    \caption{QG number density as a function of redshift: Similar to \figref{fig:SMF_tot_SMF_z} and \ref{fig:SMF_SFG_SMFz_SFG}, these panels depict the galaxy number density per mass bin as a function of redshift for the sample of \qg{}s for the three model flavours. The plot is divided into six panels representing the six mass bins from $10^{9}\:{\rm M}_{\odot}$ to $10^{11.5}\:{\rm M}_{\odot}$, with each mass bin having a width of 0.2~dex. In each plot, the number density of \qg{}s is then split for centrals and satellites represented in black and purple, respectively. The simulation results are also shown with one-sigma Poisson errors. It includes the corresponding observational data points and errors from \citet{Tomczak2014, Santini2022} and \citet{Weaver2023}.
    }
    \label{fig:SMFz_QG}
\end{figure*}

As described in \secref{sec:nuvrj}, the selection of \qg{}s is based on the $\rm NUVrJ$ criterion (Eq. \ref{eq:NUVrJ}). The \qg{} \smf{} for \aXXI{} is presented in~\figref{fig:SMF_QG} (see~\figref{fig:SMFs_H20_H15} and \figref{fig:SMF_QG_H20} for the results of \hXV{} and \hXX{}). It is derived similarly to the method described in \secref{sec:TotalSMF}, and a comparison is made with the observational results of \citet{Weaver2023} including the quoted errors. The \smf{}s outside of the $M_\star=10^{9.5}\:{\rm M}_{\odot}$ transition threshold of the MRI and MRII simulations are represented as fainter lines, highlighting the convergence of the two runs. This shows good agreement at $z\sim0.5$, although the quality of the alignment degrades at higher redshifts.

The three \lgal{} model flavours achieve reasonable agreement with the observed low-redshift ($z\lesssim 2$) \qg{} \smf{} with small 0.1-0.2~dex deviations, reproducing the double Schechter function appearance noted by e.g., \citet[][]{Santini2022}. 
Note that the `knee' of the \smf{} -- the point marking a significant downturn in the number density of galaxies -- occurs at a slightly lower mass in the simulations compared to observations. 
At redshifts, $z\sim 0.5 - 1$, the discrepancy between the `knee' of the \smf{} in observations and simulations is about $0.3-0.4$~dex. While \aXXI{} reproduces a double Schechter-like profile, including a characteristic upturn at lower masses due to updated environmental effects, this feature is less pronounced in \hXX{} and absent in \hXV{}. In \aXXI{}, the double Schechter function-like shape is also seen at larger~$z$, suggesting that the updated prescription for environmental effects is important for producing the population of quenched low-mass galaxies. However, at $z\gtrsim2$, the observational data may suffer from incompleteness issues, leading to an exponential decline in the \qg{} \smf{} at lower masses ($M_\star\lesssim10^{10}\:{\rm M}_{\odot}$). This contrasts with the mass buildup observed in low-mass quenched systems in the \aXXI{} model flavour, also aligning closely with some observational findings \citep[e.g.,][]{Santini2022}.

A notable deficit of \qg{}s is evident in all three \lgal{} flavours at $z\gtrsim2$. \figref{fig:SMFz_QG} presents a more detailed exploration of the
evolution of the quiescent galaxy population, with the simulated
galaxy population decomposed into contributions from central and satellite galaxies. The evolution of the \qg{}  number density per mass bin is shown in a series of mass bins, each 0.2~dex wide, as a function of redshift. Here, centrals are marked in black and satellites in purple, with the cross-hatched region indicating one-sigma Poisson errors. Different line styles are used for the different \lgal{} flavours. Observational data from \citet{Tomczak2014, Santini2022}, and \citet{Weaver2023} with their errors and mass completeness limits are also shown. It is important to note that \citet{Weaver2023} quote large error bars at $z \gtrsim 4$ due to mass incompleteness, observing only 200 quenched systems in this regime and only 1 object at $z\sim 6$, which limits confidence in their \smf{} fit to the range $3.5 \lesssim z\lesssim4.5$.

There is a slight excess of the lowest mass galaxies in the simulations at all redshifts in all three \lgal{} model flavours. A similar discrepancy was noted by \citet{Harrold2024_overabundance}  when comparing \lgal{} predictions to UKIDSS Ultra-Deep Survey data \citep[UDS;][]{Lawrence2007UKIDSS}.  The excess found in our work is smaller.

We note that satellite galaxies are the dominant contributors to the number density of very low mass \qg{}s at all redshifts, whereas central \qg{}s become more prevalent in the higher mass bins. The dominance of centrals over satellites occurs at $M_\star\approx 10^{10.2}-10^{10.3}\: {\rm M}_{\odot}$ in the models.

In the intermediate mass bins of $M_\star\sim10^{10}\: {\rm M}_{\odot}$ and $M_\star\sim10^{10.5}\: {\rm M}_{\odot}$, the simulation results align well with the observations at lower redshifts ($z\lesssim2-2.5$). At higher redshifts, the number density of simulated \qg{}s decreases more rapidly than observed.

In the high mass bin ($M_\star\sim10^{11} \:{\rm M}_{\odot}$), there is a considerable shortfall in the number density of simulated \qg{}s relative to observations, with the gap widening to a factor of 60 by $z\approx3$. In this range, massive quenched centrals are the primary contributors.  The final mass bin ($M_\star\sim10^{11.5} \:{\rm M}_{\odot}$)  suffers from large variability in the observational data, and a precise conclusion is difficult to draw. Nevertheless, the number density of quenched galaxies is often underpredicted in the \lgal{} simulations.

The results presented in this section suggest potential model limitations in capturing early quenching processes. The dominance in these plots shifts from satellites to centrals around $M_\star\sim 10^{10.2}-10^{10.3}\: {\rm M}_{\odot}$, coinciding with the proposed critical mass threshold for `mass quenching' suggested by \citet{Peng2010}. This threshold delineates where the influence of a galaxy's own properties such as mass and black hole activity outweigh external factors such as environmental effects or interactions with other galaxies.

In summary, the \qg{} \smf{} offers valuable insight into the evolution of quenched systems across different epochs. In the three \lgal{} flavours, we observe a rapid evolution and significant mass buildup in massive quenched centrals starting from $z \sim 3$, a rate much faster than what observations suggest. Observational data generally indicate a milder evolutionary slope for these large mass bins ($M_\star\sim 10^{10}-10^{11}\:{\rm M}_{\odot}$).  Conversely, in the lower mass bins predominantly comprised of satellites, there is an overabundance of \qg{}s,  with the degree of this surplus varying depending on the observational data set adopted. We note that because lower-mass galaxies have lower surface brightnesses, corrections for selection effects are more difficult and thus our comparison for this specific mass range may not be as robust as for the more massive systems.

Finally, a crucial point that we find in our analysis is the complete absence of massive quenched systems beyond $ z \sim 3-3.5$ in the models. This could introduce a significant challenge to the implementation of physical processes in our galaxy formation models.

\subsection{Evolution of the star-forming main sequence}

\begin{figure}
    \centering
    \includegraphics[width=1\linewidth]{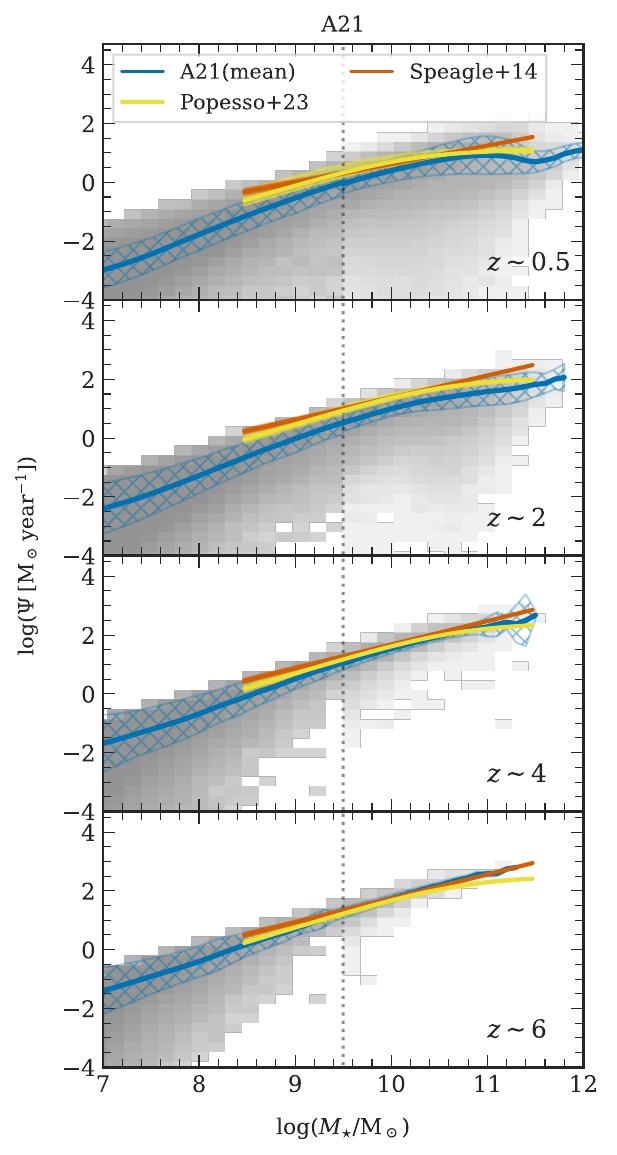}
    \caption{ ${\rm SFR}-M_\star$ relation: 
    A plot describing the main sequence of star-forming galaxies across various redshifts for the \aXXI{} flavour. The yellow line represents the observational fit from \citet{Popesso2023}, with its uncertainty indicated by the surrounding yellow shaded area, while the red line depicts the relation from \citet{Speagle2014} with its errors. The grey-shaded region in each plot denotes the distribution of galaxies from \aXXI{}. Additionally, the solid blue line traces the running mean in the mass bin of the simulated galaxies. The blue cross-hatched region represents the one-sigma scatter from the mean. The results from other flavours are shown in \figref{fig:SFR_M_H20_H15}.}
    
    \label{fig:SFR_M}
\end{figure}

\begin{figure}
    \centering
    \includegraphics[width=1\linewidth]{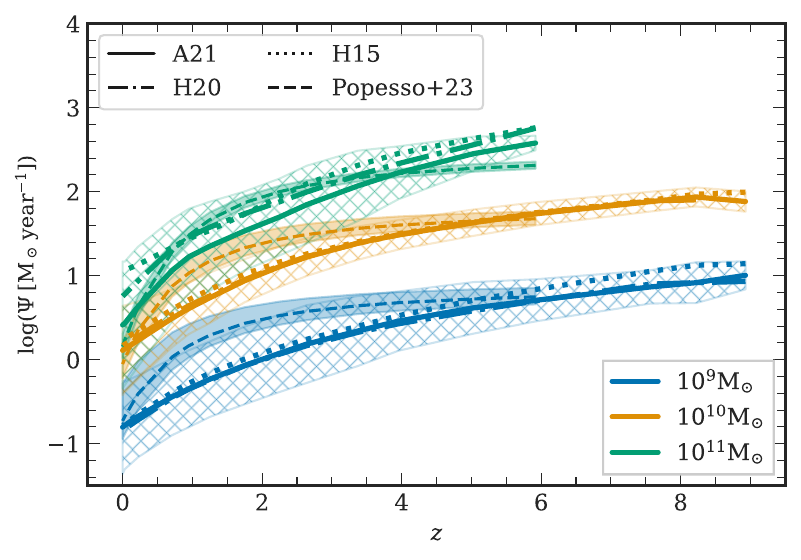}
    \caption{SFR to redshift relation: A plot showing the evolution of SFR as a function of redshift in three different stellar mass bins, colour-coded accordingly with a bin width of 0.2~dex. The line style indicates the results from different \lgal{} flavours: the solid line corresponds to \citetalias{Ayromlou2021}, the dot-dashed line to \citetalias{Henriques2020}, and the dotted line to \citetalias{Henriques2015}. Additionally, the dashed line represents the observational findings from \citet{Popesso2023}, with the corresponding uncertainties depicted by the shaded regions. The cross-hatched region represents the one-sigma scatter from the mean for each mass bin, applicable only for the results from \aXXI{}.}
    
    \label{fig:SFR_M_Lgalmodel}
\end{figure}

The main sequence of star-forming galaxies is illustrated in \figref{fig:SFR_M} for \aXXI{}. The plot presents a 2D histogram representing the simulated galaxies in the ${\rm SFR}-M_\star$ parameter space normalised by the respective volume (MRI or MRII, see \secref{sec:sim&subhalo}), thus the colour coding suggests the density of galaxies per cubic Mpc per pixel of size $0.18\: [\log({\rm M}_{\odot})] \times 0.33\:[\log({\rm M}_{\odot}/{\rm year})]$. The solid blue line traces the running mean of the distribution, while the blue cross-hatched region represents the one-sigma scatter around the mean. A comparison is made to the findings of \citet{Popesso2023} using the fitting function of \citet{Lee2015}, shown by the solid yellow line, and \citet{Speagle2014} is represented by the red line. The shaded regions around these lines indicate the quoted errors. Additionally, we account for the IMF offset in both \citet{Speagle2014} and \citet{Popesso2023}, and convert the \citet{Kroupa2001} IMF to a \citet{Chabrier2003_IMF} IMF using the relation $M_{\star,{\rm Chabrier}}=0.94 \times M_{\star,{\rm Kroupa}}$.

The three model flavours align well within one-sigma scatter with \citet{Speagle2014} across all masses and redshifts (for results from \hXV{}, see \figref{fig:SFR_M_H20_H15}).
In \aXXI{}, particularly for $0\lesssim z\lesssim4$, where the SFR is observed to plateau at higher masses near the turnover mass ($\sim10^{10.5}\:{\rm M}_{\odot}$), the SFR is suppressed and remains nearly constant for larger masses, while below this mass, the SFR is proportionally increasing with stellar mass. This is consistent with findings from \citet{Popesso2023} suggesting a flattening at the high-mass end.  This flattening trend, while absent in \hXV{}, is less pronounced in \hXX{} at lower redshifts and is also absent in \aXXI{} at larger redshift ($z\gtrsim 5$). The observed flattening at high-$z$, as reported by \citet{Popesso2023}, suggests a decrease in star formation efficiency at higher masses. This likely reflects the cumulative effects of less efficient gas cooling and increased feedback mechanisms in massive galaxies, or it could be attributed to the transition in halo mass from predominantly cold to hot gas accretion \citep{Dekel2006}.

More obvious discrepancies between the models and the data can be seen in  \figref{fig:SFR_M_Lgalmodel}, which shows the SFR as a function of redshift for three mass bins of 0.2~dex width. In this plot, we show results for all three \lgal{} flavours. The cross-hatched markings represent one-sigma scatter from the mean SFR of the \aXXI{} flavour. 

We note a reasonably good agreement with the work of \citet{Popesso2023} for all the mass bins for the three flavours, although there are some discrepancies. At lower redshift ($ z \lesssim 0.5-1$), the models align well for the $ M_\star \sim 10^{9}\: {\rm M}_{\odot}$ and $M_\star \sim 10^{10}\: {\rm M}_{\odot}$ bins. However, the mean SFR for the $M_\star \sim  10^{11}\, {\rm M}_{\odot}$ bin tends to be overpredicted by \hXX{} and \hXV{}, and to a lesser degree in \aXXI{}. At higher redshifts ($z \gtrsim 4$), all three flavours overpredict the SFR of the highest mass bin. This suggests that the model’s predictions for the SFR of high-mass galaxies need to be further suppressed to match the observed trends at the higher redshifts more accurately.

For the least massive  $M_\star \sim  10^{9}\:{\rm M}_{\odot}$ galaxies, all three models deviate from the results of \citet{Popesso2023} over the redshift range $0.5 \lesssim z \lesssim 4$,
with the SFRs in the models lying a factor of $2-3$ below the observations. This result echoes the broader discrepancy seen
in \figref{fig:CSFR} for the integrated CSFRD. The main conclusion from the $ {\rm SFR}-M_\star$ comparison is that the lower mass galaxies are responsible for the offset between the data and the models. For the redshift range from $0.5 \lesssim z \lesssim 4$ and all three mass bins, the models under-predict the mean SFR, once again suggesting that revisions in the implementation of physical processes relevant to star formation and quenching are necessary to align more closely with the observations.

\subsection{Evolution of galaxy mass-size relation}
\label{sec:MSR_model}
\begin{figure*}
    \centering
    \includegraphics[width=0.43\linewidth]{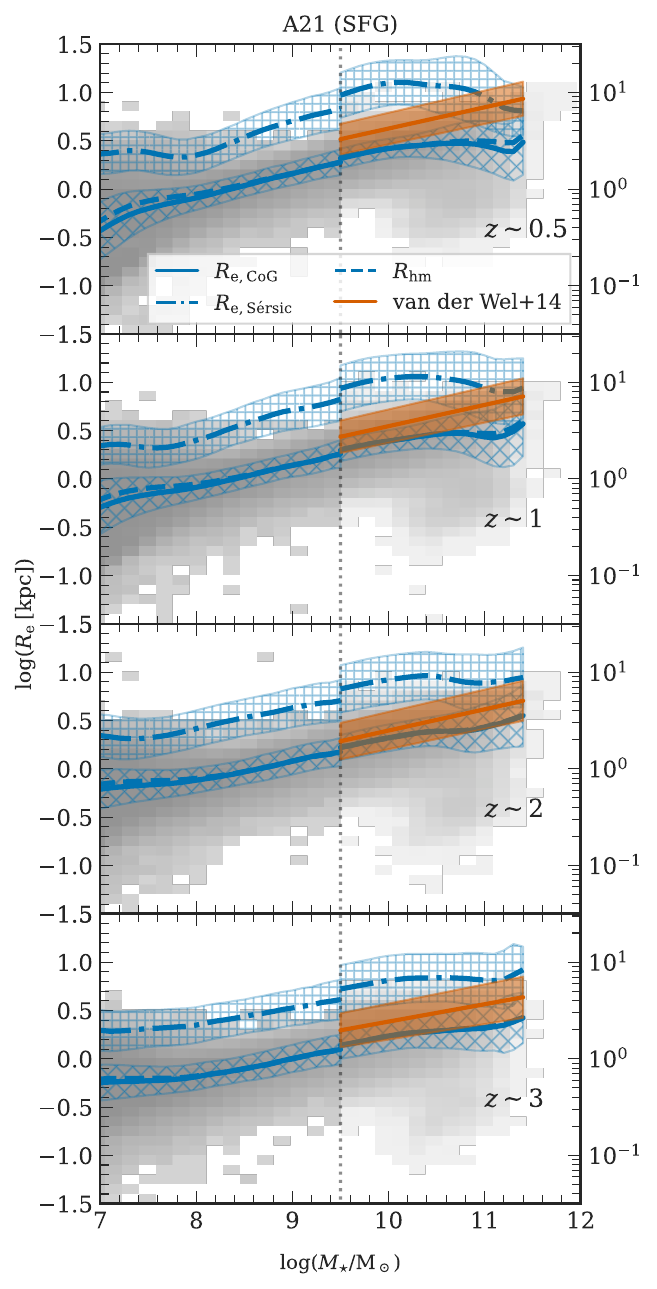}
    \includegraphics[width=0.43\linewidth]{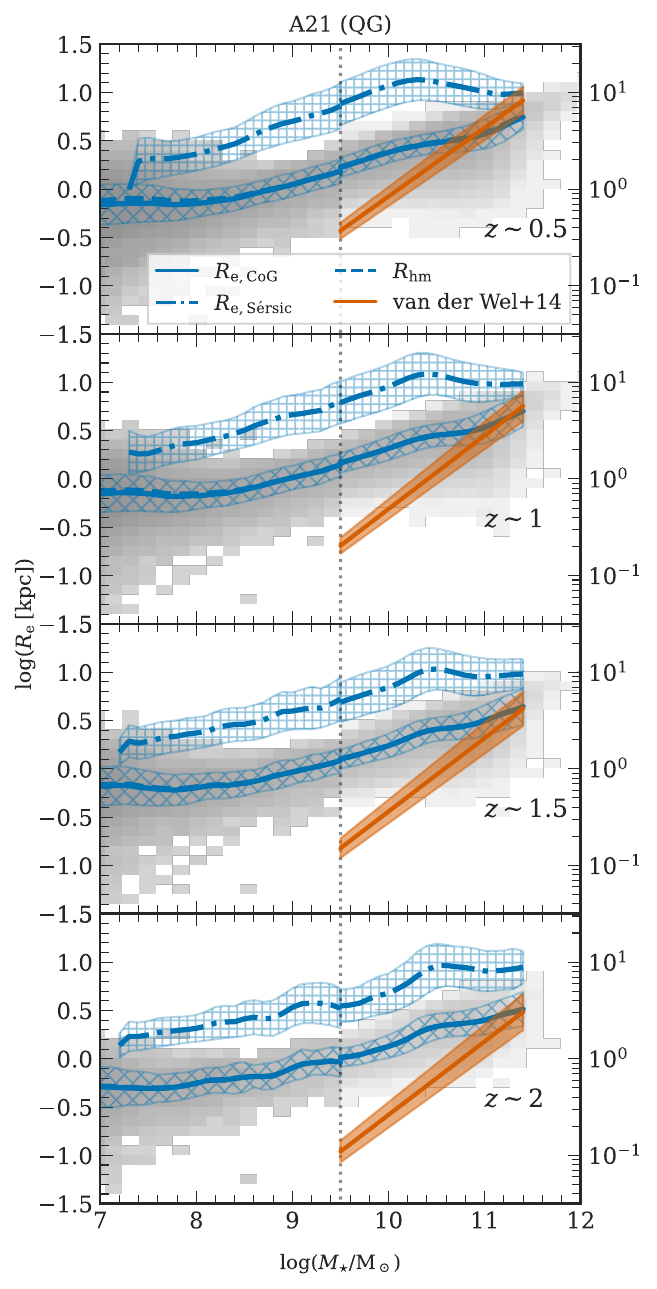}
    \caption{ Galaxy mass-size relation: 
    The plots illustrate the $R_{\rm e}-M_\star$ relation, specifically the half-light radii denoted as $R_{\rm e}$, the radius encompassing half of the total flux of the galaxy. This relation is plotted for the \sfg{}s (\textit{left}) and \qg{}s (\textit{right}) at 4 redshifts for the \aXXI{} model flavour. The 2D histogram shows the distribution of $R_{\rm e,\:CoG}$ of simulated galaxies, with the solid blue line tracing the mean of the distribution. The blue dashed line represents the half-mass radius ($R_{\rm hm}$).
    In the \lgal{} model, $R_{\rm e}$ is derived in two ways: using the curve of growth approach ($R_{\rm e,\:CoG}$, solid blue line) and by fitting a 1D S\'ersic profile ($R_{\text{e, S\'ersic}}$, dot-dashed blue line). The blue cross-hatched and plus-hatched region represents the one-sigma scatter around $R_{\rm e,\:CoG}$ and $R_{\text{e, S\'ersic}}$, respectively.
    The red line corresponds to the relation by \citet{VanDerWel2014_MSR} with one-sigma scatter. The right axis of the plot is in linear units. Mass incomplete measurements of \citet{VanDerWel2014_MSR} are not shown.
}
    \label{fig:SFGQG_MSR_M}
\end{figure*}

Estimating galaxy sizes is typically done via one of two methods: aperture fluxes, also known as the curve of growth (CoG) approach \citep[e.g.,][]{Carollo2013}, or profile fitting \citep[e.g.,][]{VanDerWel2014_MSR}. In the \lgal{} model, the half-light radii or the effective radii ($R_{\rm e}$) can be calculated only for two flavours, namely \hXX{} and \aXXI{}, as these models contain resolved properties of galaxies (see \secref{Sec:Lgal_model}). Both of these flavours employ the curve of growth, which iteratively determines the radius encompassing half of the total flux of the galaxy. In contrast, the work of \citet{VanDerWel2014_MSR} and \citet{Mowla2019} uses a S\'ersic profile fit \citep{1963Sersic} to derive this property from the observations. It is worth noting that both approaches are susceptible to systematic biases in estimating $R_{\rm e}$ \citep[e.g.,][]{Shen2003, Graham2005, Cameron2007, Whitaker2011_UVJ, Ichikawa2012, VanDerWel2014_MSR}. 

Due to the availability of resolved data for galaxy properties in \hXX{} and \aXXI{}, we derive the surface brightness profiles of each galaxy and fit a single-component 1D S\'ersic profile to obtain $R_{\rm e}$ in the rest-frame V ($0.55 \: \mu {\rm m}$) band filter, using a surface brightness limit $I_{\rm limit} = 26.6 \; [{\rm mag}\; {\rm arcsec}^{-1}]$ \citep{Papovich2012} (more details in \appref{app:SersicFit}), consistent with the observational data from \citet{vanderWel2014b, Mowla2019}. This allows for a comparison between $R_{\rm e}$ estimates from the CoG approach and those from the S\'ersic profile fit as well.

In~\figref{fig:SFGQG_MSR_M}, we present the galaxy mass-size relation for \sfg{}s (\textit{left}) and \qg{}s (\textit{right}) in the \aXXI{} flavour. Each panel includes data from different redshifts and displays a 2D histogram representing $R_{\rm e}$ derived from the CoG approach ($R_{\rm e, CoG}$), with the solid blue line and the cross-hatched region showing the running mean and the scatter, respectively.  The colour coding of the 2D histogram illustrates the number density of galaxies per cubic Mpc per pixel of size $0.18\: [\log({\rm M}_{\odot})] \times 0.10\:[\log({\rm kpc})]$.  The dashed blue line represents the half-mass radius, $R_{\rm hm}$, defined as the radius encompassing half of the galaxy's stellar mass. Additionally, the blue dot-dashed line shows $R_{\rm e}$ derived from the S\'ersic fit ($R_{\text{e, S\'ersic}}$) accompanied by the plus-hatched region depicting its one-sigma scatter from the mean.  
The S\'ersic fit results from the model can be compared directly with observational relations from  \citet{VanDerWel2014_MSR} and \citet{Mowla2019}. The solid red line represents parametrised fits from the work of \citet{VanDerWel2014_MSR}, measured along the semi-major axis of an ellipse, with the respective one-sigma scatter along the mean. The left vertical axis is in log units while the right axis of the plots is shown in linear kpc units.

In the left panel of \figref{fig:SFGQG_MSR_M}, representing \sfg{}s of \aXXI{},  the solid blue line indicates a strong correlation of  $R_{\rm e, CoG}$ with stellar mass until $M_\star\sim 10^{10.5}\:{\rm M}_{\odot}$, after which the relation flattens. The flattening is stronger at lower redshifts. The dot-dashed line, representing the S\'ersic fit half-light radius ($R_{\text{e, S\'ersic}}$), shows a similar flattening at high stellar masses. 

Overall, the galaxy sizes predicted by the model are larger than the observed sizes. The discrepancy is largest for \qg{}s and for low-mass galaxies (\figref{fig:SFGQG_MSR_M} \textit{right} panel). At stellar masses of $M_\star \sim 10^{9.5} {\rm M}_{\odot}$, the size discrepancies in the quenched population reaches a factor of 10 at low redshifts ($z\sim0.5$) and nearly a factor of 30 at $z\sim2$. The agreement between the models and the observations is best for high-mass star-forming galaxies. The size-mass relation over the stellar mass range $10^{9.5}-10^{11} \: {\rm M}_{\odot}$ is considerably flatter in the models than in the observations. The discrepancy in the slope of the relation is again worse for the quiescent population.

\begin{figure}
    \centering
    \includegraphics[width=1\linewidth]{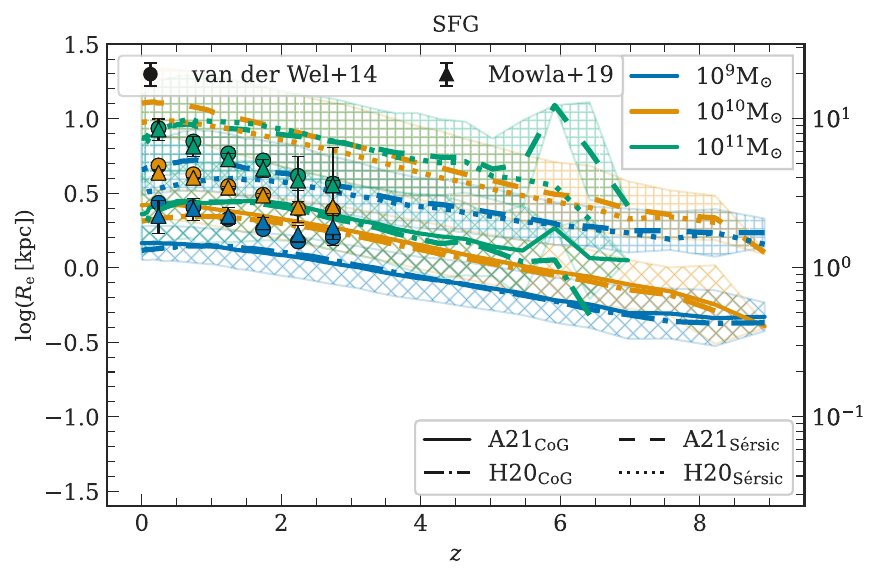}
    \includegraphics[width=1\linewidth]{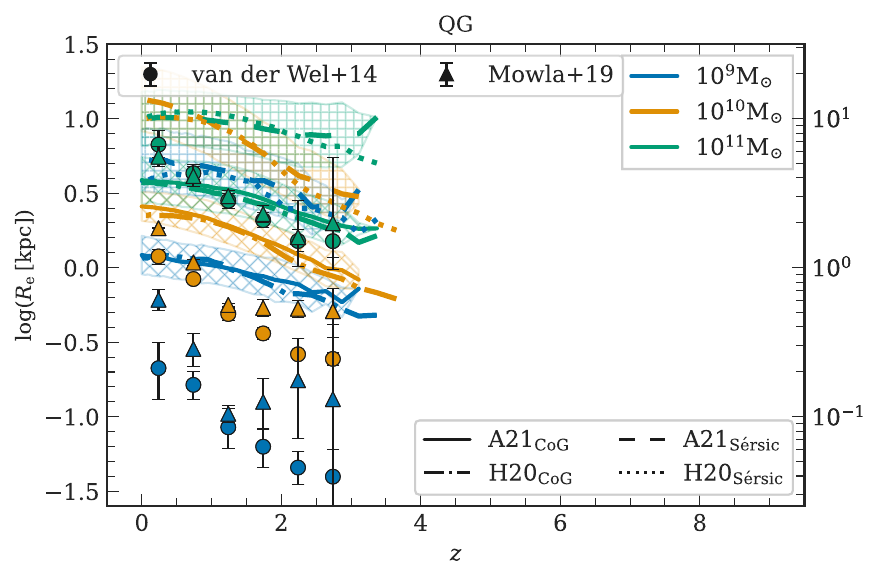}
    \caption{Galaxy size to redshift evolution: 
    These figures illustrate the evolution in  $R_{\rm e}$ as a function of redshift for three mass bins of 0.2~dex bin width for the \sfg{} sample (\textit{top}) and \qg{} sample (\textit{bottom}). The solid and dot-dashed lines represent the results from \aXXI{} and \hXX{}, respectively, using the curve of growth (CoG) approach with the cross-hatched region denoting one-sigma scatter for the results of \aXXI{}. The dashed and dotted lines represent the results from \aXXI{} and \hXX{}, respectively, for $R_{\rm e}$ derived from the 1D S\'ersic fit, with the plus-hatched region denoting scatter in the results of \aXXI{}. The results are compared with the data from \citet{VanDerWel2014_MSR} and \citet{Mowla2019}.}
    \label{fig:SFG_QG_MSR_z}
\end{figure}

In \figref{fig:SFG_QG_MSR_z}, the evolution of galaxy sizes as a function of redshift is shown in three different mass bins. The solid and dot-dashed lines represent the mean $R_{\rm e}$ of the two flavours from the CoG approach while the sparsely spaced dashed line and the dotted line show the values derived using the S\'ersic fit. Each colour corresponds to a specific mass bin, each having a 0.2~dex bin width. Additionally, the markers represent the observational data points from \citet{VanDerWel2014_MSR} and \citet{Mowla2019}. 

The results shown in the top panel of \figref{fig:SFG_QG_MSR_z} indicate that the \aXXI{} and \hXX{}  models both predict a mild evolution of $R_{\rm e}$ for \sfg{}s. However, the model results do not match the observations in detail, displaying a size discrepancy of approximately a factor of $1.5-2$. The offset between sizes derived by the two methods is also evident in the plots. Despite this, a general agreement with the slope of $R_{\rm e}$ as a function of $z$ is observed for stellar mass bins $10^{9}\:{\rm M}_{\odot}$, $10^{10}\:{\rm M}_{\odot}$, and $10^{11}\:{\rm M}_{\odot}$ when considering $R_{\rm e}$ derived from the S\'ersic fit. 

Figure~\ref{fig:SFG_QG_MSR_z} (\textit{bottom}), illustrates the evolution for \qg{}s, and reveals a noticeable discrepancy in the size evolution of galaxies between the model and the data. The very steep evolution suggested by observations is not replicated in the models. Although there seems to
be some consistency and agreement in the largest mass bin $M_\star \sim 10^{11}\:{\rm M}_{\odot}$ for the sizes derived by the CoG approach (solid and dot-dashed lines), the agreement with data is much worse for the lower mass bins. Overall, the discrepancy detected between the sizes of simulated and observed galaxies is typically larger for \qg{}s than for star-forming galaxies. This could reflect the complex set of physical processes related to size evolution such as mergers \citep{Naab2009}, host dark matter halo properties \citep{Zanisi2020} or the distribution of stars within galaxies \citep{Dabringhausen2008, VeraCiro2014, Grand2015, Okalidis2022}

\subsection{Evolution of the central surface mass density}
\label{sec:SMSD_results}
\begin{figure*}
    \centering
    \includegraphics[width=0.43\linewidth]{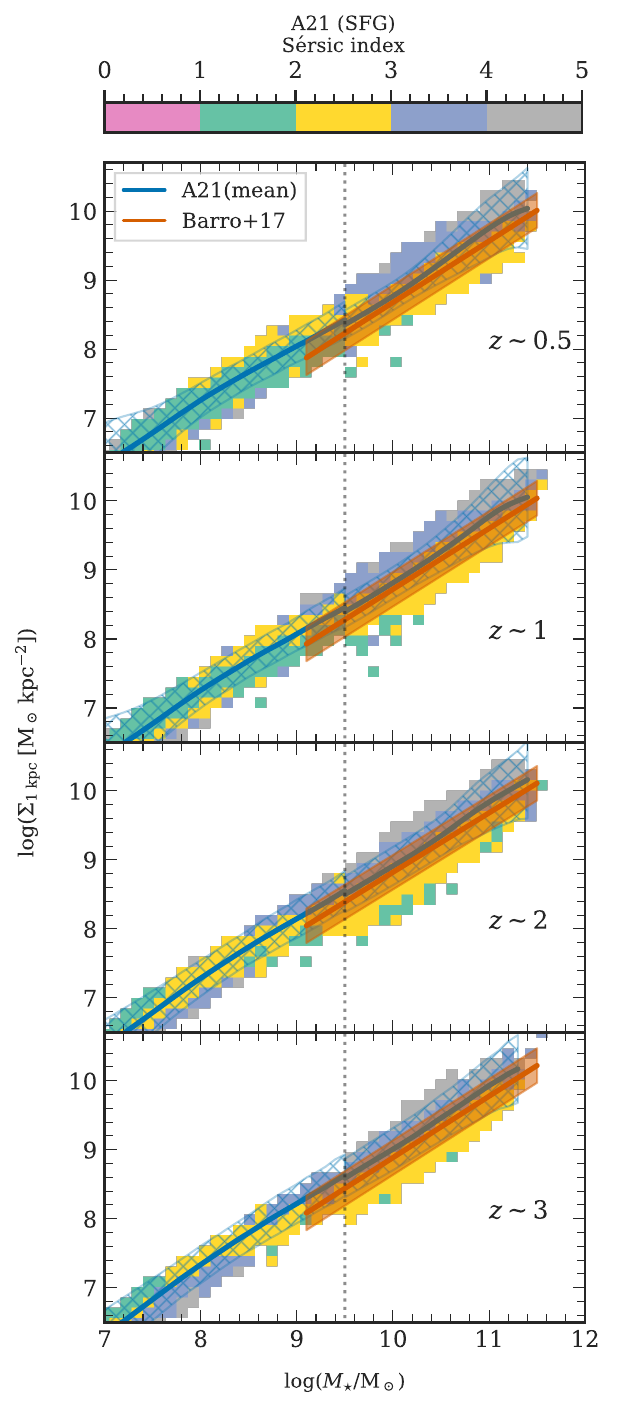}
    \includegraphics[width=0.43\linewidth]{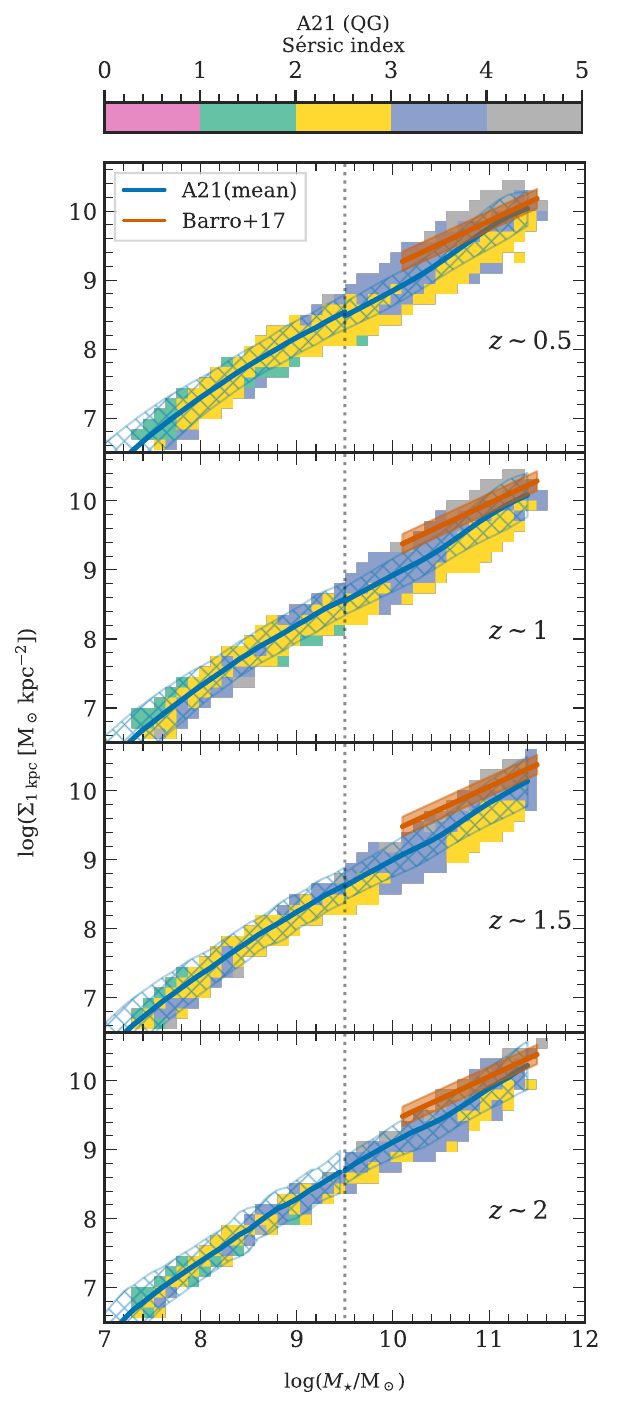}
    \caption{Galaxy core stellar mass surface density to mass relation: The plots illustrate the core ($R<1\;{\rm kpc}$) stellar mass surface density to galaxy mass relation for the \sfg{} (\textit{left}) and the \qg{}s (\textit{right}) at four redshifts presented for \aXXI{}. The 2D histogram shows the distribution of simulated galaxies with the blue line tracing the running mean and the blue cross-hatched region denoting the one-sigma scatter around the mean. The colour indicates the median S\'ersic index of each mass bin, shown if the fit is within 10\% of the residuals. The red line represents the relation derived from \citet{Barro2017} along with one-sigma scatter around the mean represented as the shaded region.
    }
    \label{fig:SMSD_1kpc}
\end{figure*}

\begin{figure*}
    \centering
    \includegraphics[width=0.43\linewidth]{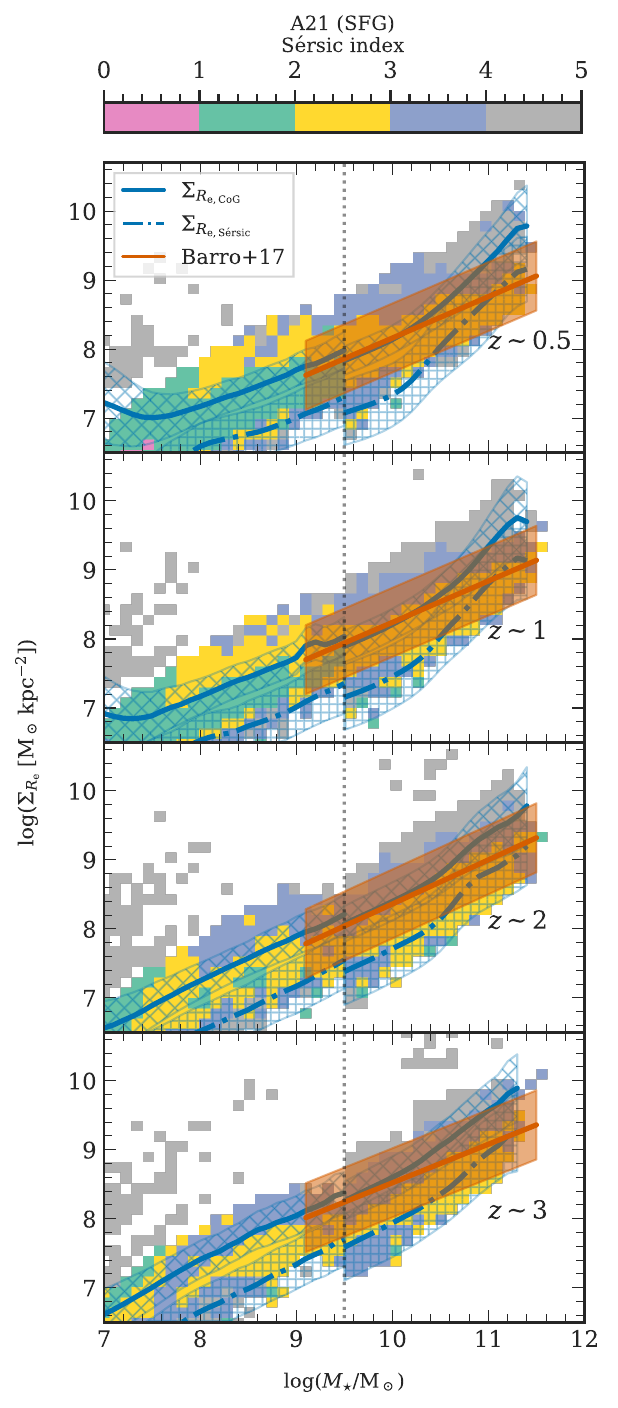}
    \includegraphics[width=0.43\linewidth]{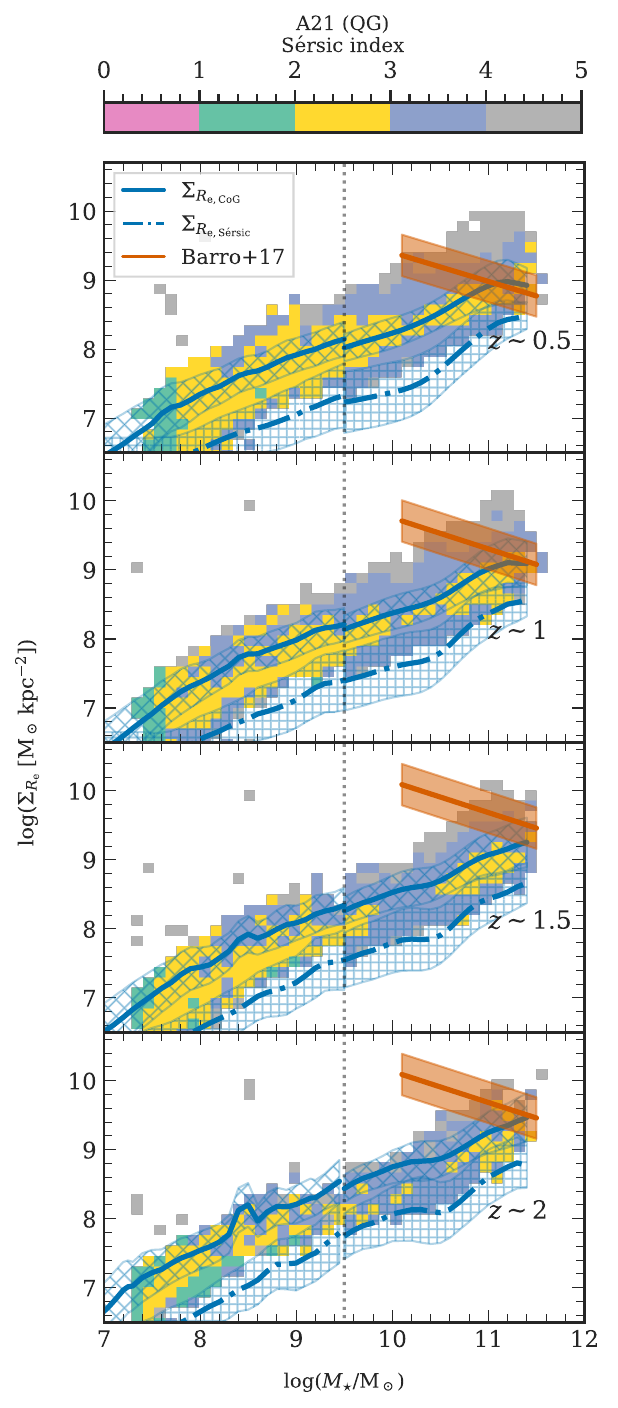}
    \caption{Galaxy effective stellar mass surface density to mass relation: Similar to \figref{fig:SMSD_1kpc} the plot represents the stellar mass surface densities within one effective radius ($\Sigma_{R_{\rm e}}$). The relation is shown for the \sfg{}s (\textit{left}) and \qg{}s (\textit{right}) for the \aXXI{} flavour. 
    In the \lgal{} model, $R_{\rm e}$ is derived in two ways: using the curve of growth approach ($\Sigma_{ R_{\rm e,\:CoG}}$, solid blue line) and by fitting a 1D S\'ersic profile ($\Sigma_{R_{\text{e, S\'ersic}}}$, dot-dashed blue line). The blue plus-hatched region represents the one-sigma scatter around $\Sigma_{R_{\text{e, S\'ersic}}}$, while the cross-matched region is for one-sigma scatter around $\Sigma_{R_{\rm e, CoG}}$. The results from \citet{Barro2017} are plotted in red. The colour coding states the median S\'ersic index of each mass bin if the S\'ersic fit is within 10\% of the residuals. }
    \label{fig:SMSD_Re}
\end{figure*}

With resolved data from the \lgal{} simulations, we can derive stellar mass surface densities within a 1~kpc radius (core) or the effective radius ($R_{\rm e}$, see \secref{sec:MSR_model} and Appendix \ref{app:SersicFit}), by utilising the model flavours with radial rings, i.e. \aXXI{} and \hXX{}. The surface mass densities include the stellar mass contribution from the bulge and the disc.

As outlined in \secref{sec:SMSD}, \qg{}s should be characterised by higher $\Sigma_{1\;{\rm kpc}}$, but this is only weakly seen in \figref{fig:SMSD_1kpc} for \aXXI{}.  Both \lgal{} flavours exhibit a tight correlation between the galaxy stellar mass and $\Sigma_{1\;{\rm kpc}}$. Figure~\ref{fig:SMSD_1kpc} presents simulation data from \aXXI{} for the \sfg{}s (\textit{left}) and \qg{}s (\textit{right}), with the 2D histogram illustrating the distribution of stellar mass surface densities simulated galaxies while the colour coding representing the median S\'ersic index of each pixel of size $0.18\: [\log({\rm M}_{\odot})] \times 0.15\:[\log({\rm M}_{\odot}\:{\rm kpc}^{-2})]$. The blue line and the cross-hatched region show the running mean of the distribution and its one-sigma scatter. The observational results from \citet{Barro2017} are plotted with a red line, including a one-sigma scatter and mass completeness limits. Both \sfg{}s and \qg{}s in the model flavours align well with observed data, showing similar trends and evolution. However, intermediate mass ($M_\star \sim 10^{10}\:{\rm M}_{\odot}$) quenched systems are not as compact as noted in the observations. 

The S\'ersic index ($n$) for galaxies is shown in the two plots only if the residual\footnote{defined as \mbox{$\Delta I_{\text{residual}}=\Sigma_{i=0}^{r_i} (I(r)_{\rm \lgal{}}/I(r)_{\text{1D S\'ersic}}-1)^2/{\rm dof}$}, where $\Sigma$ sums over the squared deviations of surface brightness ($I$) at different galactic radius ($r$) for each galaxy and $\rm dof$ is the degree of freedom.} of the single-component 1D S\'ersic fit is less than 0.1, suggesting the fits are considered good within a 10\% threshold. The scatter with high S\'ersic indices in the surface density-stellar mass plane lies outside the 1-sigma scatter represented by the cross-hatched region of the solid blue line and includes only a negligible number of galaxies. This index $n$ (detailed in Appendix \ref{app:SersicFit}), presents higher values in both \sfg{}s and \qg{}s within \aXXI{} and \hXX{} compared to those reported by \citet{Barro2017}. Typically, \qg{}s (\figsref{fig:SMSD_1kpc}, \ref{fig:SMSD_Re}, \textit{right} panels) exhibit higher S\'ersic indices than their star-forming counterparts, correlating with their more compact nature. The complex shape of surface brightness profiles and the notably high S\'ersic indices reported here underline the limitations of using a single component 1D S\'ersic fit. This mismatch suggests that a single-component S\'ersic profile may not effectively capture the structural differences of these galaxies.

Figure~\ref{fig:SMSD_Re} shows the $\Sigma_{R_{\rm e}}-M_\star$ relationship with the mean S\'ersic index denoted by the colour bar. The running mean of the $\Sigma_{R_{\rm e, CoG}}$ values derived using the CoG approach is shown as the solid blue line, while the $\Sigma_{R_{\text{e, S\'ersic}}}$ values derived using the 1D S\'ersic fit is shown as the dot-dashed blue line. The observational results from \citet{Barro2017} are shown in red. The \sfg{}s are in good agreement with the observations for almost all masses.  In contrast, the quenched systems diverge from the observed data, especially for the low mass $M_\star\sim10^{10}\:{\rm M}_{\odot}$ galaxies which appear significantly less compact deviating at most by a factor of 80 at all redshifts. The most massive $M_\star\sim10^{11}\:{\rm M}_{\odot}$ \qg{}s display a flatter $\Sigma_{R_{\rm e, CoG}}-M_\star$ relation at lower redshift, suggesting a better qualitative agreement with the observations.

\begin{figure*}
    \centering
    \includegraphics[width=0.44\linewidth]{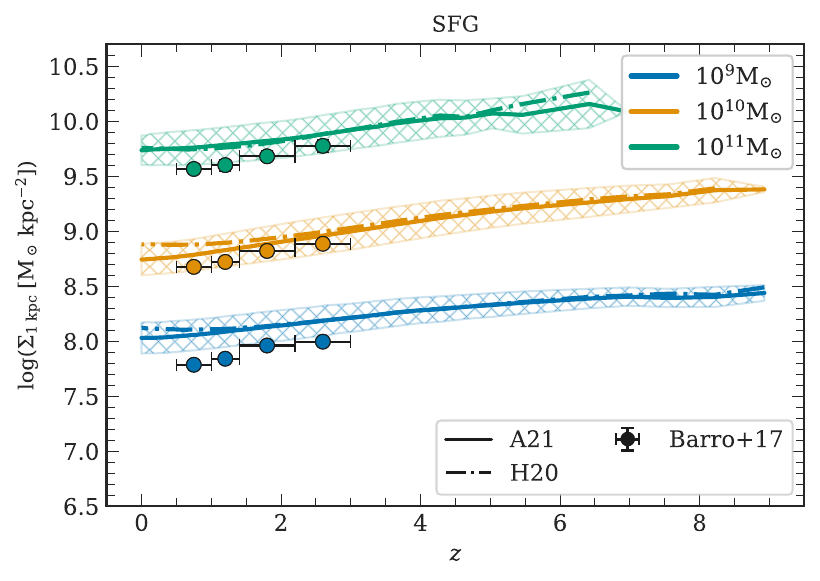}
    \includegraphics[width=0.44\linewidth]{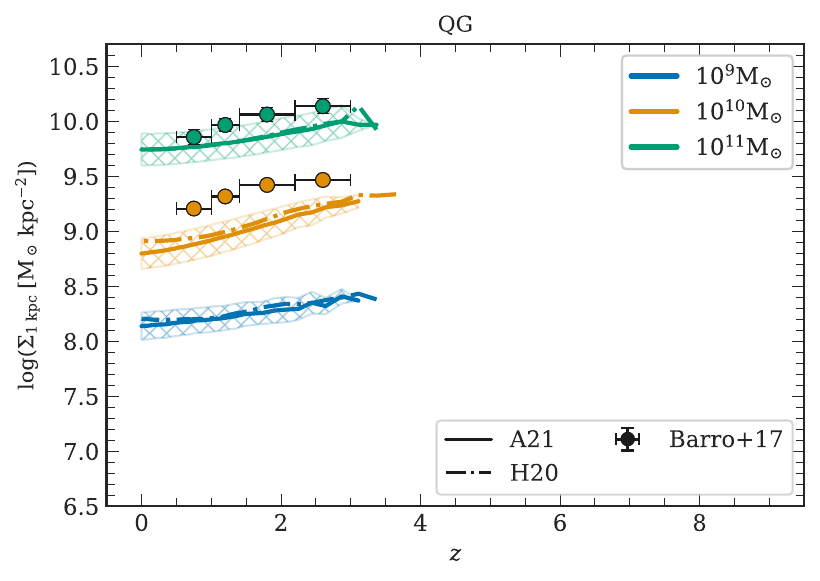}
    \includegraphics[width=0.44\linewidth]{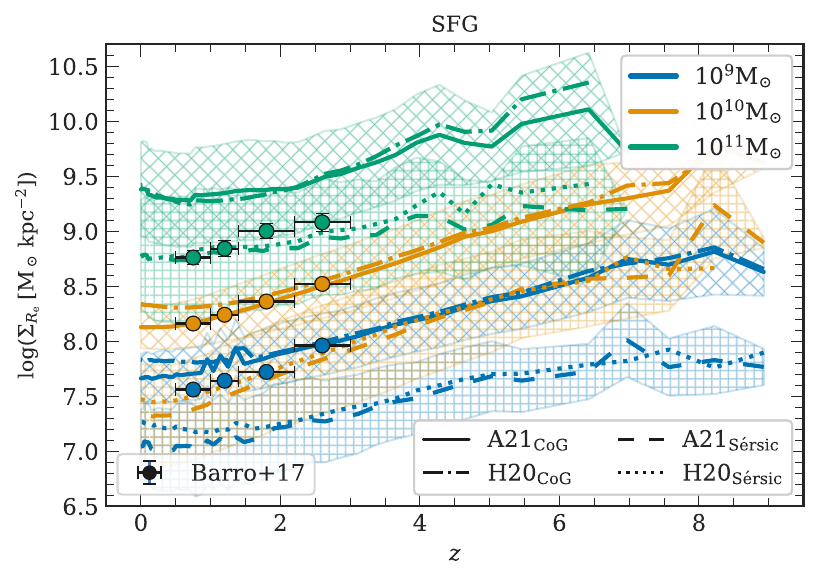}
    \includegraphics[width=0.44\linewidth]{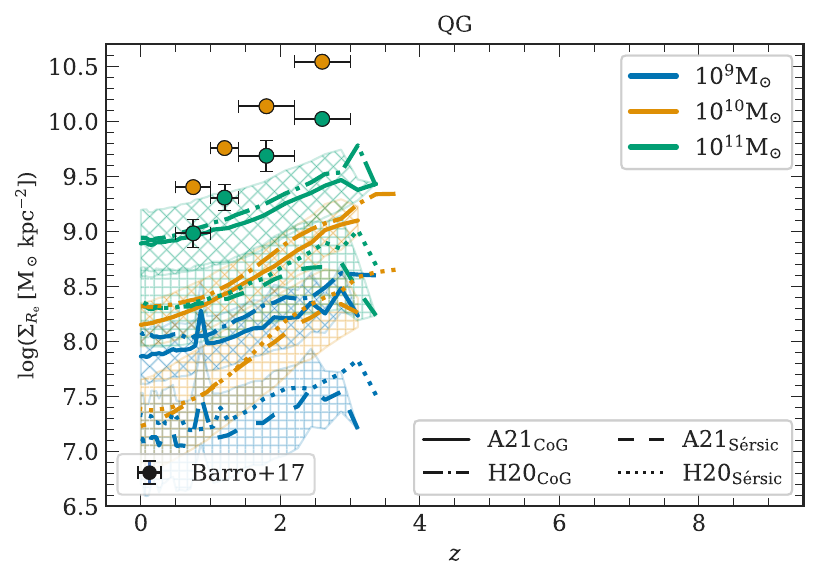}
    \caption{Galaxy core and effective stellar mass surface density to redshift evolution: This figure illustrates the evolution of the stellar mass surface density as a function of redshift for the \sfg{} fraction (\textit{left} column) and \qg{} fraction (\textit{right} column) in three mass bins (0.2~dex bin width) for two \lgal{} flavours. The upper panel shows the stellar densities with 1~kpc while the bottom panel shows the effective densities. In upper the panels, the solid line represents the data from \aXXI{}, and the dot-dashed line represents the results from \hXX{}. The cross-hatched region denotes the scatter around the mean of the \aXXI{} results. For the effective densities, the solid and dot-dashed lines have used the curve of growth (CoG) approach to derive the $ R_{\rm e}$, while the loosely dashed and dotted line use the S\'ersic fit to find the $R_{\rm e}$, where the one-sigma scatter of \aXXI{}'s S\'ersic fit and CoG fit is shown with a plus-hatched and cross-hatched region, respectively.  The markers are from \citet{Barro2017}. The $10^9\:{\rm M}_{\odot}$ mass bin is outside the completeness limits of \citet{Barro2017} for the quenched fraction, hence not included in the right column.}
    \label{fig:SFG_QG_SMSD_z}
\end{figure*}

Figure~\ref{fig:SFG_QG_SMSD_z} illustrates the redshift evolution of the stellar mass surface densities in three mass bins of 0.2~dex bin width. 
For the \sfg{}s (\figref{fig:SFG_QG_SMSD_z}, \textit{left} column), the agreement with observational data is generally robust. The most massive star-forming  systems, around $ M_\star\sim10^{11}\:{\rm M}_{\odot}$ have stellar mass densities both within 1~kpc ($ \Sigma_{1\:{\rm kpc}}$) and at effective radius ($\Sigma_{R_{\rm e}}$) that agree reasonably well with observations. The intermediate-mass star-forming systems of $M_\star\sim10^{10}\,{\rm M}_{\odot}$ agree well on the $ \Sigma_{1\:{\rm kpc}}-M_\star$ plane, though they appear about $3-4$ times denser than observed within $1\,R_{\rm e}$. The lowest mass galaxies tend to be overdense in the simulations.

The quenched systems exhibit much worse inconsistencies with the data.  As shown in \figref{fig:SFG_QG_SMSD_z} (\textit{right} column), the most massive \qg{}s agree with observed $\Sigma_{1\:{\rm kpc}}$ densities, while those around $M_{\star}\sim10^{10}\:{\rm M}_{\odot}$ are significantly less dense -- about 4 to 6 times lower than the values reported by \citet{Barro2017}.  $\Sigma_{R_{\rm e}}$ densities exhibit a strong evolution in the densities of $M_\star\sim10^{10}\:{\rm M}_{\odot}$ and $M_\star\sim10^{11}\:{\rm M}_{\odot}$ galaxies. The intermediate-mass systems are about $80-100$ times less dense than the observations. We note that there are uncertainties in the observational data due to poor statistics in the  $ 2.2\lesssim z\lesssim3$ range.

The distribution of the S\'ersic indices in the $\Sigma_{1\:{\rm kpc}}-M_\star$ and $\Sigma_{R_{\rm e}}-M_\star$ planes in the \aXXI{} and \hXX{} models do not align with \citet{Barro2017}'s observations.
Specifically, \sfg{}s in our models show a lower S\'ersic index ($n \approx 1-2$) for lower mass galaxies, indicating a disc-like structure, while more massive galaxies ($M_\star \gtrsim 10^{10}\:{\rm M}_{\odot}$) exhibit higher indices ($n \approx 2-3$), pointing to a bulge-dominated morphology. The highest mass galaxies ($M_\star \gtrsim 10^{11}\:{\rm M}_{\odot}$) in our simulation exhibit S\'ersic indices greater than 4, further indicating a bulge-dominated structure.
Moreover, in the $\Sigma_{R_{\rm e}}-M_\star$ plane, while the median \sfg{}s typically show $ n\approx 1-3$, the range extends up to $ n\approx 4-5$ at the upper edge of the distribution, deviating from \citet{Barro2017}'s findings where \sfg{}s in this plane with $\log \Sigma_{R_{\rm e}} \gtrsim 9.5$ and $M_\star \gtrsim 10^{10}\:{\rm M}_{\odot}$ are expected to have S\'ersic indices between 2 to 4. 
Overall, \qg{}s tend to show higher indices compared to the \sfg{}s, with medians $n\approx2-3$, underscoring their bulge-dominated nature. Our models demonstrate a broader spread of $n$, from approximately 1 to 5 for \sfg{}s and 2 to 5 for \qg{}s, suggesting a need for further refinement in modelling the structural parameters of galaxies.

\section{Baryonic physics controlling quenching}

\label{Sec:Baryonic_features}
The star formation rates of galaxies are controlled by a complex set of physical mechanisms that control the rate at which gas will cool and form stars in galaxies. In order to gain insight into the primary mechanisms responsible for galaxy quenching in the \aXXI{} \lgal{} flavour, here we explore the distribution of baryons within haloes and investigate the relationship between  quenching, black hole mass, hot gas mass and cold gas mass in galaxies across different cosmic epochs.

\subsection{Halo baryon and gas fractions}
\label{Sec:HaloB_gas_frac}

\begin{figure}
    \centering
    \includegraphics[width=1\linewidth]{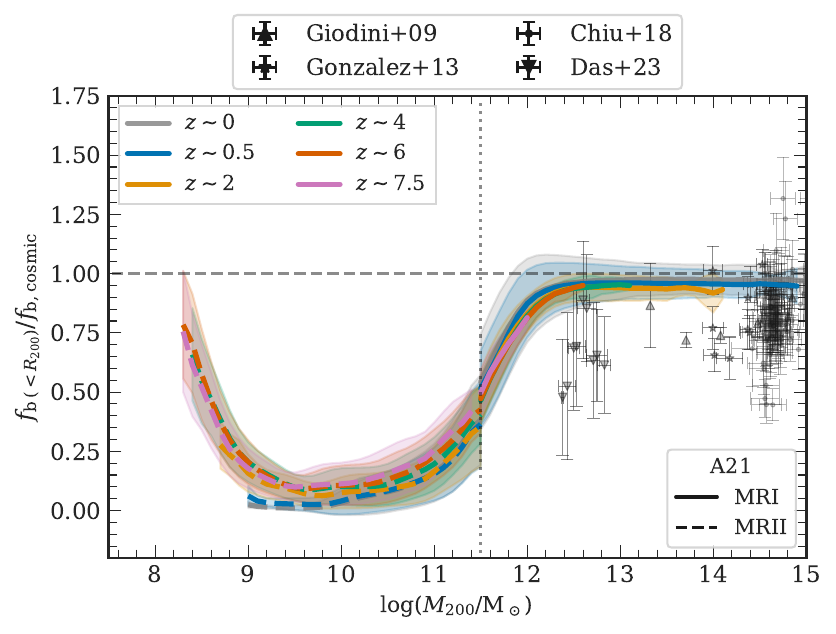}
    \includegraphics[width=1\linewidth]{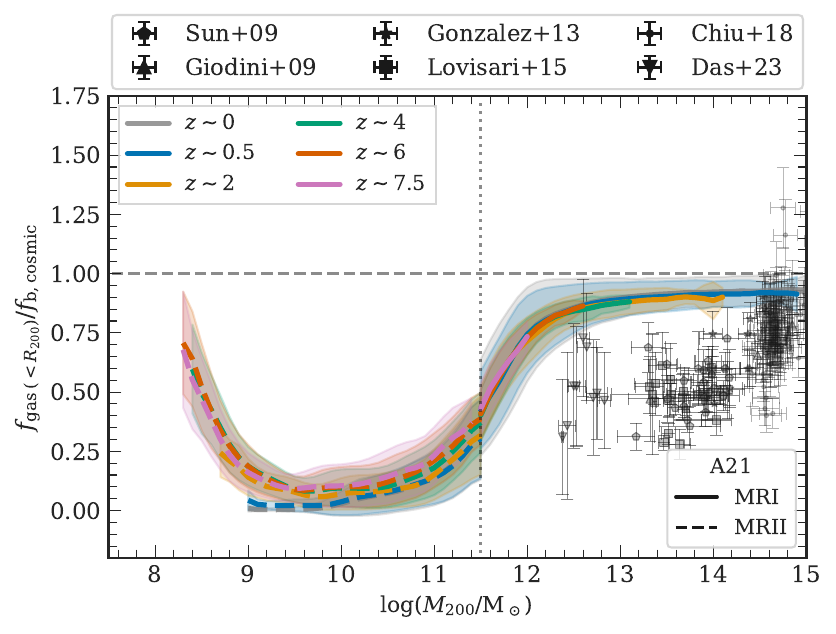}
    \caption{Halo baryon and gas fractions: The plot displays the distribution of baryon (\textit{top}) and gas (\textit{bottom}) fractions across different halo masses ($M_{200}$) for six redshift intervals. The solid and dashed lines represent the mean fractions from the MRI and MRII simulations, respectively, each normalised by the cosmic average and illustrated with a one-sigma scatter. The plot is also divided at $M_{200}=10^{11.5}\:{\rm M}_{\odot}$, equivalent to $M_\star =10^{9.5}\:{\rm M}_{\odot}$ which separates the contribution of MRI from MRII.
    Observational data points at $ z\approx0$ from \citet{Sun2009, Giodini2009, Gonzalez2013, Lovisari2015, Chiu2018} (within $R_{500}$), and \citet{Das2023} (within $ R_{200}$) are overlaid for comparison. }
    \label{fig:f_bary_gas_centrals}
\end{figure}

Figure~\ref{fig:f_bary_gas_centrals} displays the halo baryon and gas fractions in \aXXI{} across six different redshifts. The baryon fractions ($f_{\rm b}$) are calculated by summing up the contributions of all baryonic components within the virial radius ($R_{200}$)\footnote{The radius encompassing 200 times the critical density of the universe.} and dividing it by the virial mass ($M_{200}$)\footnote{Mass within $ R_{200}$.}. The resulting ratio is then normalised by the cosmic baryon fraction  $f_{\rm b, cosmic}=\Omega_{\rm b}/\Omega_{\rm m}\approx0.16$ \citep{Planck2016}. Similarly, the gas fractions ($f_{\rm gas}$) are calculated based on the combined contributions of hot and cold gas within each halo. The mean baryon and gas fractions are plotted with a one-sigma scatter from the mean, represented by the shaded region. A cut is applied to separate the MRI and MRII simulation results at $M_{200}=10^{11.5}\:{\rm M}_{\odot}$, analogous to the $M_\star=10^{9.5}\:{\rm M}_{\odot}$.

Observational measurements at $z\approx0$ of the baryon content within $R_{500}$ or $R_{200}$ are derived from X-ray data and Sunyaev-Zel'dovich \citep[SZ, ][]{Sunyaev_Zeldovich1972} effect, as reported by \citet{Giodini2009, Gonzalez2013, Chiu2018} and \citet{Das2023}. For halo gas fractions, we refer to results from \citet{Sun2009} and \citet{Lovisari2015}. It should be noted that we do not convert $M_{500}$ to $M_{200}$ (or vice versa) for the observations or simulation, as it has an insignificant impact on the halo baryon fraction.

Our analysis shows that \lgal{} does not exhibit any significant redshift evolution in the baryon and gas fractions. This indicates that physical processes responsible for altering the halo gas and baryon fractions are mostly time-independent in our models. Contrary to some hydrodynamical simulations where AGN feedback may lead to the expulsion of gas from the halo \citep[e.g., ][]{springel2018first, Harrison2018, Dave2019_simba}, AGN feedback in \lgal{} is designed to prevent the cooling of the hot gas, and does not redistribute the gas beyond the halo boundary \citep[see][]{Ayromlou2021_compare_Lgal_TNG}.

\subsection{Black hole and hot gas distribution}
As described in \secref{sec:BHprocesses}, the AGN feedback is primarily driven by the `radio mode' which relies solely on the mass of the BH, the surrounding hot gas and a constant efficiency parameter, $\kappa_{\rm AGN}=\rm 8.5\times10^{-3} \: M_{\odot}\:yr^{-1}$.

Figure~\ref{fig:MBH_HG_Cold_SM_A21} illustrates the log-ratio between the black hole (BH) masses (black lines) and the hot gas masses (purple lines) of \qg{}s relative to those of \sfg{}s. Solid and dashed lines correspond to central and satellite galaxies, respectively. The ratios are derived by calculating the median of each log-scaled quantity within bins of 0.2~dex width, organised using a sliding bin of step size 0.1~dex. These bins are iteratively adjusted by an additional 0.1~dex to ensure a minimum of 50 galaxies per bin. The shaded regions show the one-sigma scatter.

In \aXXI{}, the distribution of hot gas and BH mass across galaxy types and masses reveals trends tied to quenching processes. Low-mass quenched central galaxies ($M_\star \lesssim 10^{9.8}\:{\rm M}_{\odot}$) have on average less hot gas than their star-forming counterparts, reflecting the impact of ejective stellar feedback. On the other hand, more massive ($M_\star \gtrsim 10^{9.8}\:{\rm M}_{\odot}$) quenched centrals exhibit elevated hot gas levels, causing stronger AGN feedback (see \equref{eq:RadioMode}) and, consequently, the cessation of star formation activity. This increased feedback, however, may not entirely represent the astrophysical nature of observed AGNs. Additionally, these quenched centrals tend to have more massive BHs, a direct outcome of the implemented AGN feedback mechanism (see \equref{eq:RadioMode}). 
Similarly, quenched satellites show significantly less hot gas, highlighting gas stripping in dense environments as the main quenching mechanism. This stripping effect is more pronounced in less massive satellites  ($M_\star \lesssim 10^{10.5}\:{\rm M}_{\odot}$) and diminishes as satellite mass increases. For the most massive satellites ($M_\star \gtrsim 10^{10.5}\:{\rm M}_{\odot}$), AGN feedback is the primary quenching method rather than environmental processes. Additionally, the BH mass ratios show weaker trends but indicate that star-forming satellites, likely due to their abundant hot gas content, experience more substantial BH growth.

The hot gas dynamics in the \lgal{} models illustrate that while stellar feedback and UV background radiation are effective at altering hot gas content in galaxies up to Milky Way mass scales ($M_\star \sim 10^{10.8}\:{\rm M}_{\odot}$), they become ineffective in more massive ones. This leads to a hot gas fraction near the cosmic value among these massive systems with minimal scatter, as shown in \figref{fig:f_bary_gas_centrals}. Nevertheless, this does not align with various X-ray observations, thus motivating future improvements in the AGN feedback prescription to facilitate the ejection of gas beyond the halo boundary.

\subsection{Cold gas distribution}
\begin{figure*}
    \centering
    \includegraphics[width=0.33\linewidth]{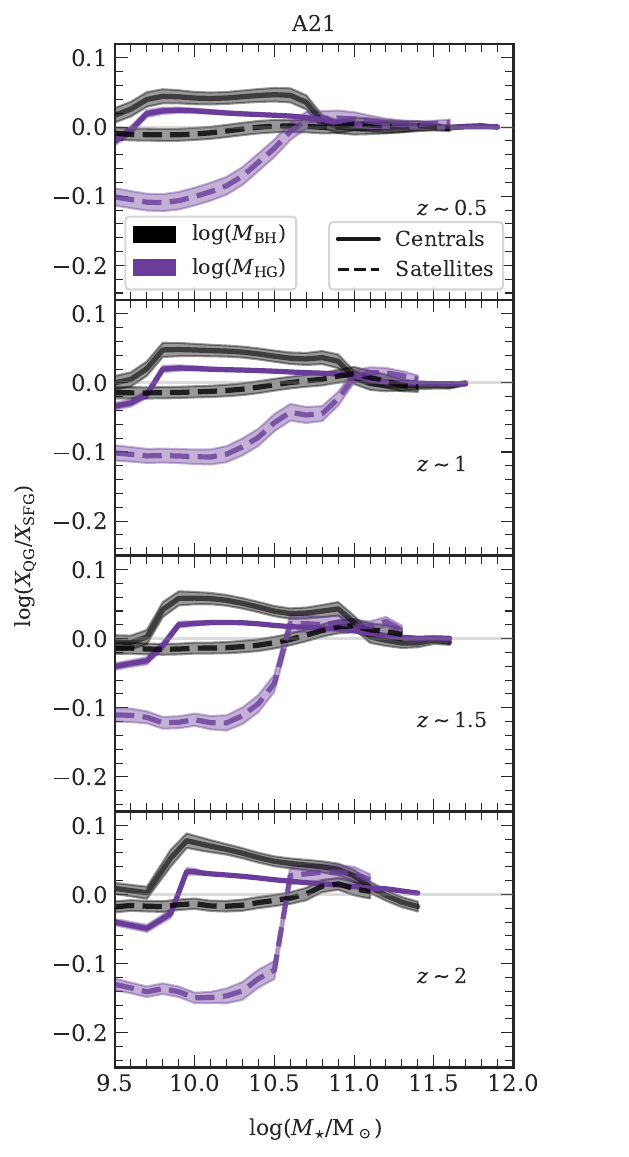}
    \includegraphics[width=0.33\linewidth]{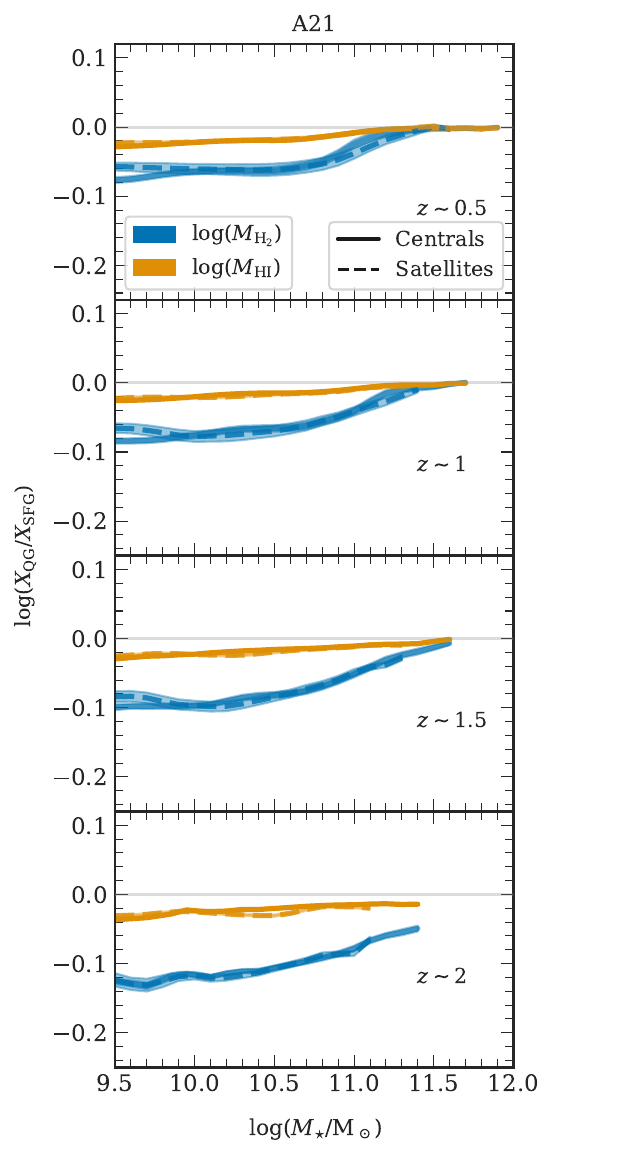}
    \includegraphics[width=0.33\linewidth]{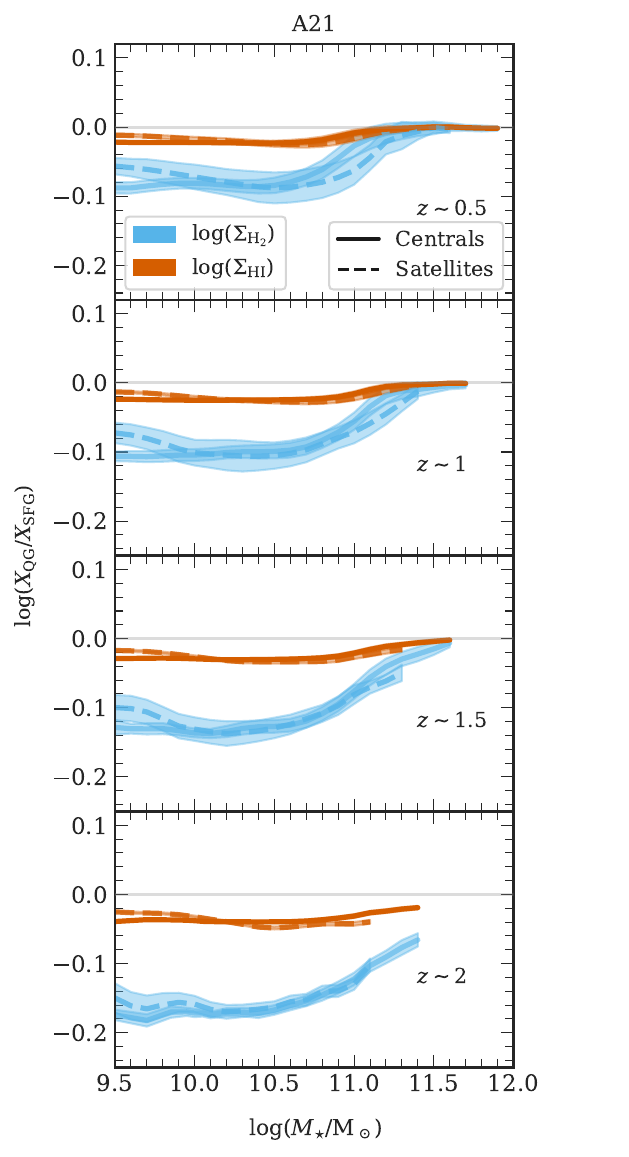}

    \caption{Black hole and hot gas (\textit{left}), and cold gas distributions (\textit{middle} and \textit{right}): The left plot illustrates the logarithmic ratio between BH masses (black lines) and hot gas (HG) masses (purple line), while the plots in the middle and on the right show the distribution of 
    the atomic ($\rm HI$) and molecular ($\rm H_2$) cold gas. These log-ratios are for the quantities in \qg{}s relative to the \sfg{}s within the \aXXI{} model flavour. The middle panel displays mass ratios, while the right panel shows ratios of surface densities of $\rm HI$ and $\rm H_2$. Solid and dashed lines represent central and satellite galaxies, respectively. 
    The scatter within the sample is depicted by the shaded areas.}
    \label{fig:MBH_HG_Cold_SM_A21}
\end{figure*}

Atomic hydrogen ($\rm HI$) and molecular hydrogen ($\rm H_2$) are critical for star formation, with \sfg{}s typically harbouring larger fractions of both. In contrast, \qg{}s often show reduced amounts of these gases due to processes like gas depletion, environmental effects, and feedback mechanisms \citep[e.g., ][]{Tacconi2010, Tacconi2018, Janowiecki2020, Guo2021}. Metallicity plays a crucial role in the formation of molecular hydrogen as it provides cooling pathways essential for converting atomic to molecular phases \citep[e.g., ][]{Tacconi2010, Kauffmann2012}. Dust grains, more prevalent in metal-rich environments, facilitate $\rm H_2$ formation by offering surfaces for hydrogen atoms to combine \citep[e.g.,][]{Cazaux2009}. Additionally, UV radiation can influence gas phases through photodissociation, impacting the balance between the cold gases. 

The middle and right panels of Figure \ref{fig:MBH_HG_Cold_SM_A21} illustrate the log-ratio between the hydrogen masses (middle) and surface density (right) of \qg{}s relative to those of \sfg{}s in the \aXXI{} model. The cold gas surface densities \mbox{($\Sigma_{\rm HI\:or \:H_2}=f_{\rm HI\:or\:H_2}(M_{\rm Cold\:Gas}/2\pi R_{\rm GasDiscRadius}^2$))} are calculated using the fraction of cold gas mass that is in $\rm H_2$ or $\rm HI$ ($f_{\rm HI\:or\:H_2}$) and the size of gas disc ($R_{\rm GasDiscRadius}$), which is dependent on the halo spin parameter, and the maximum rotational velocity of the subhalo in the simulations \citep[][]{Guo2011}. 
These distributions are separately analysed for central and satellite galaxies, providing a detailed view of the cold gas dynamics across different galactic environments.

Overall, \qg{}s have a significantly lower fraction of $\rm H_2$ gas compared to their star-forming counterparts, reflecting our $\rm H_2$-dependent star-formation mechanism. This trend typically intensifies toward lower-mass galaxies and at higher redshifts. Quenched galaxies also contain slightly less HI gas than star-forming ones, although the difference is not as marked as with $\rm H_2$ gas. There is no significant distinction between central and satellite galaxies in terms of the cold gas content, primarily due to the absence of cold gas stripping in our model \citep[e.g., ][]{Brown2017}. This process will be incorporated into future versions of \lgal{}. 

\section{Discussion}
\label{sec:discussion}

\subsection{Comparison with other work}
Recent observations of the high-redshift universe, especially those focusing on massive quenched systems and the detailed characteristics of the UV luminosity function (UVLF) as observed by the JWST \citep[e.g.,][]{Carnall2023, Bouwens2023}, pose significant challenges for galaxy formation simulations. Numerous studies have been struggling to replicate these phenomena accurately. A pivotal challenge is promoting early galaxy quenching, which would significantly impact the high redshift number density of passive galaxies in simulations.

\citet{Somerville2015_review} compared the efficacy of three hydrodynamic simulations and five semi-analytical models, including an older \lgal{} model variant \citep{Henriques2013winds}. This assessment revealed that while the models generally adhered well to established calibration benchmarks, they struggled to produce massive galaxies at higher redshifts. Since then, a number of authors have attempted to solve these problems by changing the treatment of certain physical processes in the simulation.

In a recent study, \cite{Lagos2023_shark2.0} introduced new AGN wind and jet modes based on the properties of the SMBH in the SHARK semi-analytic model \citep[][]{Lagos2018_SHARK}. Similar to our work, they compared their results with data from \citet{Weaver2023} for the galaxy stellar mass function across different redshifts. They showed that these updates improved the model's ability to suppress star formation in massive galaxies at later times. Although the number density of quenched galaxies increased by about 0.8~dex at $z\sim3$ for $M_\star\gtrsim10^{10.5}\:{\rm M}_{\odot}$ systems, the models underpredict the number density of massive galaxies at higher redshifts. 

We note that unlike our approach, which solely utilises the NUVrJ criterion, \citet{Lagos2023_shark2.0} applied both NUVrJ and sSFR cuts to classify galaxies as quenched. They found that alignment with the observational data from \citet{Weaver2023} was more accurate when similar selection criteria (i.e., NUVrJ) were used. Furthermore, the authors introduced a random error of 0.3~dex to account for systematics and uncertainties in the SHARK semi-analytic model's \smf{}, an approach we did not adopt in our analysis.
 
Despite the inclusion of AGN outflows from the cold interstellar medium -- a feature not present in the \lgal{} models -- SHARK underpredicts $M_\star\sim 10^{11}\:{\rm M}_{\odot}$ quenched systems by about a factor of $2-3$ compared to the \aXXI{} \lgal{} model at $z\sim2$, signifying that the challenge of the quenching mechanisms is still not solved. Additionally, SHARK overestimates the number density of low-mass quenched systems at lower redshifts in comparison to both the observational data and the \lgal{} models, a common issue in semi-analytical models as is discussed in \citet{Harrold2024_overabundance}. 

Finally, we note that \citet{Lagos2023_shark2.0} also updated the SHARK semi-analytic model with a new prescription for galaxy angular momentum, influenced by star formation laws and circular velocity profiles. This update facilitates more accurate predictions of the galaxy mass-size relationships for both bulge and disc components, showing improvements over previous versions and better alignment with observational data at $z=0$. Their analysis focused on comparing direct simulation output with observations. Nevertheless, there are significant complexities associated with comparing galaxy sizes with observations. As demonstrated in our comparison of galaxy sizes from \lgal{} with observational data in \secref{sec:MSR_model}, the results can substantially change once mock galaxy sizes are determined using the same methodology as applied to observational data, for instance by fitting a single component S\'ersic profile (\figref{fig:SFGQG_MSR_M}), compared to using raw simulation output. For reference, currently the \lgal{} models follow \citet{Mo1998}, assuming that the angular momentum of infalling material aligns with that already present in the galaxy's disc.

\citet{DeLucia2024QGPorperties} recently evaluated three different versions of the GAEA semi-analytic model \citep[][]{DeLucia2007}, including a version with a new AGN-driven outflow mechanism and updated treatment for low mass satellites, and compared these to findings from \citet{Weaver2023}. The GAEA model was calibrated using the stellar mass function up to $z\sim3$ and the AGN luminosity function up to $z\sim4$.
While the updated GAEA model showed improvements in simulating total and quenched systems up to $z\sim4$, discrepancies remained, albeit smaller than those in earlier versions of their models. It is important to note that they used a sSFR cutoff to classify galaxies as quenched, in contrast to the $\rm NUVrJ$ criterion used by \citet{Weaver2023} and our paper, potentially introducing systematic biases such as overestimation of the quenched fraction in the simulations. 
The NUV band captures star formation history over the past billion years, offering a longer-term view of the past star formation history than the instantaneous SFR in the model, which only reflects recent activity on the timescale of a single snapshot. This could lead to an overestimation of quenched galaxies when using sSFR criteria in the model \citep[e.g., ][]{Haydon2020}. 
Including these biases and Gaussian distributed uncertainties in the \smf{}s, the model predictions fell slightly short of matching the observations at the massive end. The models produced about 2.5 times more $M_\star\sim 10^{11}\:{\rm M}_{\odot}$ quenched systems at $z\sim2$ than the \aXXI{} \lgal{} model, likely a consequence of the AGN-driven outflows.

In a study by \citet{Zoldan2019}, galaxy sizes in an older version of the GAEA model \citep{Xie2017} were evaluated against observational data from \citet{vanderWel2014b}. The analysis showed good agreement with the properties of \sfg{}s but less agreement with \qg{}s, yet better than \lgal{}. The model reproduced the steep slope of the galaxy mass-size relation noted by \citet{vanderWel2014b}. However, the study employed a sSFR cutoff to define quenched systems, differing from \citet{vanderWel2014b}'s selection criteria. This likely would have led to an overestimation of the number of quenched systems affecting their median sizes. Moreover, they did not attempt to derive galaxy sizes by fitting a S\'ersic profile, an exercise conducted both in \cite{vanderWel2014b} and in our study.

Parallel efforts by \citet{Yung2024_SCSam}, showed that the updated Santa Cruz semi-analytic model \citep{Somerville1999} underpredicted the UVLF at high and ultra-high redshifts ($z\gtrsim 8$) compared to observations. While the \lgal{} models currently lack  the capability to probe these ultra-high redshift galaxies due to fewer time snapshots in the MRI and MRII simulations, running \lgal{} on newer simulations such as MillenniumTNG \citep{Pakmor2023, Barrera2023_LGAL_MTNG} could provide enhanced resolution and better temporal coverage for exploring these early epochs in more detail.

\subsection{Interpretation of the results}
In light of the interactions between hot gas, black holes, and cold gas across different galactic environments, detailed in \secref{Sec:Baryonic_features}, our findings suggest several avenues for refining future astrophysical prescriptions in the \lgal{} framework. 

The relationship between baryon or gas fraction and halo mass ($f_{\rm b \:or \: gas} -M_{200}$) is now becoming much better constrained at low redshifts. However, a discrepancy exists with current observational data suggesting that more baryons need to be expelled in galaxy groups. This inconsistency arises because the \lgal{} model predicts a redshift-independent relationship for these fractions, owing to the scaling of SN and AGN feedback efficiencies with halo-mass rather than internal galactic properties (\figref{fig:f_bary_gas_centrals}). Future observations \citep[e.g.,][]{Ade2019JCAP_SimonObs, Kraft2022arXiv_LEM, Zhang2024arXiv} may provide further light on this matter, potentially driving refinements in feedback mechanisms within the \lgal{} framework.

Our investigation of the differences between hot gas, black holes and the masses of cold gas in atomic and molecular form in \sfg{}s and \qg{}s yields the most insightful results (\figref{fig:MBH_HG_Cold_SM_A21}). 
The findings highlight strong differences primarily in low-mass satellite galaxies, where differences in the hot gas distribution is more pronounced. In contrast, central galaxies, which are the highest-mass galaxies at all redshifts, exhibit almost no differences in these parameters. However, a notable distinction is observed in quenched galaxies at high redshifts, where the molecular gas content of both centrals and satellites has been strongly depleted, indicating the important role of the internal gas reservoir in regulating the star-forming state of these high-$z$ galaxies.

These findings may support the AGN feedback model enhancements implemented into the SHARK semi-analytical model. Combining these enhancements with the \citet{Ayromlou2019_rampressure} gas stripping prescriptions could alleviate some of the discrepancies in the SHARK model, particularly concerning the fitting of low-mass quenched galaxies. However, if the inflow of gas into the molecular reservoir and its consumption rate into stars has not been accurately parameterised at very high redshift, changes to the AGN feedback recipes may be spurious.

Studies on galaxy growth mechanisms, such as those by \citet{Genel2018} and \citet{Ferrero2021_Eagle_TNG}, corroborate that quenched galaxies expand predominantly through dry mergers and tidal interactions following their quenching episodes. 
Furthermore, studies by \citet{Taylor2016, Trayford2019}, and \citet{DOnofrio2023} agree that galaxy merging is crucial in galaxy size evolution, with AGN feedback playing a substantial role in reducing central gas densities and influencing the compactness of galaxies’ central regions. These studies, on the other hand, highlight the significant role of galaxy morphology and mass in these processes, and suggest the necessity of incorporating more explicit mechanisms for modelling the impacts of merging and feedback on galaxy sizes and structures within \lgal{} models.

In contrast, the \lgal{} models currently lack advanced prescriptions for modelling the detailed impacts of merging on galaxy sizes and structures, and AGN outflows. The \lgal{} models currently utilise the prescription from \citet{Guo2011}, which relies on energy conservation and the virial theorem to compute the changes in galaxy sizes during mergers. In addition, the bulge component of the galaxies follows a \citet{Jaffe1983} distribution. Overall, addressing the complexities in the galaxy mass-size relations remains challenging and requires careful implementation of various physical processes.

Finally, observational studies of MgII absorbers in the vicinity of radio-loud AGN provide evidence that AGN influence the gas around galaxies even outside their dark matter haloes \citep{Kauffmann2018MgIIabsorbers}. This redistribution of the gas within and beyond the halo boundary, out to the closure radius, due to AGN feedback adds substantial complexities to the physics of galaxy evolution \citep{Ayromlou2023}. Further exploration of these mechanisms and their long-term effects on galaxy evolution is essential and reserved for future studies.

In forthcoming work, we plan to explore a variety of physical parameter changes that could help us to match the high redshift data with better accuracy. Monte Carlo Markov Chain (MCMC) and more modern machine learning parameter optimisation techniques are expected to be very useful tools in this endeavour.

\section{Summary and conclusion}
\label{sec:summary_conclusion}
The study of galaxy evolution across cosmic epochs is essential for assessing cosmological galaxy formation models against observational data, thereby guiding future model refinements. 
In this study, we present a rigorous evaluation of three calibrated versions of the \lgal{} cosmological galaxy evolution semi-analytical model; namely \aXXI{}, \hXX{} and \hXV{}, across a broad range of masses ($10^7\: {\rm M}_{\odot}\lesssim M_\star\lesssim 10^{12}\:{\rm M}_{\odot}$) and redshifts ($0\le z \lesssim10$). A key novel aspect of this work is the incorporation of high-redshift observations and an in-depth analysis of galaxy size and stellar mass surface density evolution. The analysis systematically compares simulated galaxy properties with observational data to assess the fidelity of these models in reproducing key aspects of galaxy evolution, including stellar mass functions, cosmic star formation rate densities, galaxy size-mass relations, and the evolution of core $(R<1\,{\rm{kpc}})$ and effective $(R<R_{\rm{e}})$ stellar mass surface densities, both for star-forming and quenched galaxies. Here are our key findings:

\begin{itemize}

    \item Total SMF: The \lgal{} flavours align well with observations up to $z\sim7.5$ for all mass ranges with small deviations (\figref{fig:SMF_tot_SMF_z}).

    \item CSFRD: While the simulations capture the general trend up to $z\sim10$, they exhibit a slightly suppressed SFR at $z\sim2$ by a factor of 1.2 - 1.5 (\figref{fig:CSFR}).

    \item SFG SMF: The three model flavours align well with the observations at low to intermediate redshifts $(z \lesssim 2)$, however, at higher redshifts, small discrepancies emerge for $M_\star\sim10^{10}\:{\rm M}_{\odot}$ galaxies (\figref{fig:SMF_SFG_SMFz_SFG}).
    
    \item QG SMF: The \aXXI{} model flavour shows a closer agreement and predicts a more accurate match to the observed \smf{} and the number density of \qg{}s at lower redshifts. However, despite \aXXI{}'s qualitative agreement with observations in the evolution of the \smf{} of \qg{}s, all \lgal{} flavours underpredict the abundance of \qg{}s at $z \gtrsim1.5-2$ for $M_\star\gtrsim10^{10.5}\:{\rm M}_{\odot}$ galaxies. This discrepancy grows from a factor of 1.5 at $z \sim2$ to a factor of 60 by $z \sim3$ for the $M_\star\sim10^{11}\:{\rm M}_{\odot}$ quenched systems (\figref{fig:SMF_QG} and \ref{fig:SMFz_QG}).

    \item QG number density$-z$: The \lgal{} flavours exhibit a transition in dominance from satellites to centrals in the number density of \qg{}s around $M_\star \sim10^{10.2-10.3}\:{\rm M}_{\odot}$, highlighting a shift in the primary quenching mechanisms at these mass scales (\figref{fig:SMFz_QG}). Below this threshold, environmental effects predominantly drive quenching, whereas, above it, AGN feedback becomes the main quenching mechanism.

    \item ${\rm SFR}-M_\star$:  All three model flavours reproduce the star-forming main sequence relationship effectively up to $z\sim 6$. However, they often miss the flattening of this relation for the most massive ($M_\star\sim 10^{11}\:{\rm M}_{\odot}$) galaxies at $z\gtrsim 5$, also showing a suppressed SFR at $z\sim2$ (\figref{fig:SFR_M} and \ref{fig:SFR_M_Lgalmodel}).

    \item $R_{\rm e}-M_\star$: While massive \sfg{}s and \qg{}s show consistency with the observations up to $z\sim3$, \sfg{}s and \qg{}s with mass $M_\star\lesssim10^{10.5}\:{\rm M}_{\odot}$ and  $M_\star\lesssim10^{11.2}\:{\rm M}_{\odot}$, respectively, tend to have larger sizes by a factor of up to 15 than observed, with the greatest deviations occurring in quenched systems (\figref{fig:SFGQG_MSR_M} and \ref{fig:SFG_QG_MSR_z}).

    \item $\Sigma-M_\star$: For the \sfg{}s the core ($R< 1\:{\rm kpc}$) and effective ($R<R_{\rm e}$) stellar mass surface densities are generally in alignment up to $z\sim 3$ for all masses with small deviations. 
    At $z \sim 0.5$, \qg{}s with $M_\star \sim 10^{10}\:{\rm M}_{\odot}$ are significantly less compact than expected, under-compact by a factor of $\sim 4$ in the $\Sigma_{1\:{\rm kpc}}-M_\star$ plane and by $\sim 80$ in the $\Sigma_{R_{\rm e}}-M_\star$ plane (\figref{fig:SMSD_1kpc}, \ref{fig:SMSD_Re} and \ref{fig:SFG_QG_SMSD_z}). 

  \end{itemize}

A central focus of this work is the discrepancies observed from $z\gtrsim 2$. Without high-redshift comparisons, it is not possible to determine the direction for future improvements in the \lgal{} model, emphasising the importance of high-z data in shaping our understanding of galaxy formation. Notably, our results highlight that the primary discrepancy arises not from advanced gas stripping prescription in low-mass galaxies, but rather from the galaxy quenching mechanisms, likely the AGN feedback and galaxy merging mechanism, in more massive galaxies. This issue becomes increasingly apparent beyond $z \gtrsim 1.5$, suggesting that future iterations of \lgal{} must revisit galaxy quenching implementations.

In conclusion, our investigation indicates that while significant strides have been made in physically simulating realistic galaxy populations, critical discrepancies persist, particularly at high redshifts and in representing the number density, size and surface density distributions of quenched galaxies. Our analysis has identified several avenues for potential improvements in the \lgal{} framework. These include revising galaxy quenching mechanisms -- for example AGN feedback -- as well as implementing a more physical treatment for galaxy mergers, which are crucial for accurately modelling the size and surface density distribution of quenched systems. To navigate these issues and effectively generate new models and prescriptions, we aim to develop tools using machine learning, that can more efficiently probe the parameter space.  To mitigate calibration biases, we also plan to recalibrate the model using an MCMC approach with the latest high-redshift observational data, which were not available during the last calibration of the \lgal{} model.  Incorporating data from new observational facilities like JWST and utilising the results of the most recent hydrodynamical simulation projects such as MillenniumTNG \citep[][]{Pakmor2023} and FLAMINGO \citep[][]{Schaye2023_FLAMINGO} are expected to provide fresh insights that can guide the design of future versions of the \lgal{} model, further enhancing the reliability and accuracy of its predictions.

\section*{Acknowledgements}
The analysis herein was carried out on the compute cluster of the Max Planck Institute for Astrophysics (MPA) and systems at the Max Planck Computing and Data Facility (MPCDF). AV wants to thank Bo Peng, Silvia Bonoli, Rachel Somerville, Greg Bryan, Rob Yates, Stephanie Tonnesen, and Anshuman Acharya for the useful discussions and assistance. MA extends gratitude to Moein Mosleh for the insightful conversations. MA also acknowledge funding from the Deutsche Forschungsgemeinschaft (DFG) through an Emmy Noether Research Group (grant number NE 2441/1-1). This work has made use of: \texttt{Astropy} \citep{AstropyCollaboration2018}, \texttt{TOPCAT} \citep{Taylor2005_TOPCAT}, \texttt{Scipy} \citep{Virtanen2020_SciPy}, \texttt{Seaborn}/\texttt{Matplotlib} \citep{Hunter2007_Matplotlib, Waskom2021_Seaborn}, and the National Aeronautics and Space Administration (NASA) Astrophysics Data System (ADS).

\section*{DATA AVAILABILITY}
The full outputs of the three \lgal{} flavours for
all the Millennium-I and Millennium-II snapshots are publicly available at \url{http://lgalaxiespublicrelease.github.io/}.



\bibliographystyle{mnras}
\bibliography{Ref} 




\appendix

\section{Results from H20 and H15}
\label{app:H20_H15}

\begin{figure*}
    \centering
    \includegraphics[width=1\linewidth]{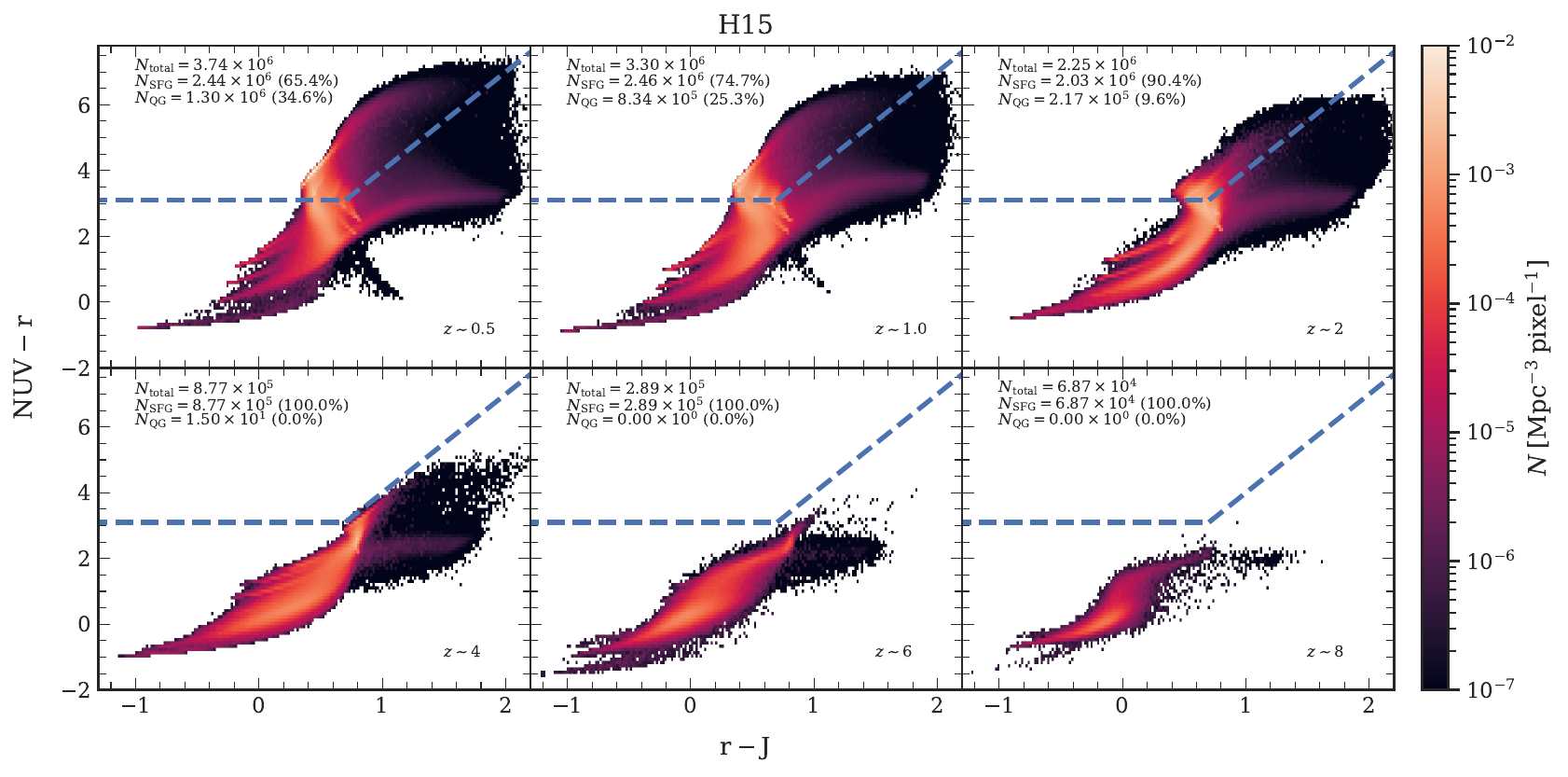}
    \caption{
    $\rm NUVrJ$ colour-colour plot: The plot, combined for MRI and MRII, illustrates the rest frame $\rm NUVrJ$ colour-colour selection criteria used to classify galaxies in \hXV{} as either star-forming or quiescent across various redshifts. The plot is similar to \figref{fig:NUVRJ}.}
    \label{fig:NUVRJ_H20_H15}
\end{figure*}

The results of classifying \sfg{}s and \qg{}s in the \hXV{} flavour, using the $\rm NUVrJ$ colour-colour selection criterion (refer to Eq. \ref{eq:NUVrJ}), are shown in \figref{fig:NUVRJ_H20_H15}. Notably, the \hXV{} flavour exhibits red-ward scatter in the $\rm NUVrJ$ plot, which could be attributed to its older star formation prescription which maintains a fixed ratio of atomic to molecular hydrogen. This trend towards redder colours is independent of the dust models, which are consistent across all \lgal{} flavours \citep{Devriendt1999_extinction, Charlot2000extinction}.  Both the \hXX{} and \hXV{} flavours exhibit a deficit of quenched systems starting at $z \gtrsim 2-3$, similar to the trends observed in the \aXXI{} flavour.

\begin{figure}
    \centering
    \includegraphics[width=1\linewidth]{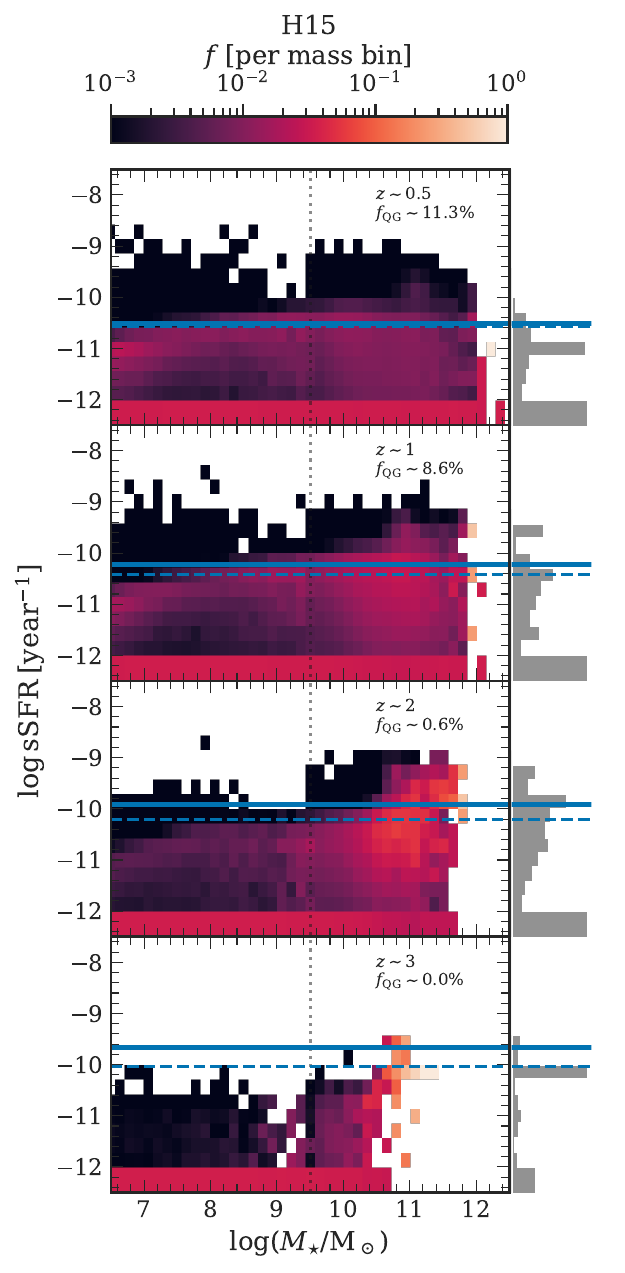}
    \caption{${\rm sSFR}-M_\star$ plot for \qg{}s: The plot illustrates the occupation of the ${\rm sSFR}-M_\star$ plane by \qg{}s produced by the \hXV{} model flavour.
    The plot is similar to \figref{fig:sSFR}.
    }
    \label{fig:sSFR_H20_H15}
\end{figure}

Figure~\ref{fig:sSFR_H20_H15} illustrates the ${\rm sSFR}-M_\star$  plot for the chosen quenched systems modelled by the \hXV{} flavours, showing trends and distributions similar with those observed in \aXXI{}. Notably, there is a minor degree of contamination from galaxies with slightly higher sSFRs than the established cutoff threshold ($f_{\text{QG}}$), as highlighted in \secref{sec:nuvrj}. These `contaminating' galaxies are predominantly massive satellite galaxies, with the fraction of such contaminants decreasing with increasing redshift. 

\begin{figure}
    \centering
    \includegraphics[width=0.95\linewidth]{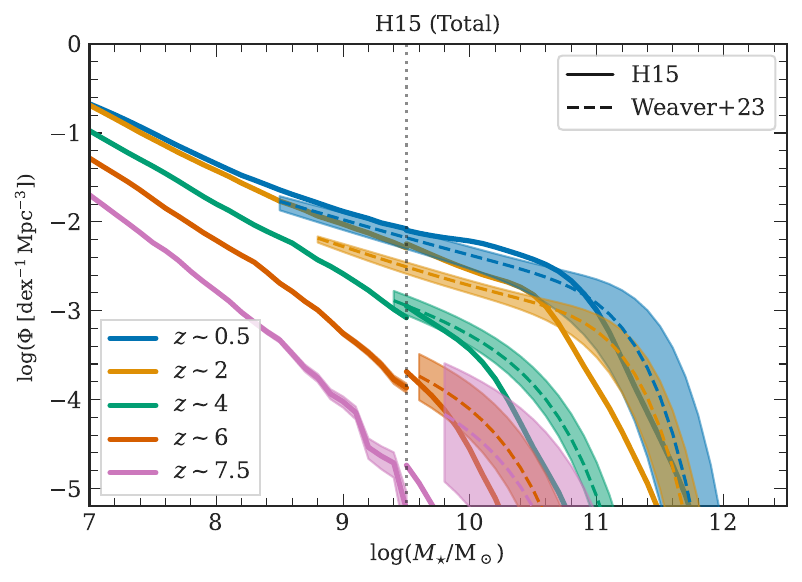}
    \includegraphics[width=0.95\linewidth]{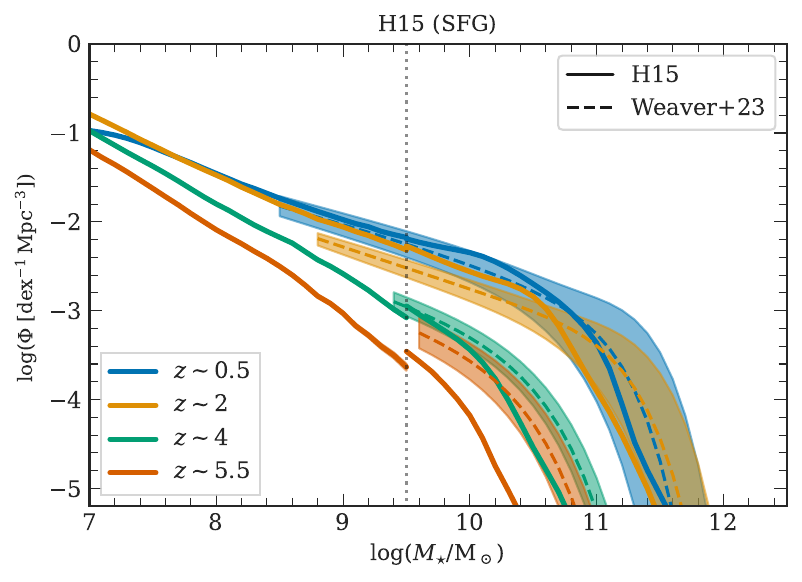}

    \includegraphics[width=0.95\linewidth]{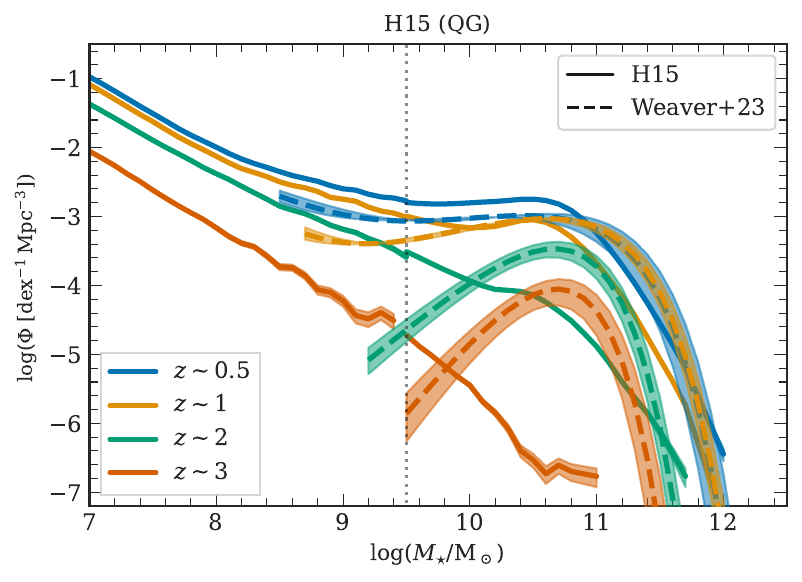}

    \caption{Stellar mass functions: 
    These plots display the total \smf{} (\textit{top}), the star-forming galaxy \smf{} (\textit{middle}) and the quenched galaxy \smf{} (\textit{bottom}) for the \hXV{} model flavour.
    In these plots, the solid line shows the evolution of the stellar mass functions from the simulations across 5 redshifts, and the dashed line is from the observational findings of \citet{Weaver2023}. The plot is similar to \figsref{fig:SMF_tot_SMF_z}, \ref{fig:SMF_SFG_SMFz_SFG} and \ref{fig:SMF_QG}.}
    \label{fig:SMFs_H20_H15}
\end{figure}

\begin{figure}
    \centering
    \includegraphics[width=0.95\linewidth]{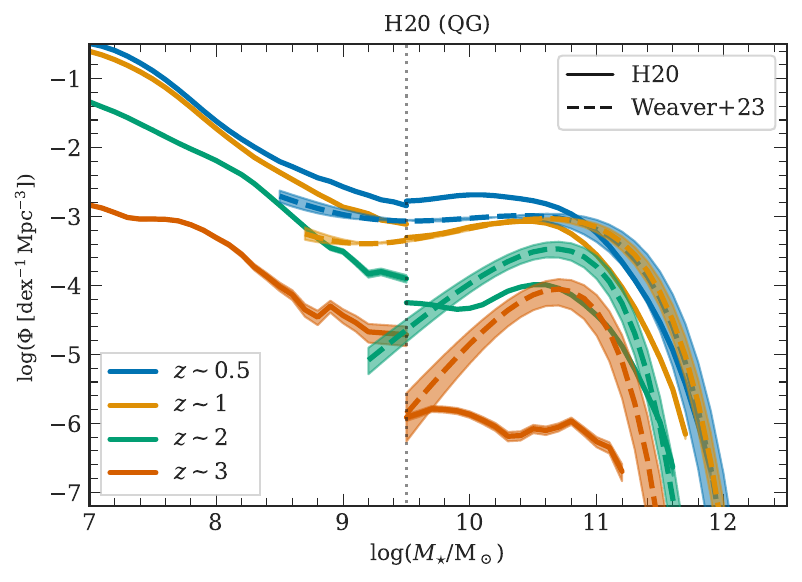}
    \caption{Stellar mass function of QG: This plot displays the \qg{} \smf{} for the \hXX{} model flavour. The plot is similar to \figref{fig:SMFs_H20_H15} and \ref{fig:SMF_QG}.}
    \label{fig:SMF_QG_H20}
\end{figure}

Figure~\ref{fig:SMFs_H20_H15} shows the total \smf{} (\textit{top} panel), the \sfg{} \smf{} (\textit{middle} panel) and \qg{} \smf{} (\textit{bottom} panel) for the \hXV{} with the dashed line representing the observational fits from \citet{Weaver2023} along with its one-sigma errors. Similarly, \figref{fig:SMF_QG_H20} shows the \qg{} \smf{} for the \hXX{} model flavour. Overall the conclusions are similar to the ones mentioned for \aXXI{} with small deviations as mentioned in \secref{sec:TotalSMF},  \ref{sec:SFG_SMF} and \ref{sec:QG_smf}.

\begin{figure}
    \centering
    \includegraphics[width=1\linewidth]{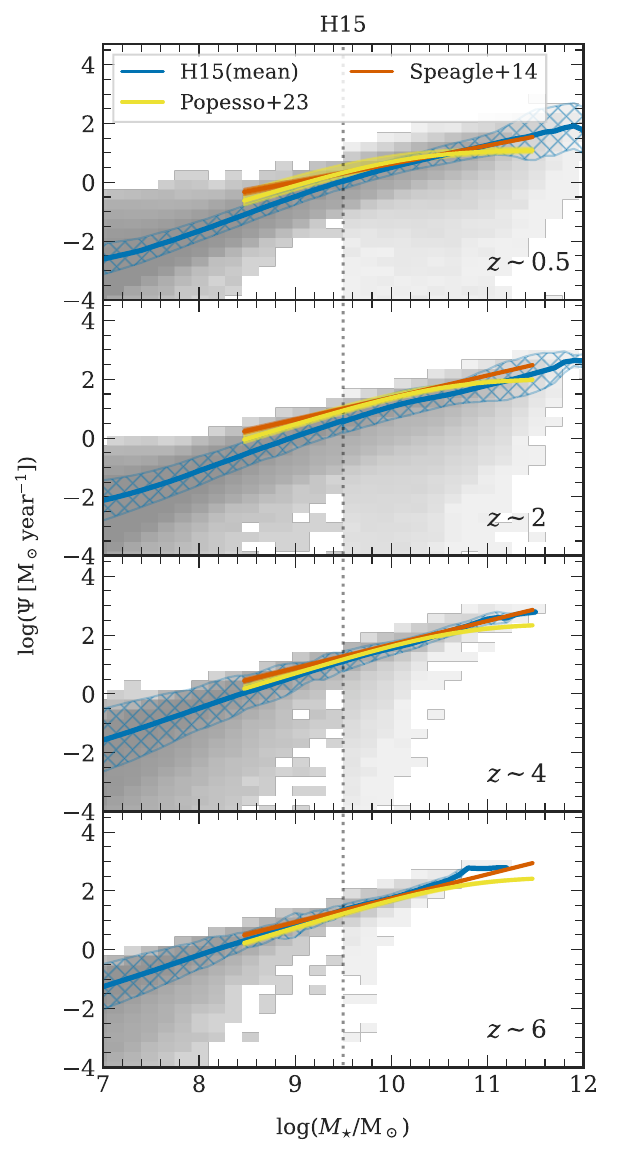}

    \caption{ ${\rm SFR}-M_\star$ relation: 
    A plot describing the main sequence of star-forming galaxies across various redshifts for the \hXV{} model flavour. The plot is similar to \figref{fig:SFR_M}.}
    
    \label{fig:SFR_M_H20_H15}
\end{figure}

Figure~\ref{fig:SFR_M_H20_H15} shows the main sequence of star-forming galaxies for \hXV{} with observational fits from \citet{Popesso2023} and \citet{Speagle2014} in red and yellow solid lines. 

\section{Convergence of MRI and MRII}
\begin{figure}
    \centering
    \includegraphics[width=0.95\linewidth]{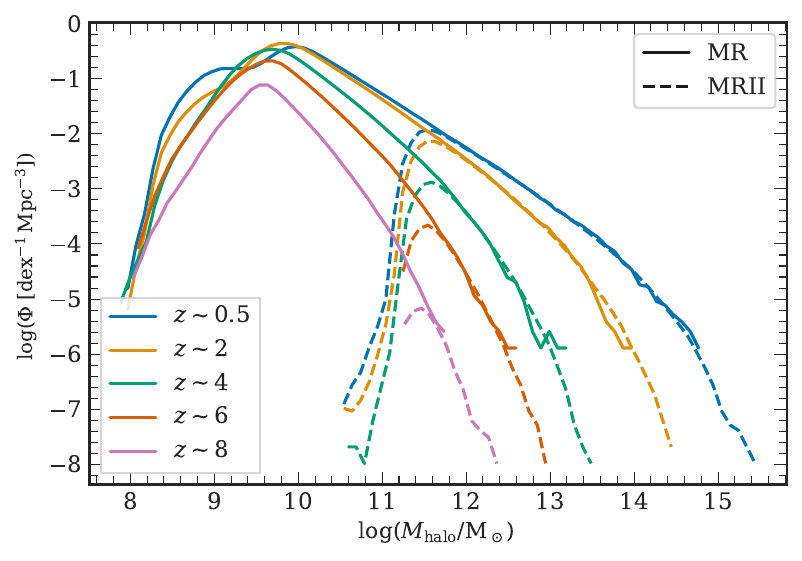}
    \includegraphics[width=0.95\linewidth]{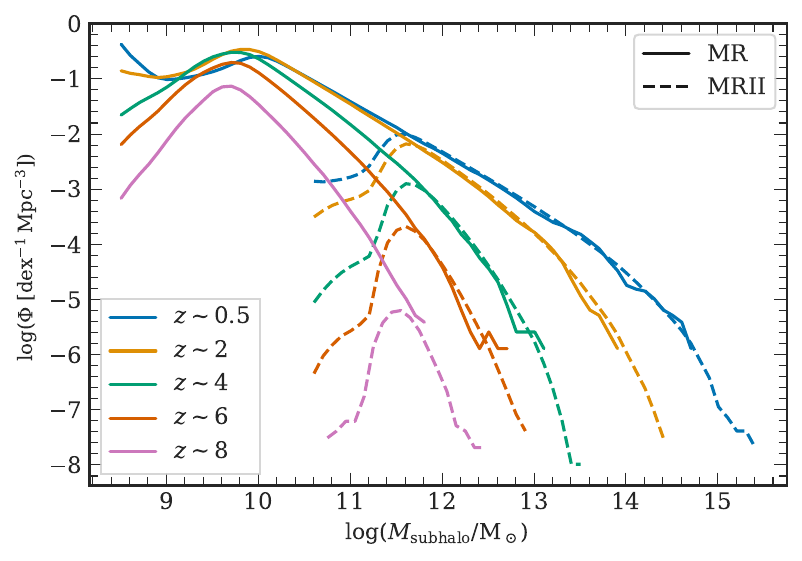}

    \caption{These plots show the halo mass function (\textit{left}) and the sub-halo mass function (\textit{right}). The colour coding represents the redshift bin for which the mass function is plotted. The solid and dashed lines represent the MRI and MRII simulations.}
    \label{fig:HMF}
\end{figure}

As is extensively discussed in \citet{Boylan-Kolchin2009MilII} the halo mass function and the sub-halo mass function exhibit excellent convergence between MRI and MRII.  For completeness, these two relations are reproduced in \figref{fig:HMF} for five representative ranges of redshift bins. The results from MRI are shown by the solid lines, while the dashed lines represent the smaller box of MRII.

Any observed deviation in the galaxy scaling relation plots discussed in this study between the MRI and MRII boxes arises from the divergence of the physical models at the resolution limit boundary. The choice of the transition region between MRI and MRII and among different scaling relations could result in better or more unsatisfactory convergence of these relations.  
Using a different transition mass for each \lgal{} flavour does not change any of our results or conclusions significantly (e.g., refer to \figref{fig:SMF_QG}).

\section{S\'ersic profile fitting}
\label{app:SersicFit}
The two \lgal{} flavours with radial rings (\aXXI{} and \hXX{}) use a curve of growth approach, iteratively finding the radius that contains half of the total flux of the galaxy, to calculate the half-light radii ($R_{\rm e}$). However, the observational works of \citet{VanDerWel2014_MSR} and \citet{Mowla2019}, which we use for comparison, employ the \citet{1963Sersic} profile to derive the effective radius.
Thus, to facilitate a meaningful comparison between observations and simulation data, we derive the $R_{\rm e}$ from a single-component 1D S\'ersic profile fit. The S\'ersic profile is expressed as:
\begin{equation}
    I(r) = I_0 \exp\left(-b_n \left[\left(\frac{r}{R_{\rm e}}\right)^{1/n} - 1\right]\right),
\end{equation}
where $I_0$ is the intensity or brightness at the centre of the galaxy ($r=0$), $R_{\rm e}$ is the half-light radius, $n$ is the S\'ersic index, determining the shape of the profile, and $b_n$ is a constant related to $n$.

\begin{figure}
    \centering
    \includegraphics[width=0.95\linewidth]{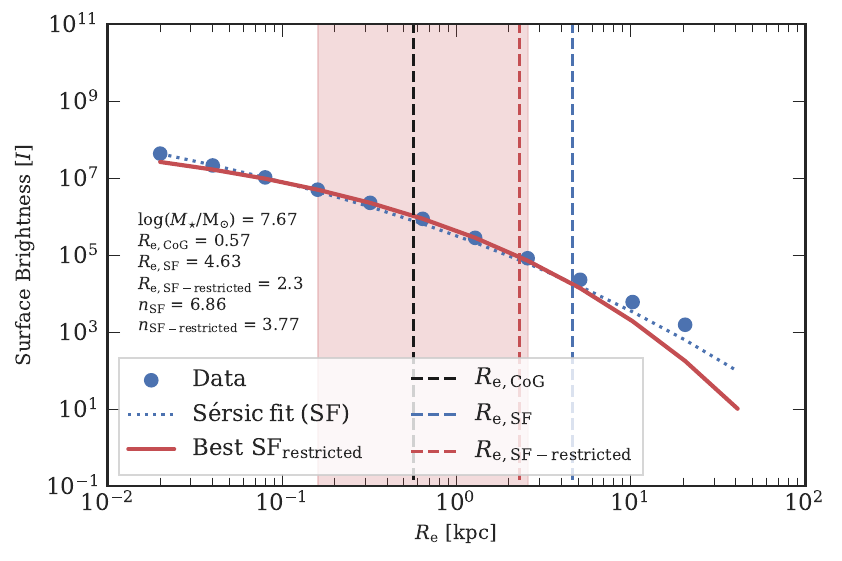}
    \includegraphics[width=0.95\linewidth]{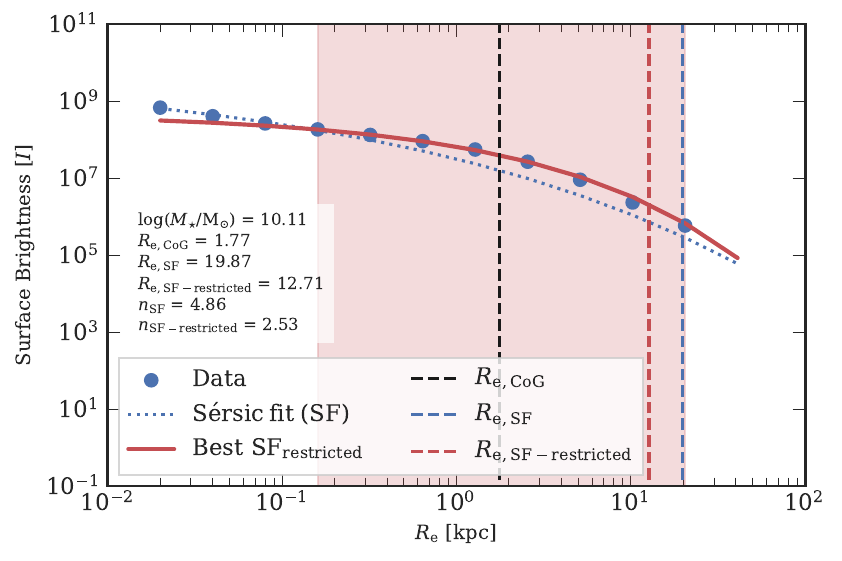}
    \caption{These two plots illustrate the surface brightness profile ($I$) of two randomly selected galaxies in the \aXXI{} model flavour. The one on the left represents the surface brightness profile of a $M_\star=10^{7.67}\:{\rm M}_{\odot}$ galaxy from MRII, while the plot on the right shows the surface brightness profile of a $M_\star=10^{10.11}\:{\rm M}_{\odot}$ galaxy from MRI. In both plots, the blue dots represent the data points from simulated galaxies, and the blue dotted line is the 1D S\'ersic fit (SF) for these data points, with the $R_{\rm e, SF}$ indicated by the vertical blue dashed line. The red-shaded region in each plot represents a restricted region used for fitting the 1D S\'ersic profile. The restricted fit is shown as the solid red line, with the $R_{\rm e, SF-restricted}$ denoted by the vertical red dashed line. The $R_{\rm e, CoG}$ derived by the curve of growth approach in \lgal{} is given by the black dashed vertical line.
    }
    \label{fig:SersicFit}
\end{figure}

Assuming the mass-to-light ratio to be unity, we calculate the cumulative light contribution from the disc and bulge regions. Furthermore, the surface brightness profile is determined by normalising the cumulative light contribution by the area of the respective ring. Using this data, we fit a 1D S\'ersic profile, as shown in \figref{fig:SersicFit}. The blue data points represent the data from the \lgal{} model. These data points exhibit a complex shape, and fitting a single power law may not capture this complexity well. Different galaxies have complex surface brightness profiles; for example, they could have a flat inner bulge or a flared outer disc. The blue dotted line in \figref{fig:SersicFit} represents the S\'ersic fit (SF) for these data points from simulated galaxies, resulting in $R_{\text{e, SF}}$ indicated by the blue dashed vertical line, while the black dashed vertical line provides $R_{\text{e, CoG}}$ derived by the curve of growth approach used by the \lgal{} model.

Upon inspection, we realised that fitting a S\'ersic profile in a restricted galactocentric radial range simplifies the complex profile, making it resemble a single power-law profile. This is achieved by excluding the bulge region and the faint end of the outer disc of the galaxy, by incorporating a limiting surface brightness cut, ${ I_{\rm limit}} = 26.6 \; [\rm mag\; arcsec^{-1}]$ \citep{Papovich2012}. Thus, we fit the surface brightness profile within $0.2 \:{\rm kpc}<R<R_{\rm I,limit}$, where $R_{\rm I,limit}$ is the galactocentric radius at which the galaxy reaches the surface brightness limit. This is highlighted as the red-shaded region, and the fit is represented by the solid red line. The red dashed vertical line denotes the effective radius ($R_{\text{e, SF-restricted}}$) for this fit. It is ensured that the red-shaded region includes at least three data points for fitting the S\'ersic profile. The plots also display the mass of the galaxy and the other fitting parameters.


\bsp	
\label{lastpage}
\end{document}